\documentclass[final,5p,times,twocolumn]{elsarticle}
\usepackage{lineno}
\modulolinenumbers[5]
\usepackage{amssymb}
\usepackage{subfigure}
\usepackage{placeins}
\usepackage{mathtools}
\usepackage{amsmath}
\usepackage{gensymb}
\journal{Journal of \LaTeX\ Templates}

\newcommand{\labr}{LaBr$_{3}$}
\newcommand{\labrs}{LaBr$_{3}$ }
\newcommand{\lacl}{LaCl$_{3}$(Ce)} 
\newcommand{\lacls}{LaCl$_{3}$(Ce) }
\newcommand{\fwhm}{\textsc{fwhm}}
\newcommand{\fwhms}{\textsc{fwhm} }
\newcommand{\knn}{k-\textsc{nn}}
\newcommand{\knns}{k-\textsc{nn} }

\journal{Nucl. Instr. and Meth. in Phys. Res. A}
\begin{document}

\begin{frontmatter}

\title{Machine Learning aided 3D-position reconstruction in large LaCl$_{3}$ crystals}
\author{J.~Balibrea-Correa\footnote{email:javier.balibrea@ific.uv.es}, J.~Lerendegui-Marco, V.~Babiano, L.~Caballero, D.~Calvo, I.~Ladarescu, P.~Olleros-Rodr{\'{\i}}guez\footnote{Present address: IMDEA Nanociencia, Madrid, Spain.}, C.~Domingo-Pardo}

\address{Instituto de F{\'\i}sica Corpuscular, CSIC-University of Valencia, Spain}
%\author{F.~Calvino, A.~Casanovas, A.~Tarife\~no-Saldivia}
%\address{Universitat Politecnica de Catalunya (UPC), Spain}

\begin{abstract}
We investigate five different models to reconstruct the 3D $\gamma$-ray hit coordinates in five large \lacls monolithic crystals optically coupled to pixelated silicon photomultipliers. These scintillators have a base surface of 50 $\times$ 50~mm$^2$ and five different thicknesses, from 10 mm to 30 mm. Four of these models are analytical prescriptions and one is based on a Convolutional Neural Network. Average resolutions close to 1-2~mm \fwhms are obtained in the transverse crystal plane for crystal thicknesses between 10~mm and 20~mm using analytical models. For thicker crystals average resolutions of about 3-5~mm \fwhms are obtained. Depth of interaction resolutions between 1~mm and 4~mm are achieved depending on the distance of the interaction point to the photosensor surface. We propose a Machine Learning algorithm to correct for linearity distortions and pin-cushion effects. The latter allows one to keep a large field of view of about 70-80\% of the crystal surface, regardless of crystal thickness. This work is aimed at optimizing the performance of the so-called Total Energy Detector with Compton imaging capability (i-TED) for time-of-flight neutron capture cross-section measurements.
\end{abstract}


\end{frontmatter}

%\linenumbers

\section{Introduction}

One of the key ingredients for the study of the formation of chemical elements heavier than iron in the stars are neutron capture cross-sections~\cite{RevModPhys.29.547,RevModPhys.83.157,CAMERON1957AJ629C}. In this regard, the direct measurement of ($n$,$\gamma$) cross sections on some specific isotopes is especially important for the understanding of the so-called slow nucleosynthesis or s-process~\cite{RevModPhys.83.157}. The two main methodologies are used to determine ($n$,$\gamma$) cross-sections, in the relevant stellar energy range, are neutron activation and time-of-flight (TOF) measurements~\cite{RevModPhys.83.157}. For TOF measurements, one of the most widely used techniques is based on Total Energy Detectors (TEDs)~\cite{PhysRev.159.1007,ABBONDANNO2004454,BORELLA2007626}. This technique has been used for many years for this kind of cross-section measurements~\cite{ABBONDANNO2004454,BORELLA2007626}. In this context and in the framework of the HYMNS project~\cite{HYMNS}, we are developing an array of $\gamma$-cameras with electronic collimation, referred to as i-TED, for ($n$,$\gamma$) cross-section TOF measurements~\cite{DOMINGOPARDO201678,BABIANO2020163228}. The bottom line in this development is to use the $\gamma$-ray imaging capability of i-TED to suppress spatially localized $\gamma$-ray backgrounds, thus enhancing the sensitivity to the neutron capture channel of interest.

i-TED is composed of four Compton modules, each of them consisting of two-planes of Position Sensitive Detectors (PSDs) operated in time-coincidence mode. Aiming at highest possible efficiency, each PSD is based on largest commercially available LaCl$_{3}$(Ce) monolithic scintillation crystals with a square base surface of 25~cm$^2$, optically coupled to Silicon Photomultipliers (SiPMs) of 8 $\times$ 8 pixels~\cite{DOMINGOPARDO201678}. Because of their high average atomic number $Z$ and high photon yield, these crystals exhibit good intrinsic efficiency and energy resolution~\cite{Olleros2018}. On the other hand, the attainable spatial resolution depends on the methodology used to derive the 3D coordinates of the $\gamma$-ray hit in the crystal. Typically, the spatial resolution and linearity response in the transverse crystal plane ($x$- and $y$-axis) deteriorate toward the edges of the detector. This bias or pin-cushion effect is reflected in a shift of the reconstructed positions toward the center of the crystal~\cite{8871159}. This effect is rather prominent for scintillation crystals where the walls are covered with reflector, thus in the peripheral region the major contribution to the optical light distribution registered by the SiPM is due to reflections in the walls.

To overcome such difficulties, different methodologies have been developed and applied in the recent years. Some of these new methodologies include maximum likelihood algorithms~\cite{4782175}, \knns algorithms~\cite{5783323,Schaart_2009,7012118}, and Voronoi diagrams~\cite{8871159}. Characterizations based on the \knn-algorithm have demonstrated great accuracy in LSO and LYSO crystals, gaining popularity in the last years. For medium-size LYSO crystals of 9~cm$^{2}$ size \knns algorithms yield rather good resolutions of around 1.1~mm \fwhms~\cite{5783323}. However, for larger crystals of 50 $\times$ 50~mm$^{2}$ as those used in this work, the best resolutions reported for the \knns method are of about 4.5~mm \fwhms at 662~keV~\cite{Liprandi2017}.

Voronoi diagrams have demonstrated effective for position calibrations in large crystals (50$\times$50$\times$15~mm$^3$)~\cite{8871159}, reporting resolutions of $\sim$2 \fwhms in the center of the LYSO crystal at 511 keV. In a previous work, we applied the analytical model of Li~\cite{Li2010} to 50 $\times$ 50~mm$^2$ \lacls crystals of different thicknesses, 10 mm, 20 mm and 30 mm~\cite{BABIANO20191}. We found resolutions ranging from 1.2 mm to $\sim$1.4 mm \fwhm. We observed that the attainable resolution deteriorates with increasing crystal thickness because of the light transport in the crystal. For this reason, studies involving large and thick crystals are difficult to find in the literature.

In the same way, many efforts have been made to determine the $z$-coordinate of the $\gamma$-ray hit or Depth of Interaction (DoI) for single-ended read detectors~\cite{BABIANO2020163228,Li2010,PANI2016,LERCHE2009624,8069405,4774303,PANI2011324,LERCHE1487684,LERCHE2005326,Bettiol_2016}. It is expected that, from the nature of the problem, much higher sensitivity and precision is achieved in the $x$,$y$ plane than in its perpendicular axis because the DoI is  deduced from the second moment of the light distribution registered by the SiPM. For instance, in~\cite{PANI2011324} the authors report 2~mm DoI resolution for a 50 $\times$ 50 $\times$ 4~mm$^{3}$ \labrs crystal using MC simulations. For a large LSO crystal (42 $\times$ 42 $\times$ 10~mm$^{3}$) resolutions of 1.9~mm are reported in~\cite{LERCHE2009624}. The DoI resolution also worsens with increasing crystal thickness. Thus, 5~mm DoI resolution is reported in ~\cite{PANI2016} for a 50 $\times$ 50 $\times$ 20~mm$^{3}$ LYSO crystal. Besides, the DoI estimation based on light width distribution~\cite{PANI2016,LERCHE2009624,8069405,6152614,Scrimger_1967} for large monolithic crystals saturates in the region close to the entrance of the crystal or small DoI values~\cite{PANI2016,Bettiol_2016}. In~\cite{Bettiol_2016}, using a 51~$\times$~51~$\times$~6~mm$^3$ \labr crystal the saturation point is reached at a DoI$\sim$2 mm. For thicker crystals, as it is described in~\cite{PANI2016}, the saturation is reached at DoI values of about 3-4 mm corresponding to 15-20\% of the crystal thickness. 

In the recent years machine learning techniques have been also applied to this problem. Their increasing popularity is due, to some extent, to the increasing computational power of modern CPUs and GPUs. Apart from the aforementioned \knns and Voronoi diagram methods, Gradient Tree Boost has been applied to LYSO crystals 32~$\times$~32~$\times$~12~mm$^{3}$, achieving 1.4 mm \fwhms in the $x$,$y$ plane and 2.15~mm \fwhms in DoI~\cite{8360486,8554136} at 511~keV. In the same way, Neural Networks have been applied with great success in different crystal shapes and thicknesses with position reconstruction resolutions close to 2 mm \fwhms in large crystals~\cite{BABIANO20191,4545078,1344371,Wang_2013,Iborra_2019,9036979}.

In this work, we explore the performance of four analytical models and one Convolutional Neural Network (CNN)~\cite{NIPS1989_293} technique in terms of resolution, linearity and compression or pin-cushion effects. We apply these five methods to five large \lacls crystals with a base surface of 50~$\times$~50~mm$^{2}$ and thicknesses of 10~mm, 15~mm, 20~mm, 25~mm and 30~mm. The motivation for such study is to optimize the background rejection capability of i-TED in ($n,\gamma$) experiments, by means of improving its imaging capability. As it is shown in Sec.~\ref{sec:Motivation}, the performance in $\gamma$-ray imaging of a Compton camera depends, with a different measure, on the accuracy in the determination of the $\gamma$-ray hit position along the transverse $x$,$y$ plane, and in DoI. Therefore, it is important to ensure the best possible spatial resolution on the relevant coordinates and eventually correct for non-linearity distortions particularly in the peripheral region of the crystal volume. In turn, an optimized imaging performance will translate, for the present application, in a better background rejection and overall detection sensitivity. 

The detectors and the experimental apparatus for the characterization measurements are described in Sec.~\ref{sec:Exp_setup}. The five models investigated here are described in Sec.\ref{sec:Light_Yield}. The results obtained for the linearity and resolution are discussed in Sec.\ref{sec:XYaxis} and Sec.\ref{sec:resolution}, respectively. Several correction methods have been studied for the non-linearity of the 3D coordinates. For the $x$ and $y$-axis, the correction is based on  a Support Vector Machine (SVM) algorithm described in Sec.\ref{sec:Machine_Learning}. On the other hand, an empirical calibration is applied to the $z$-coordinate of the crystal as described in Sec.~\ref{sec:Z_axis}. Finally, Sec.~\ref{sec:Conclusions} summarizes the main conclusions and outlook of this work.

\section{Impact of the intrinsic spatial sensitivity on the Compton imaging performance}

\label{sec:Motivation}

\begin{figure}[htbp!]
\begin{center}
\begin{tabular}{c}
  \includegraphics[width=\columnwidth]{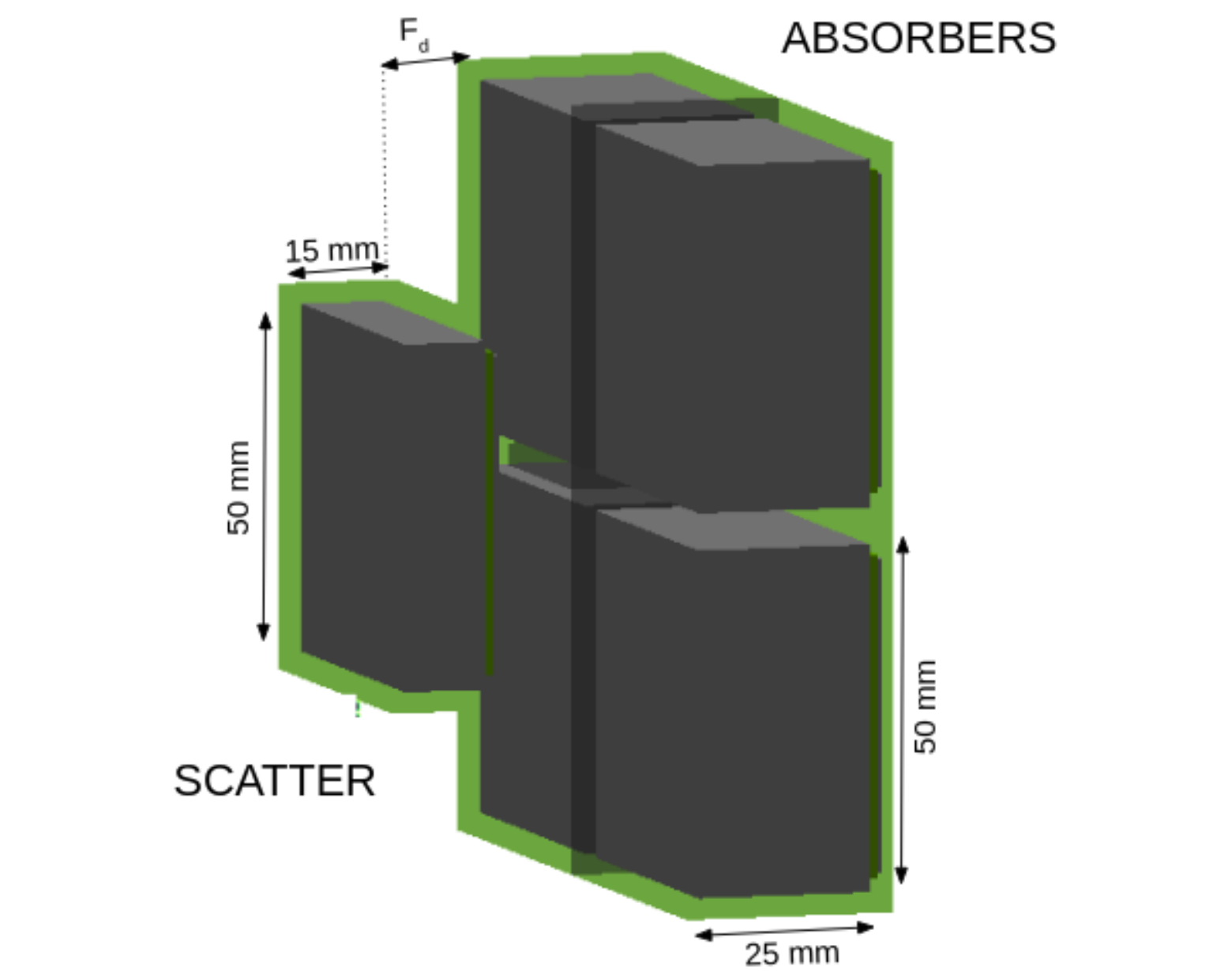}
 \end{tabular}
 \end{center}
\caption{Schematic drawing of one i-TED Compton module composed of 5 PSDs as used in the \textsc{Geant4} MC study. The dimensions of the \lacls crystals are indicated.}
\label{fig:iTED}
\end{figure}

In order to illustrate the impact of the spatial accuracy of the PSDs on the Compton imaging performance, simulations with the detailed MC model of the final i-TED detector~\cite{ProofProspectsiTED2020} were carried out using the \textsc{Geant4}~\cite{ALLISON2016186} toolkit. In the simulated setup, shown in Fig.~\ref{fig:iTED}, the focal distance (F$_{d}$) between the two detection planes, so-called absorber and scatter, was set to 15~mm. This is a reasonable value that ensures a good balance between detection efficiency and angular resolution in the upcoming neutron capture experiments~\cite{BABIANO2020163228}. Point-like $\gamma$-ray sources of 511~keV and 1 ~MeV with an incident angle of 0$\degree$ and 50$\degree$ were simulated in this MC sensitivity study. 

\begin{figure*}[hbtbp!]
\begin{center}
\begin{tabular}{c}
  \includegraphics[width=0.666\columnwidth]{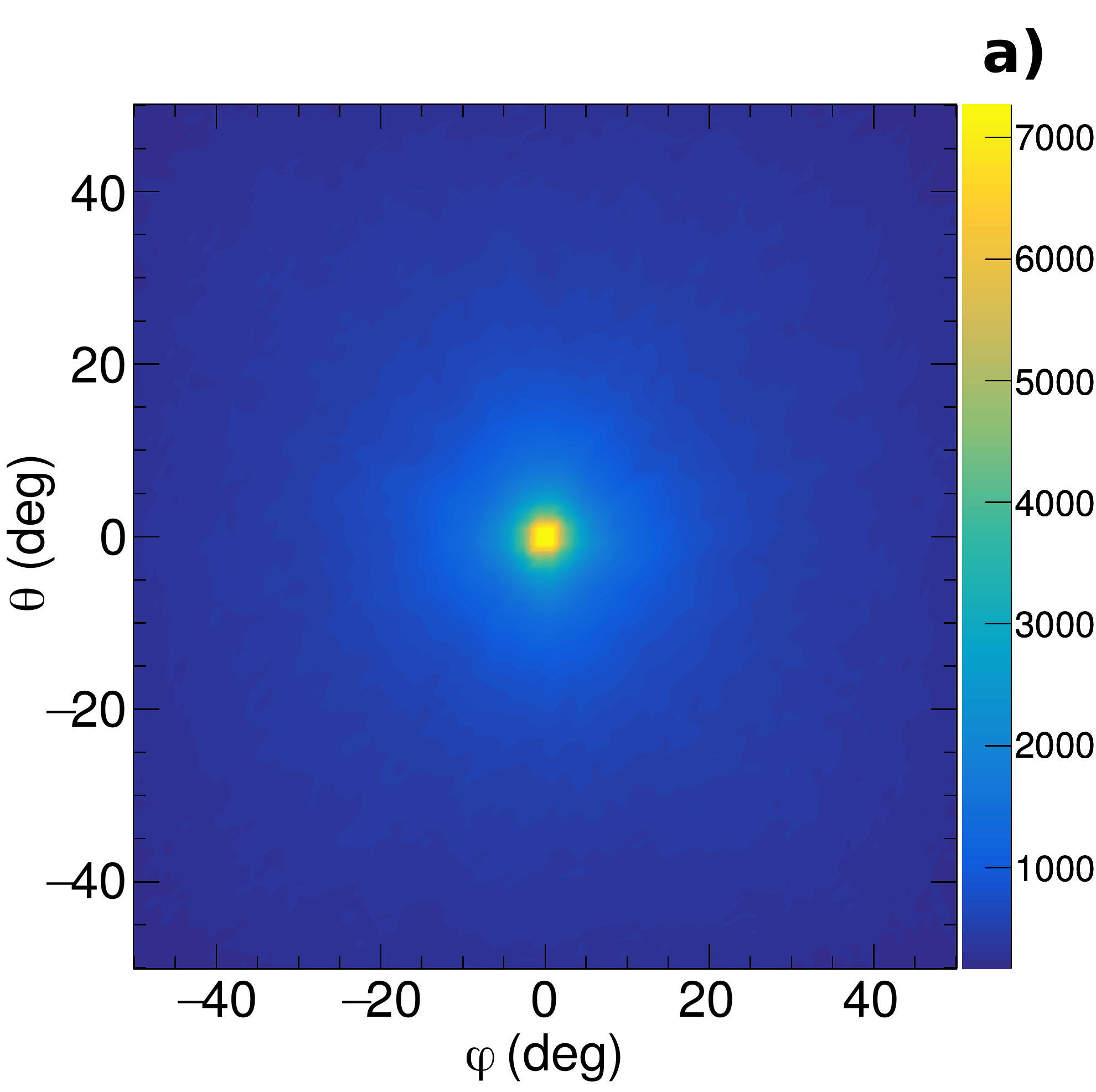} 
  \includegraphics[width=0.666\columnwidth]{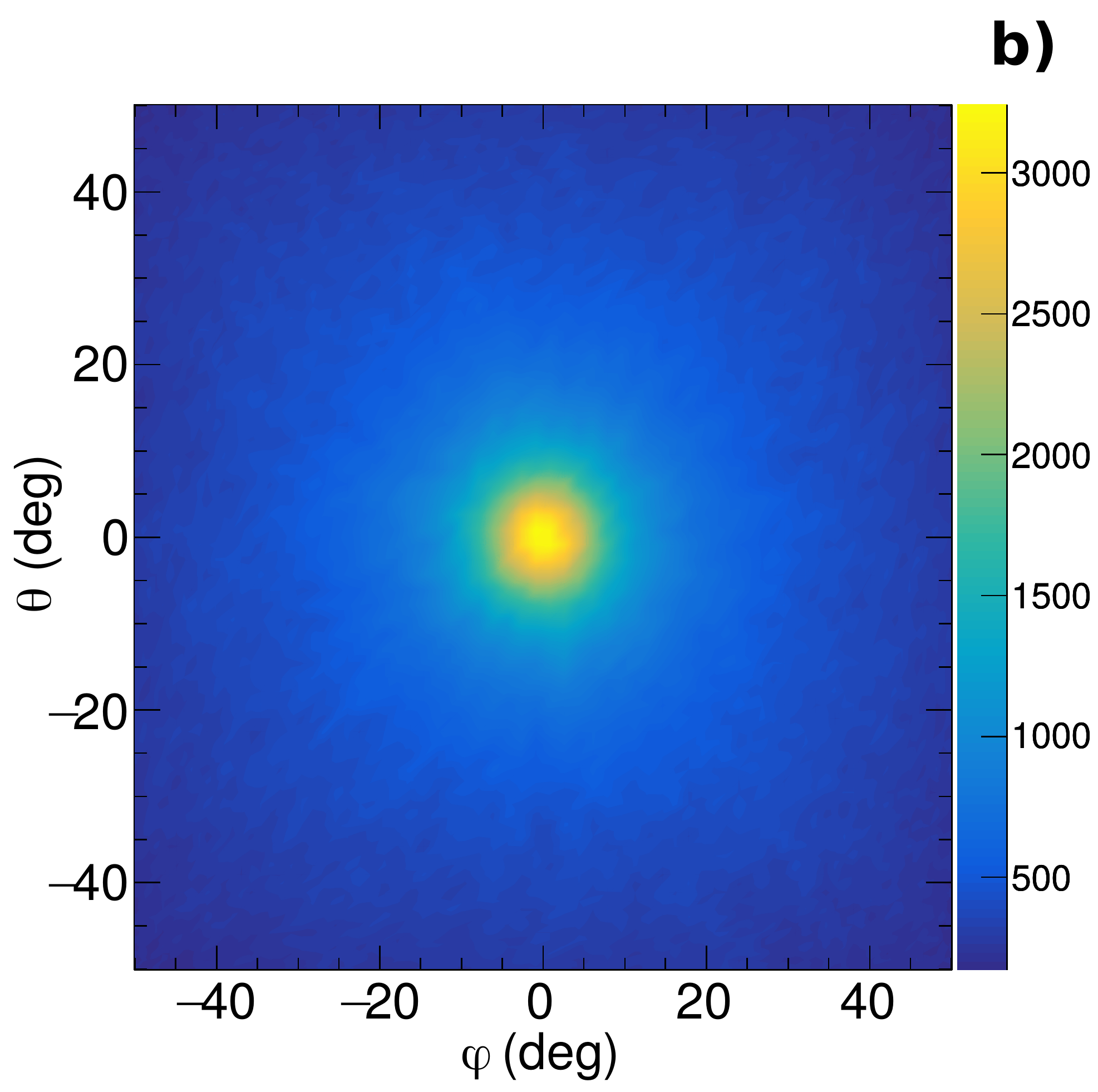}
  \includegraphics[width=0.666\columnwidth]{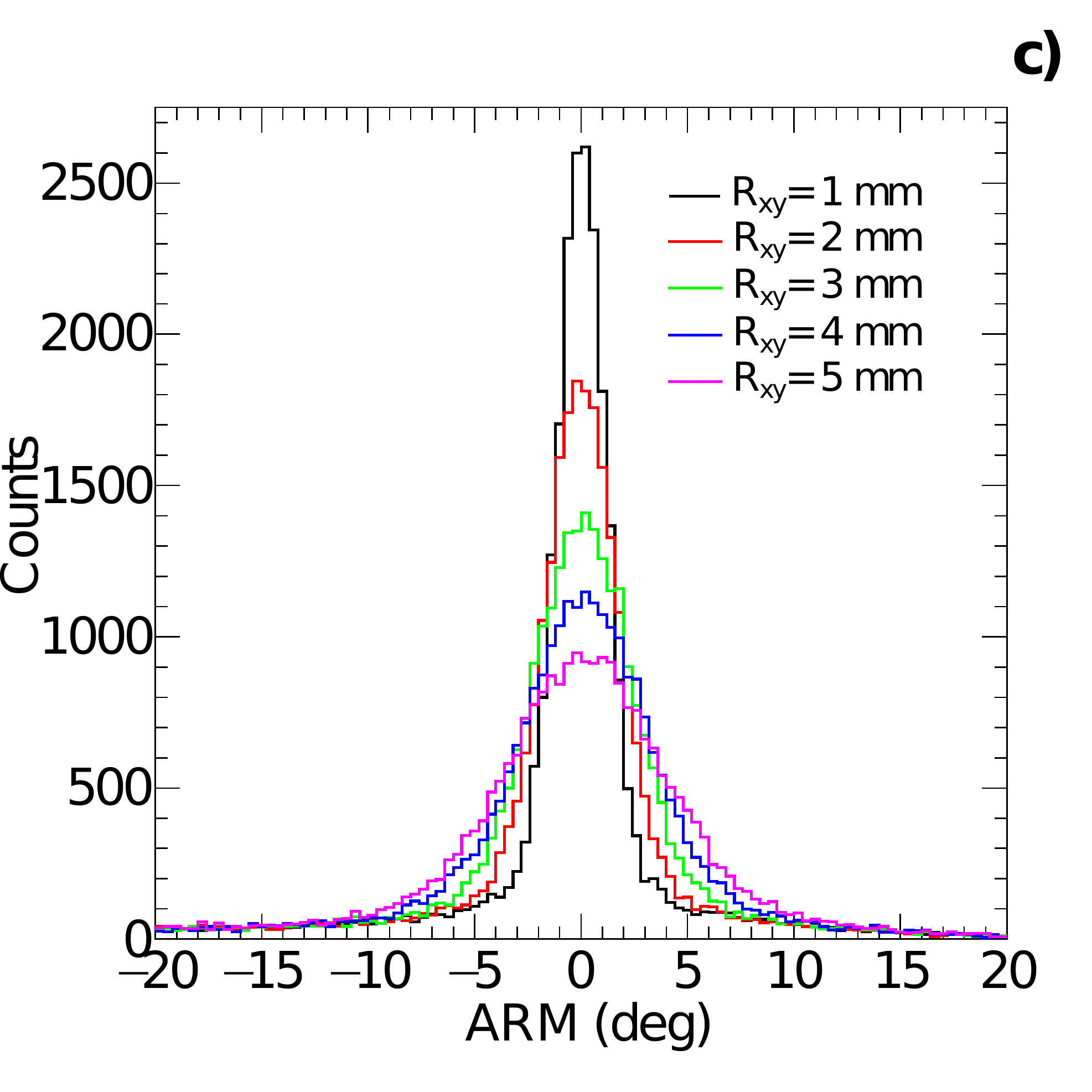}
\end{tabular}
 \end{center}
\caption{Compton images in spherical coordinates of a 1~MeV gamma-ray source at an incident angle of 0$\degree$ with R$_{xy}$=1~mm (panel a)), R$_{xy}$=5~mm (panel b)). The corresponding ARM distributions as a function the resolution in the xy plane R$_{xy}$ are shown in panel c). }
\label{fig:Images_iTED}
\end{figure*}

In order to model the experimental resolution in the position reconstruction, the real $\gamma$-ray interaction points from the MC simulations were distorted with a varying Gaussian resolution. An energy resolution of 4.5\% \fwhms at 500 keV, representative of high-resolution inorganic scintillation crystals~\cite{BABIANO2020163228}, was included in the modelling of the experimental effects. Last, a realistic low energy threshold of 100 keV was applied for the deposited energy recorded in each crystal. After the inclusion of these experimental effects, the Compton images were reconstructed with the simple back-projection method~\cite{WILDERMAN1998} selecting only those events in which the $\gamma$-ray energy is fully deposited between the scatter and absorber planes. To illustrate the effect of the position resolution of the PSDs, Fig.~\ref{fig:Images_iTED} shows the Compton images of the 1~MeV source at 0$\degree$ reconstructed with a resolution (\fwhm) of the PSD in the xy-plane, R$_{xy}$, of 1~mm and 5~mm. In both cases, the resolution in DoI was fixed to R$_{doi}$ = 1~mm.

To evaluate the impact of the position reconstruction in the final angular resolution of i-TED, we used the Angular Resolution Measure (ARM)~\cite{Hosokoshi2019}. The ARM is defined as the difference between the geometrical scattering angle, determined from the interaction point in the two detection planes and the source position, and the scatter angle calculated using the Compton formula:
\begin{equation}
    \cos(\theta)=1-\frac{m_{e}c^{2}E_{s}}{(E_{s}+E_{a})E_{a}}.
\end{equation}
In the formula, $m_{e}c^{2}$ is the invariant mass of the electron and $E_{s}$ and $E_{a}$ are the deposited energies in the scatter and absorber detection planes of the Compton camera, respectively. The ARM distributions for a centered 1 MeV $\gamma$-ray source are displayed in  panel c) of Fig.~\ref{fig:Images_iTED} for different spatial resolutions in the $x$ and $y$ coordinates.
\begin{figure}[htbp!]
\begin{center}
\begin{tabular}{c}
  \includegraphics[width=\columnwidth]{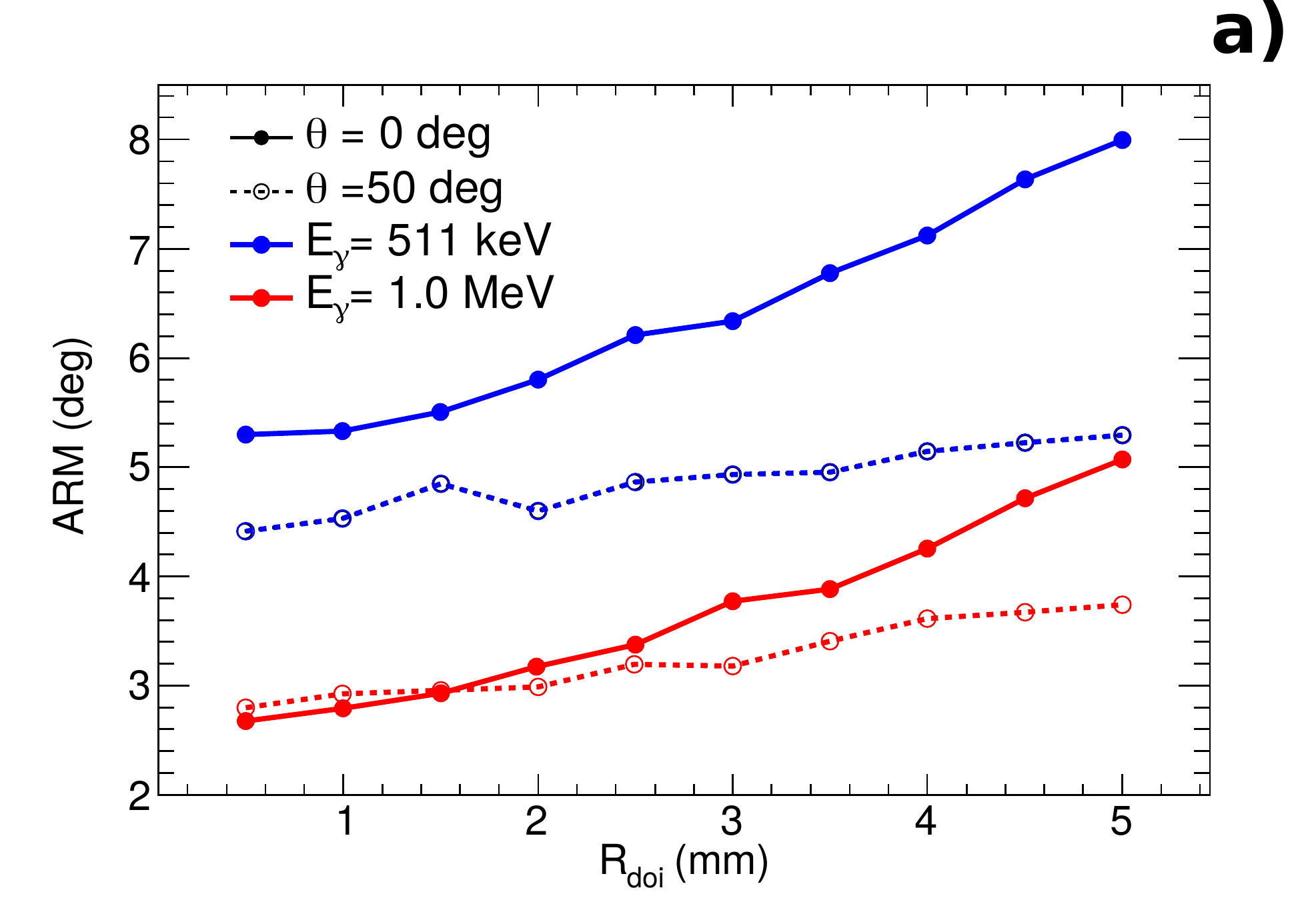} \\
  \includegraphics[width=\columnwidth]{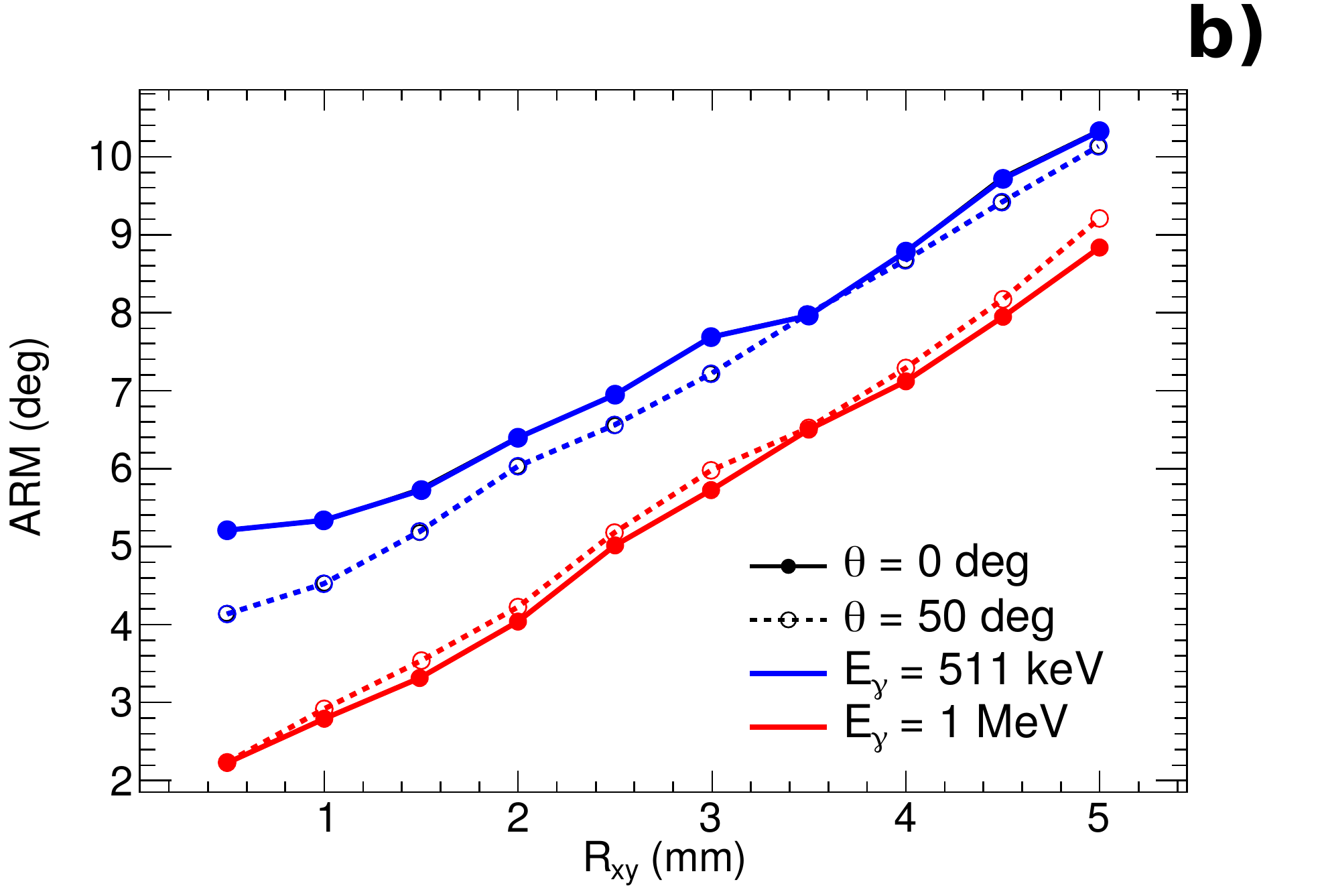} \\
\end{tabular}
 \end{center}
\caption{Impact of the position reconstruction resolution in the angular resolution of i-TED for 511 keV (blue) and 1 MeV (red) sources at $\theta$~=~0$\degree$ (solid) and 50$\degree$ (dashed). Panel a): ARM (\fwhm) as a function of the resolutions in the coordinates parallel to the SiPM (R$_{xy}$). Panel b): ARM (\fwhm) as a function of the resolutions in DoI R$_{doi}$.}
\label{fig:ARM_iTED}
\end{figure}
 The Figure of Merit (FoM) to quantify the angular resolution of i-TED is the width (\fwhm) of the ARM distributions. The impact of the position reconstruction on this FoM is presented in Fig.~\ref{fig:ARM_iTED}. This figure shows the results obtained by varying the resolution in DoI on one hand, and the resolution in the $xy$-plane on the other hand. These results demonstrate the large influence of the PSD resolution in the transverse crystal plane, $R_{xy}$, on the imaging performance of i-TED. Our quantitative study indicates that variations of the spatial $x,y$ resolution by a factor of five, have a similar impact on the ARM for 1~MeV $\gamma$-rays. For 511~keV $\gamma$-rays the impact on the ARM is of a factor of two. On the other hand, the DoI resolution $R_{doi}$, for sensible values between 1~mm and 5~mm, has a much less significant effect on the ARM. Finally, it is worth mentioning the larger influence of the DoI resolution for centered sources than for peripheral $\gamma$-ray events. Also, the angular resolution for 1~MeV is always better than for 511~keV as a consequence of the better energy resolution.
 
Using the same sensitivity study, we have analyzed the worsening of the final i-TED performance associated to the pin-cushion or non-linear reconstruction in the crystal edges. The results indicate that the compression effect leads to a significant worsening of the angular resolution, being especially critical for large angles of incidence and low $\gamma$-ray energies. The simplest solution to avoid the compression is discarding the non-linear range of the crystal for the analysis. However, this has a large cost in terms of efficiency, which is a critical magnitude for the final purpose of i-TED. For this reason we have developed an alternative solution based on a SVM correction, that will be presented in Sec.~\ref{sec:XYaxis} and further discussed in Sec.~\ref{sec:Conclusions}.
 
\section{Experimental setup}\label{sec:Exp_setup}

 All \lacls crystals are encapsulated in a 0.5 mm thick aluminum housing, isolated from the crystal itself with a 1 mm thick seam gum. The base of the crystal is optically coupled to a fused-silica glass window of 2~mm thickness while the rest of the crystal surface was sealed in a diffusive reflector made from polytetrafluoroethylene (PTFE). For further details on the crystals the reader is referred to Refs.~\cite{BABIANO20191,Olleros2018}. Each scintillation crystal is coupled with silicon grease (BC-630) to a 8 $\times$ 8 pixels, 50.44 $\times$ 50.44 mm$^{2}$ silicon photomultiplier array (SiPM) from SensL (ArrayJ-60035-65P-PCB)~\cite{SENSL2020}. The \lacls crystals, together with the SiPM, are encapsulated into a light-tight black polylactic acid (PLA) plastic housing of 2.80~mm thickness. The SiPMs are biased and read out by means of the PETsys Front-End Board D version 2 (FEB/D-1024)~\cite{DIFRANCESCO2016194}. The reverse bias voltage chosen is of +4 V beyond the nominal breakdown value of +25~V, which yields a quantum photo-detection efficiency of around 50\% at 420~nm. The SiPM is connected to a 64-channel front-end ASIC chip named TOFPET2, which performs the readout, digitization and signal processing continuously. Every time one of the 64 channels in each SiPM exceeds certain discriminator threshold, a pixel event is created containing the corresponding channel number, timestamp, and integral of the signal. The pixel event information is sent via Samtec EQCD High-Speed flat cables to the FEB/D$\_$v2 motherboard for further processing. At this point, all the information is sent via a fast Gigabit Ethernet link to the acquisition computer running the control system. The acquired data is stored in binary files for posterior event-building and time-coincident event selection.

The individual pixel events for the same \lacls crystal are grouped in $\gamma$-ray hit events applying a time-window of 100~ns, as described in previous works~\cite{BABIANO2020163228}. The resulting light-yield pixel-distribution, for each detected gamma-ray event, is used for the position reconstruction associated to that $\gamma$-ray interaction by means of the algorithms described in the next section.
\begin{figure}[htb!]
\begin{center}
  \includegraphics[width=\columnwidth]{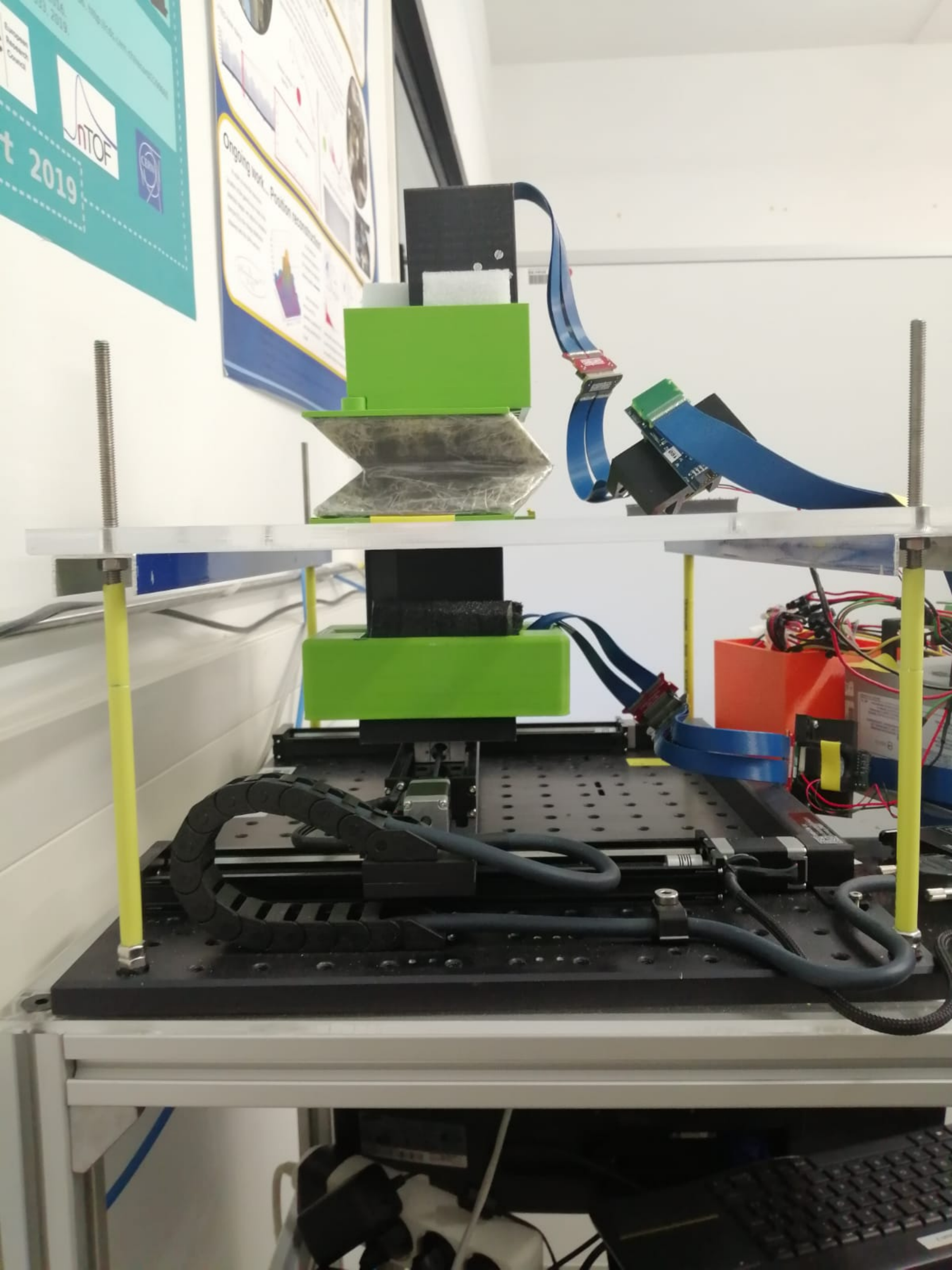}
 \end{center}
\caption{Experimental setup used in this work for the spatial characterization of the PSDs. On top of the experimental setup an ancillary \lacls crystal is placed in a subsidiary fixed position. The $^{22}$Na radioactive source is sandwiched between two tungsten collimators (top green box), which is sitting on top of a 5~cm thick lead collimator. The structure was accommodated on a Plexiglas table of 10~mm thickness. Beneath this assembly, the \lacls crystal under study is placed over a small movable platform (bottom green box) attached to the motorized $xy$-gantry system.}
\label{fig:experimental_setup}
\end{figure}

The systematic spatial characterization in the $x$-$y$ transverse crystal plane and along its $z$ axis (depth) is carried out using a collimated beam of $\gamma$-quanta in combination with an $x,y$-positioning table. This experimental setup is shown in the picture of Fig.~\ref{fig:experimental_setup}. Two \lacls crystals are operated in time-coincidence mode with a 10~ns coincidence window. For the analysis we use only full-energy deposition of the 511~keV annihilation $\gamma$-rays emitted from a point-like $^{22}$Na-source (2~MBq activity). The position sensitive detector under characterization is sitting in a small movable platform at the bottom of the experimental setup, whereas the collimating system, the $^{22}$Na-source, and the subsidiary detector are placed and fixed on top of a Plexiglas table. The $x,y$-gantry which drives the movable platform with the detector under study has been upgraded with respect to the one used in our previous work~\cite{BABIANO20191}. The new system corresponds to the T-G-LSM200A200A model from Zaber Technologies~\cite{Zaber_Gantry}, which has a microstep precision of 0.047625~$\mu$m and repeatability of 3~$\mu$m~\cite{Zaber_Gantry}.
The $^{22}$Na-source is sandwiched in a collimating structure under the subsidiary detector to achieve a pencil beam in the direction perpendicular to both crystal surfaces. The upper part of this collimator is made of a tungsten parallelepiped with a central hole of 1~mm and 30~mm thickness. Under the radioactive sample a circular tungsten collimator is placed with a central hole of 1~mm and 24~mm effective thickness. An additional lead collimator of 5.5~cm thickness is placed beneath the collimating structure (see also Fig.~\ref{fig:MC_setup} for details). The aperture of this lead collimator is of 3~mm diameter. In total, accounting for all the collimator's thicknesses and distance from the bottom detector to the plexiglas table, the distance between the front face of both \lacls crystals in the experimental setup is 26~cm.

Two different experimental configurations were used for the characterization of the position response in the $xy$-transverse plane, and across the crystal depth. In the first one, the front face of the bottom \lacls crystal was looking to the top of the experimental setup. A mesh of individual irradiations with 1~mm pitch was used, covering a total area of 60 $\times$ 60 mm$^{2}$ that corresponds to 3600 individual measurements per crystal. Such scan surface is significantly larger than the crystal size, which allowed us to reliably determine the effective transverse crystal size and its center. In the second scan configuration the bottom crystal under study was placed with one of its lateral sides looking to the top of the experimental setup. This enabled a $z$-axis or DoI characterization. In this case, only the central part of the crystal was irradiated through a mesh of individual irradiations, spaced on a pitch of 0.5~mm. 

\section{Light yield models for $\gamma$-ray interaction position reconstruction}\label{sec:Light_Yield}

From the simplest center of gravity algorithms~\cite{doi:10.1063/1.1715998,4324123} to the sophisticated solid angle based models~\cite{Li2010,PANI2016,LERCHE2005326,SHI2019117} there exist many analytical light-yield models available in the literature for the position reconstruction of a $\gamma$-ray interaction in a monolithic PSD. In this work, we have selected and compared a few of them. In all these analytical models, the scintillation light collected at a given detection pixel $ij$, L$_{ij}$, is described by an analytical formula, which is a function of the pixel position (x$_{i}$,y$_{j}$), the 3D coordinates of the $\gamma$-ray interaction position ($x_{\circ},y_{\circ},z_{\circ}$) and other dependencies specific of each individual model. It is important to note that $z_{\circ}$ refers to the perpendicular distance starting from the SiPM surface to the $\gamma$-ray interaction position. This coordinate is related to the DoI by ${DoI}$ = $t - z_{\circ}$, where $t$ refers to the crystal thickness. The analytical models used in this work are listed in the following.

\begin{itemize}
    \item {\bf{Solid angle model}}: Assuming that the light collected in a specific pixel is proportional to the subtended solid angle, the detected light can be described as~\cite{Li2010}:
    \begin{equation}\label{eq:sam}
        L_{ij}=A_{\circ}\Omega_{i,j} + \sum{L^{\prime}_{ij}} + C
    \end{equation}
    where A$_{\circ}$ represents the total amount of scintillation photons produced in a single gamma-ray interaction, $\Omega_{i,j}$ is the exact solid angle subtended by the rectangular slit of the pixel~\cite{GOTOH1971485}, $L^{\prime}_{ij}$ are the specular sources produced by the reflection of the light in the detector walls and C is a term representing the diffusive light arriving in all the detection cells~\cite{Li2010}. After a first exploration of this model with our data, it was found that for the \lacls crystals used in this work the contribution from the specular sources was negligible and thus, only the term for the diffusive light was kept. The detected light yield was therefore described as
   
    \begin{equation}
        L_{ij}\simeq A_{\circ}\Omega_{i,j} + C.
    \end{equation}
    
    This result was in contrast with the crystal design specifications of diffusive reflector on the base and specular reflector on the walls. Also, in our previous spectroscopic study for the same crystals~\cite{Olleros2018} we found a 70\% contribution of specular sources and a 30\% diffusive contribution. However, in terms of position reconstruction, the latest approximation clearly led to better results than the expression~(\ref{eq:sam}).
   
    Hereafter this model will be referred to as {\it{Solid angle}} model.
   
    \item {\bf{Gaussian Model}}: In this common empirical model~\cite{8069405,4774303} the light collected at a specific pixel $ij$ is given by
   
    \begin{equation}
        L_{i,j}=A_{\circ}\exp{[-f]},
    \end{equation}
    %(x_{i}-x_{\circ},y_{j}-y_{\circ},\sigma_{x},\sigma_{y})
    where $f$, is defined as:
   
    \begin{equation}
        f(x_{i}-x_{\circ},y_{j}-y_{\circ})=\frac{1}{2}\left[\frac{(x_{i}-x_{\circ})^{2}}{\sigma^{2}_{x}}+\frac{(y_{j}-y_{\circ})^{2}}{\sigma^{2}_{y}}\right].
    \end{equation}
   
   In the function, the $\sigma_{x}$ and $\sigma_{y}$ variables are the $x$ and $y$ variances of the detected light distribution, respectively. Note that for this specific model, the $z_{\circ}$ coordinate is not obtained directly from the fit. As a result, this quantity has to be derived and calibrated from the width of the distribution, i. e. using a function of the type $z_{\circ}$~=~$F\left(\sigma^{2}_{x},\sigma^{2}_{y}\right)$ as it is described in~\cite{Mikiko2010}.
   
    In the next sections this approach will be referred as {\it{Gauss}} model.

    \item {\bf{Gaussian model with correlations in both $x$- and $y$-axis}}: Here we introduce a Gaussian model similar to the one described above, which includes also correlations between the $x$- and $y$- axis to account for the contribution of reflections in the walls of the detector. This effect is especially notorious for interactions in the peripheral region of the crystals. The light yield registered by the SiPM in this model is described by:
   
    \begin{equation}
        L_{i,j}=A_{\circ}\exp{[-g]}
    \end{equation}
    where $g$ is defined as
    \begin{equation}\label{eq:g}
        g(x,y)=\frac{1}{2}\left[\frac{(x_{i}-x_{\circ})^{2}}{\sigma^{2}_{x}}+\frac{(y_{j}-y_{\circ})^{2}}{\sigma^{2}_{y}}-\frac{\rho(x_{i}-x_{\circ})(y_{j}-y_{\circ})}{\sigma_{x}\sigma_{y}} \right].
    \end{equation}
   
    In the model, $\rho$ is the so-called correlation coefficient defined as the ratio between the covariance and the variance of the detected light distribution, $\rho=Cov(x,y)/\sigma_{x}\sigma_{y}$. As in the previous model, the z$_{\circ}$ coordinate cannot be directly obtained from the fit. Thus, it has to be deduced from complementary methods, such as the width of the distribution: $z_{\circ}$=$F\left(\sigma^{2}_{x},\sigma^{2}_{y},\rho \right)$, which requires a further calibration. In the next sections this approach will be named {\it{Gauss($\rho$)}} model.
   
    \item {\bf{Cauchy model with correlations in both $x$- and $y$-axis}}: As described by Scrimger and Baker in 1967, the spatial distribution for the scintillation light yield in a monolithic crystal can be described by a Cauchy distribution~\cite{Scrimger_1967}. This function can be used to derive an empirical formula for a DoI calibration~\cite{PANI2016,6152614}. In the present work we have modified the original expressions in order to introduce correlations in the $x$ and $y$-axis to account for peripheral light-reflection effects. In this model, the light yield is described as:
    \begin{equation}
        L_{i,j}=\frac{A_{\circ}}{[1+g]^{3/2}}
    \end{equation}
 where $g$ is the aforementioned function~(\ref{eq:g}). This model will be referred to as {\it{Lorentzian($\rho$)}} model.
\end{itemize}

It is worth to mention that, in order to reconstruct the position interaction, the light-yield distribution registered by the SiPM was normalized to the QDC-integral of all the pixels fired in coincidence for the $\gamma$-ray event, independently of the model used. 

As the model complexity increases, larger computational power is required to reconstruct the $\gamma$-ray interaction position in a reasonable amount of time. For this reason, the aforementioned models were included in the Gpufit framework. In particular, we used a GPU-accelerated CUDA~\cite{10.1145/1365490.1365500} implementation of the Levenberg-Marquardt algorithm~\cite{PRZYBYLSKY2017}. In this way, the performance of the fitting procedure was enormously improved reaching event processing rates that go from $\sim$3.5~kFit/s to $\sim$35~kFit/s for NVIDIA GeForce 920M and NVIDIA GTX 1080 Ti, respectively.   
When compared to a more conventional single-thread CPU based approach, this represents an improvement in reconstruction-time efficiency of a factor between 3.5 and 35, depending on the GPU model. The speed-up factor obtained here is comparable to other works based on the Maximum Likelihood approach~\cite{LERCHE2009359} for the position reconstruction. Such enhancements in processing efficiency and speed can be of interest for applications, such as medical imaging, where an online real-time position monitoring becomes relevant.

\begin{figure}
\begin{center}
  \includegraphics[width=\columnwidth]{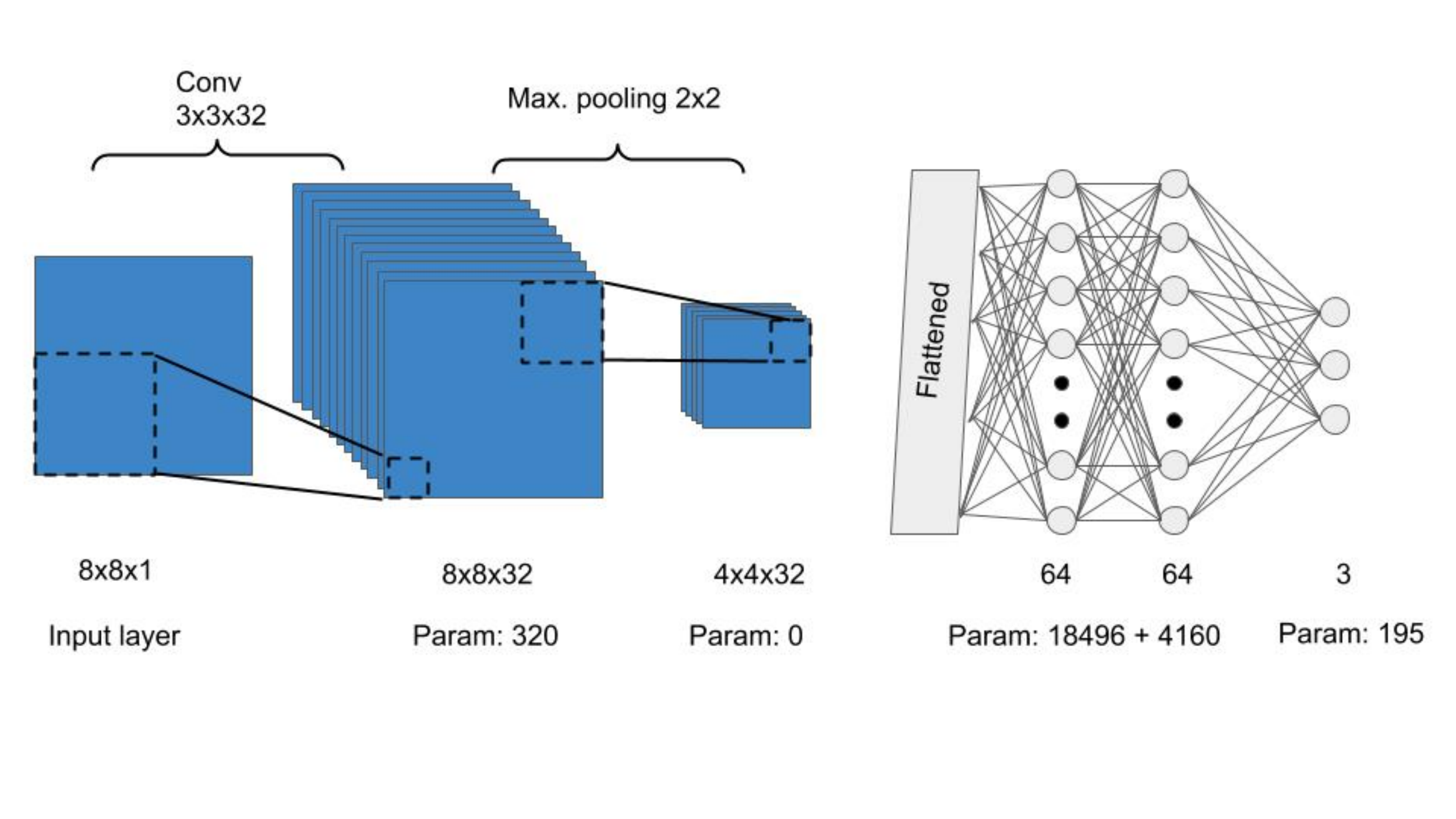}
 \end{center}
\caption{Summary of the CNN architecture used in this work. Beneath each individual layer is displayed the output tensor shape and the number of trainable parameters.}
\label{fig:Convolutional}
\end{figure}

In addition to the analytical models listed before, a deep learning algorithm was also investigated in this work and its performance benchmarked against the analytical models. In the last years, a growing interest toward the use of this type of algorithms for the $\gamma$-ray position reconstruction in monolithic crystals has become apparent~\cite{Iborra_2019,BRUYNDONCKX2007304,MATEO2009366}. In our previous work~\cite{BABIANO20191} we also investigated and compared the performance of analytical models and a linear neural network (LNN) method. 
%In the literature, CNN has been successfully applied to 16x16 pixelated LYSO crystals with size 1.4×1.4×20 mm$^{3}$ crystals coupled to SiPM~\cite{9036979}. They have demonstrated that outperform the classical Anger logic algorithm for the position reconstruction in the plane paralell to the SiPM, especially at the edge of the crystal. 
In the present study we have implemented a more sophisticated CNN approach that outperforms LNN in image processing. In our case the image corresponds to the charge-distribution registered in the pixelated SiPM. The CNN was coded in the {\it{tensorflow}} and {\it{Keras}} deep-learning API~\cite{tensorflow2015-whitepaper,chollet2015keras}. The architecture for the CNN is sketched in Fig.~\ref{fig:Convolutional}, which represents the result of several trials and errors optimizing the performance in accuracy. An {\it{Input}} layer takes the 8 $\times$ 8 normalized charge distribution from the SiPM on an event-by-event basis. This layer is followed by a 3 $\times$ 3 {\it{Convolutional}} layer with 32 filters. A 2 $\times$ 2 {\it{Maximum pooling}} layer is connected to the {\it{Convolutional}} layer, reducing the size of the images produced down to 4 $\times$ 4 pixels. This layer is then flattened and fully connected to two {\it{Dense}} layers made of 64 neurons each. The output is given by a {\it{Dense}} layer with three neurons for the reconstructed ($x_{\circ},y_{\circ},z_{\circ}$) coordinates of the registered $\gamma$-ray interaction. 

To account for the possible non-linearity in the SiPM response Rectified Linear Unit (ReLU) activation functions are used in the the {\it{Convolutional}} {\it{Dense}} layers, except for the output layer. As a result, the CNN model has 23171 trainable parameters. The model was compiled using as loss function the {\it{mean squared error}} and the stochastic {\it{Adam}} optimizer~\cite{kingma2017adam}.

The training and data validation data used for the CNN were made based on MC simulations of the light production and transport in the \lacls crystal, which is described in our previous work~\cite{Olleros2018}. The simulation was enclosed in the Multithread version of \textsc{Geant4} to reduce the large computational time required for such simulations. The MC application was run in an Intel(R) Xeon(R) Gold 6134 CPU (3.20~GHz) which boosted the simulation capability to $\sim$28000~events/s. This allowed us to achieve sufficient statistics for the CNN training, while reducing the amount of time needed. More than three million 511~keV $\gamma$-ray events were used for each \lacls crystal studied in this work. The true interaction position ($x_{\circ},y_{\circ},z_{\circ}$) from the MC simulations was used for the training of the model. For several-hit interactions, the average value for the $\gamma$-ray interactions was used instead. The MC simulated positions were normalized to the 50~mm crystal basis axis. Training the deep-learning model in this way, the weight for fitting the three-axis of the crystal are equally accounted for. This deep-learning model will be labeled as {\it{Keras}} in the following sections.

\begin{figure}
\begin{center}
    \includegraphics[width=\columnwidth]{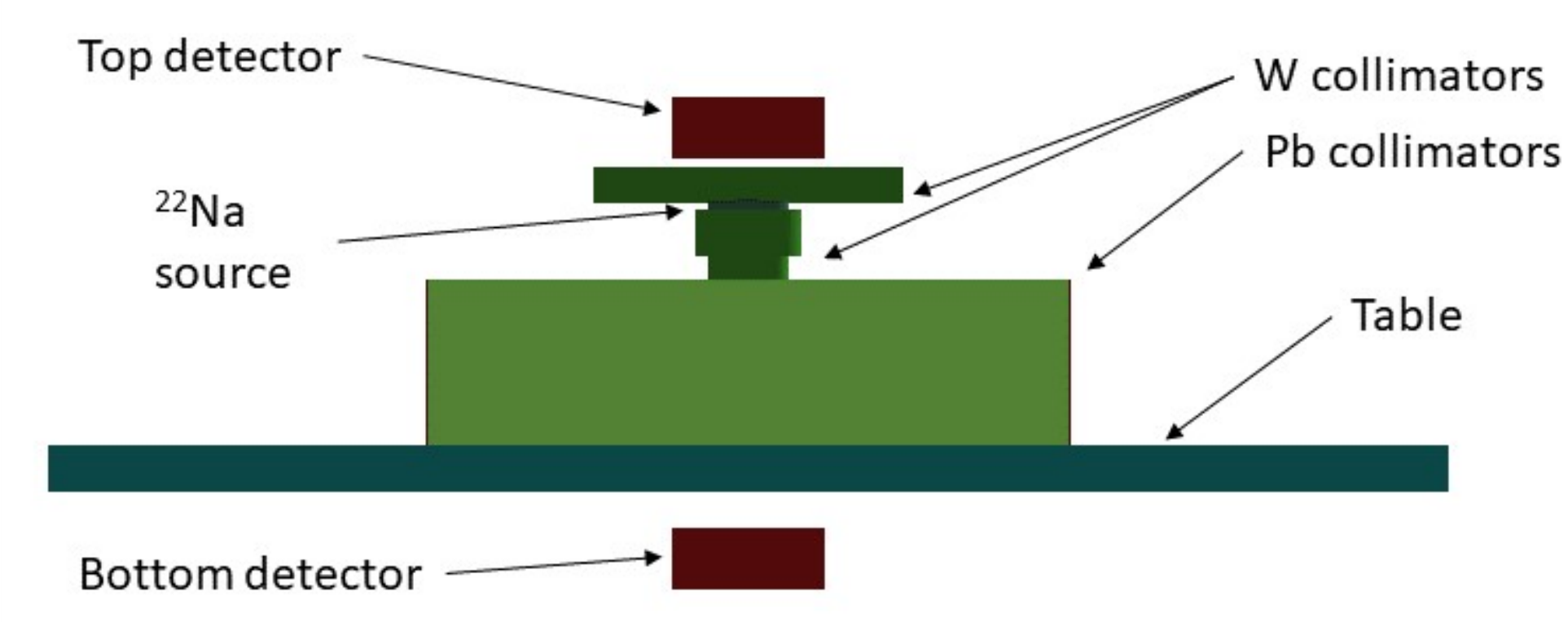}
    \caption{MC realization of the experimental setup implemented using the \textsc{Geant4} toolkit~\cite{ALLISON2016186}. The different collimators are shown by the green color pieces. The \lacls crystals are displayed in red.}
    \label{fig:MC_setup}
\end{center}
\end{figure}

Apart from this first MC light-transport study for training the CNN, a second MC work was carried out. In this second MC study a detailed realization of the experimental setup was implemented using again the \textsc{Geant4} toolkit~\cite{ALLISON2016186} as it is shown in the sketch of Fig.~\ref{fig:MC_setup}. In this case, instead of simulating the scintillation photons only the electromagnetic interactions of the collimated $\gamma$-ray beam in the crystals were calculated. Still, due to the large collimation length, the \textsc{Geant4} multithread version was preferred to ensure enough statistics in a reasonable amount of time. The aim of this MC work is to determine the interplay between the intrinsic detector spatial resolution and the overall (measured) width, as it has been performed in a previous work~\cite{BABIANO20191}. The results of these MC calculations will be used in Sec.~\ref{sec:XYaxis} in order to deconvolve the beam divergence and determine the intrinsic spatial resolution.
%\section{Results from the x and y axis characterization}\label{sec:XYaxis}

\section{Linearity in the transverse crystal plane}\label{sec:XYaxis}

In order to perform a precise spatial characterization of the \lacls crystals, including the true linearity and the real sensitive volume of the detector, it is important to determine the effective base surface of each crystal. In turn, this measure will determine the real size of the active volume. This quantity was determined from the 60 $\times$ 60~mm$^2$ scan (see Sec.\ref{sec:Exp_setup}) by defining the effective crystal edge as the position where the counting statistics in the photopeak reach 50\% of the maximum value registered in coincidence by the detectors of the experimental setup. These values are reported in Tab.~\ref{tab:effective area}. The last column shows the ratio between the effective and the ideal area of 50 $\times$ 50~mm$^{2}$.
\begin{table}
    \centering
    \begin{tabular}{c c c}\hline
        Crystal thickness & Effective area & Ratio \\
         (mm) & (mm$^{2}$) & (\%) \\ \hline \hline

        10 & 48 $\times$ 48 & 92\\
        15 & 48 $\times$ 48 & 92\\
        20 & 47 $\times$ 47 & 88\\
        25 & 49 $\times$ 48 & 94\\
        30 & 48 $\times$ 49 & 94\\ \hline
    \end{tabular}
    \caption{Effective area measured for all five \lacls crystals studied in this work (see text for details).}
    \label{tab:effective area}
\end{table}
In all cases the effective areas have reasonable values, with minimum and maximum ratios of 88\% and 94\% for the 20~mm and 25-30~mm thick crystals.
\begin{figure*}
\begin{center}
\begin{tabular}{c c}
  \includegraphics[width=\columnwidth]{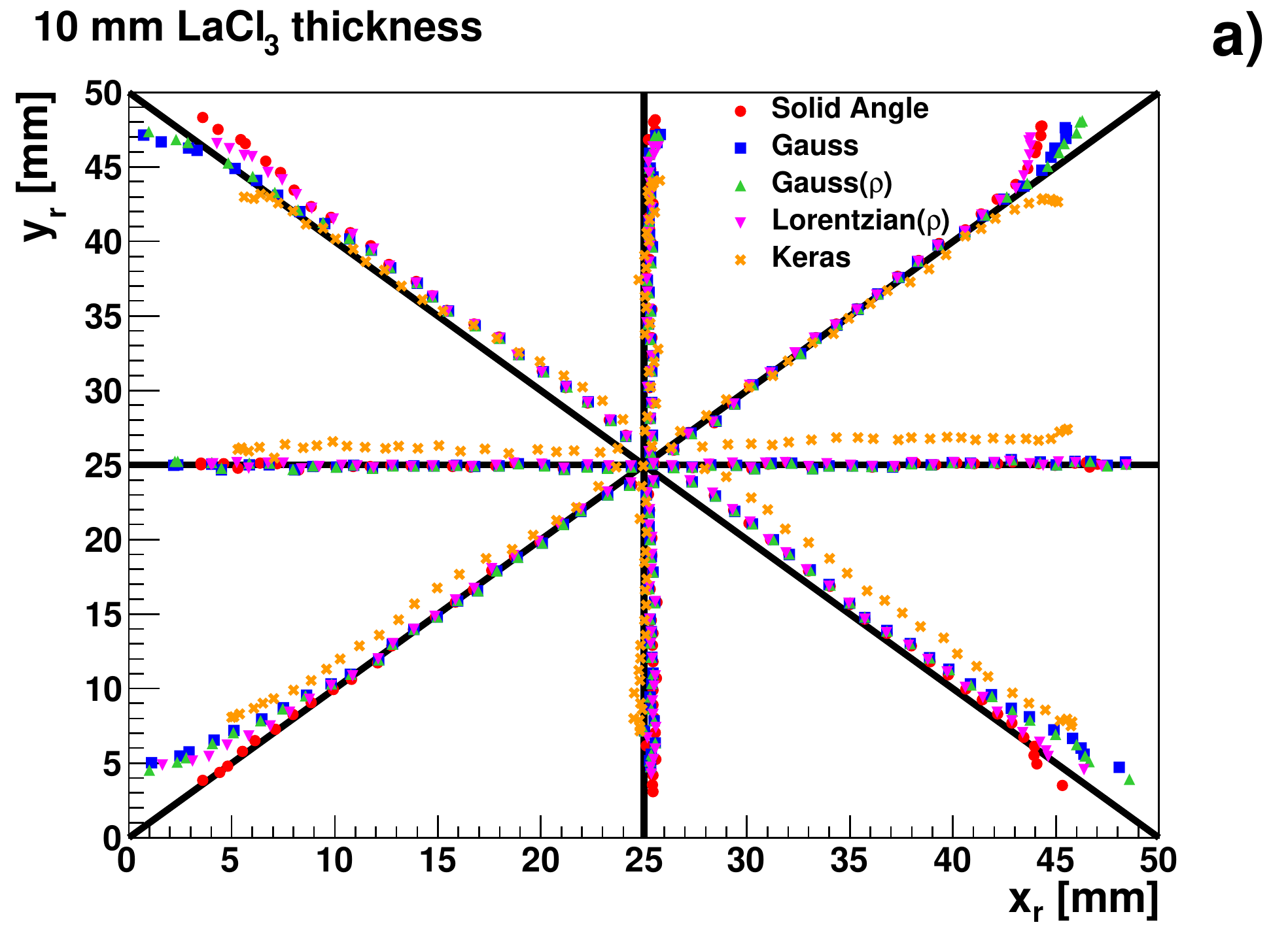} &
  \includegraphics[width=\columnwidth]{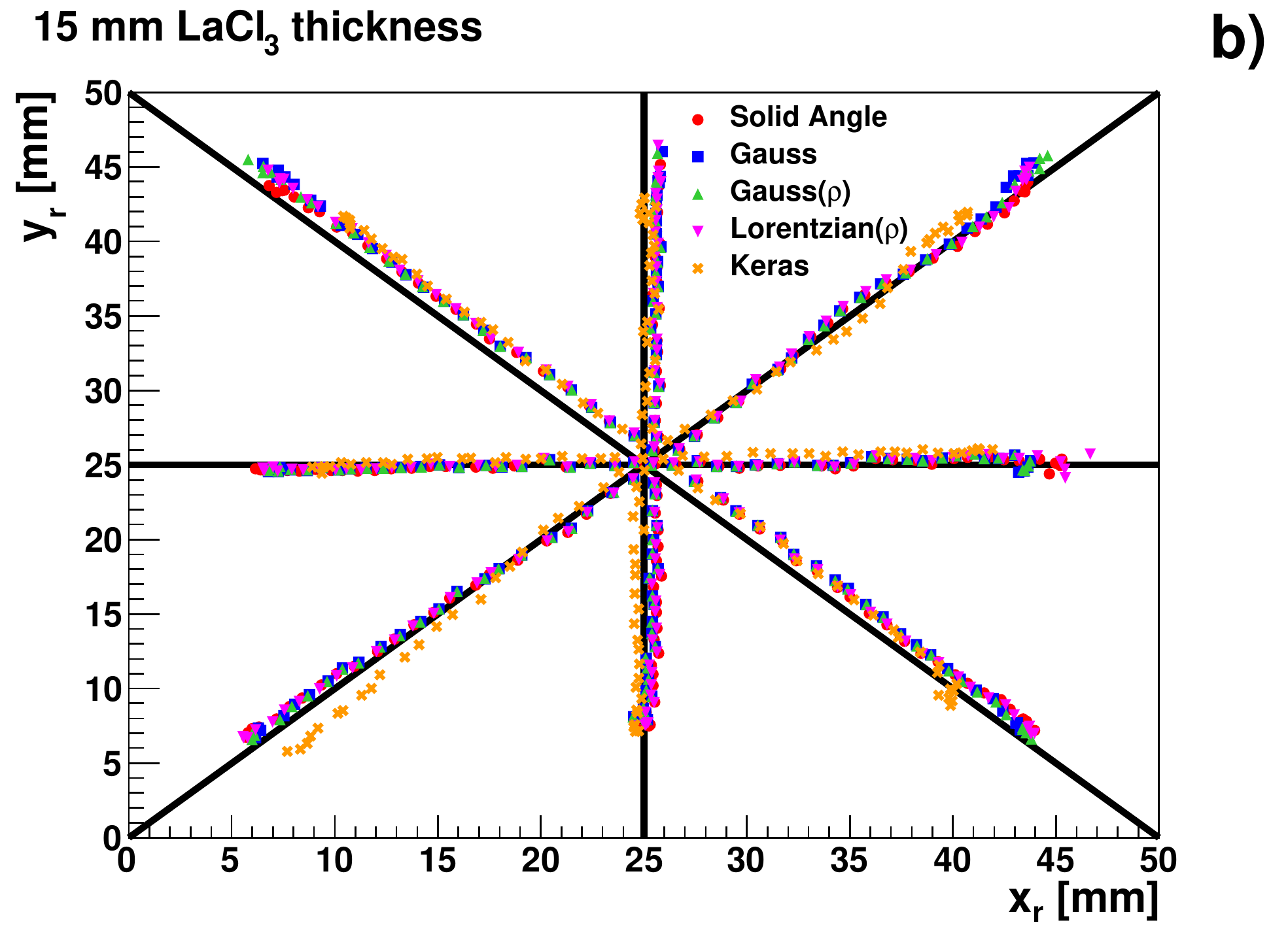} \\
  \includegraphics[width=\columnwidth]{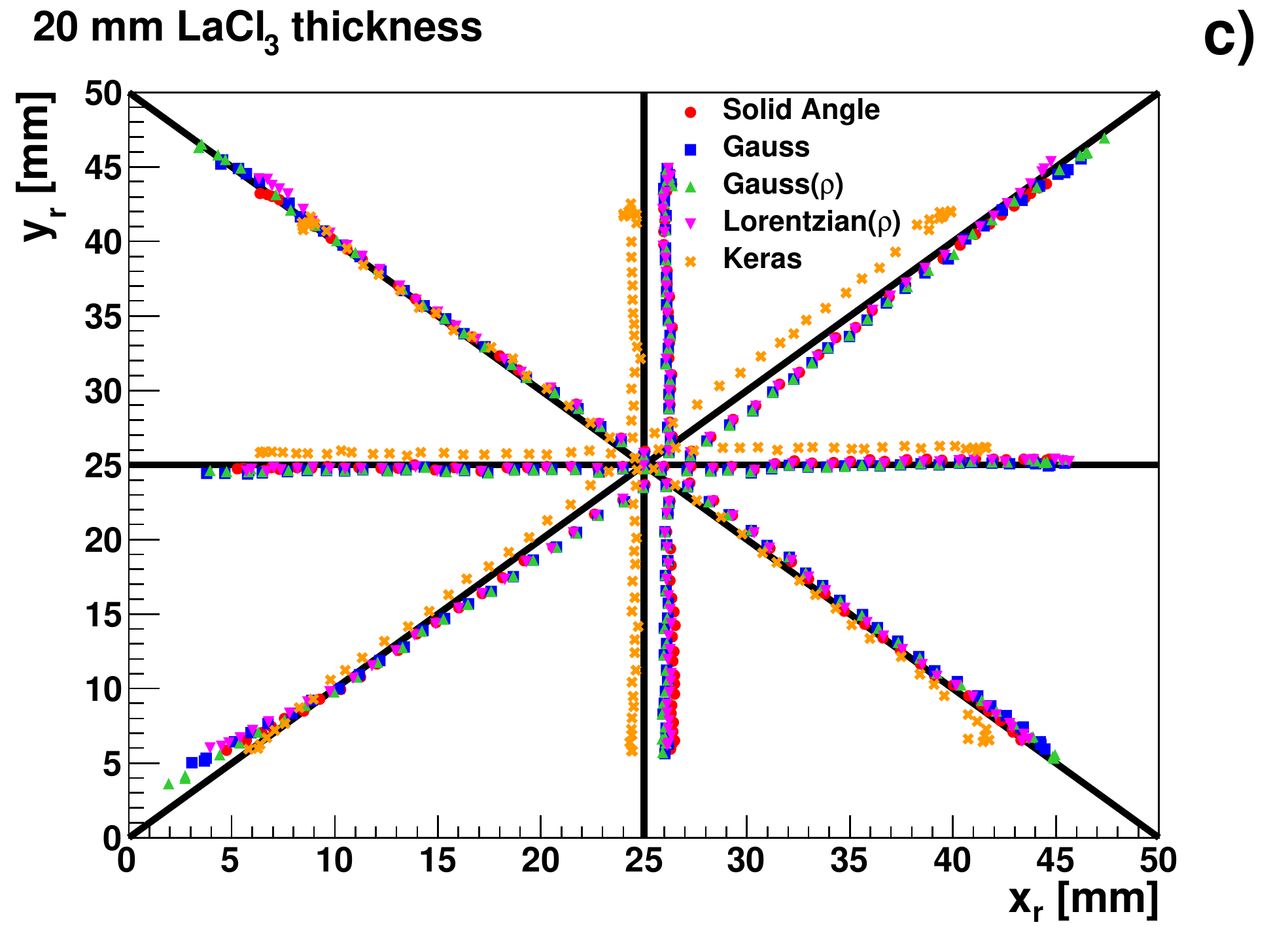} &
  \includegraphics[width=\columnwidth]{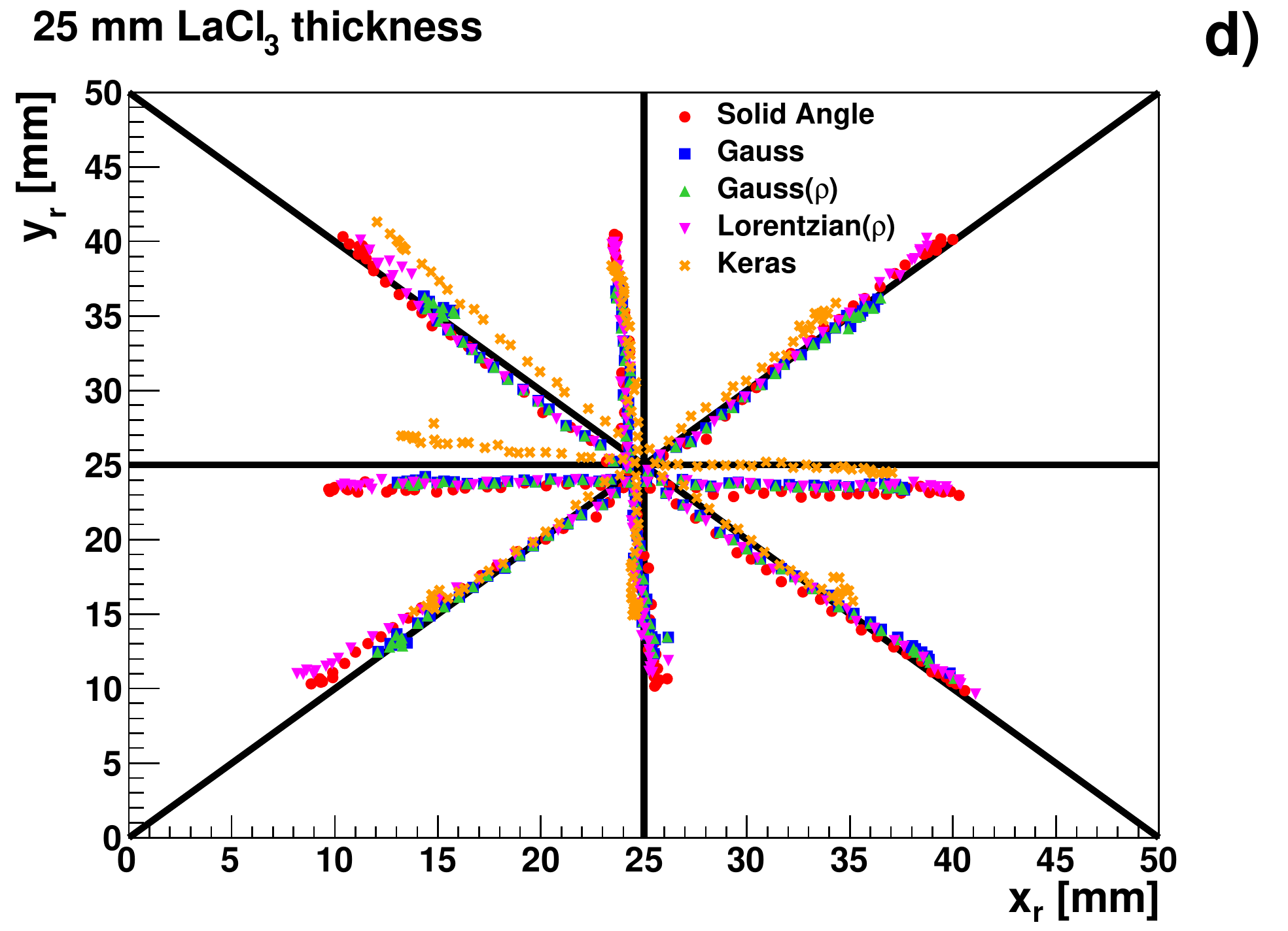} \\ 
  \includegraphics[width=\columnwidth]{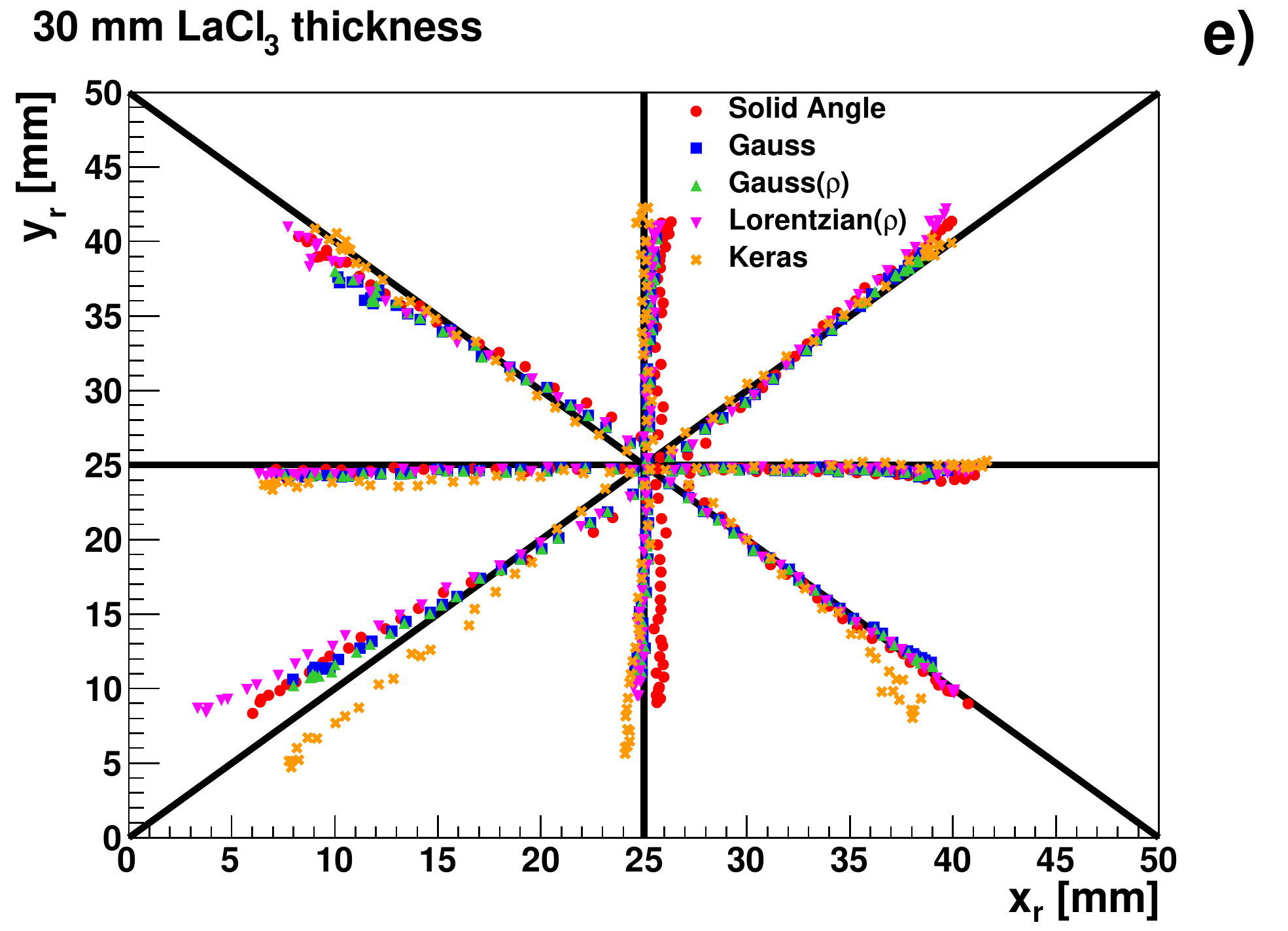} &
   \\ 
\end{tabular}
 \end{center}
\caption{Linearity diagram for the five different \lacls crystal thicknesses. The results shown in panels a), b), c), d) and e) correspond to thicknesses of 10 mm, 15 mm, 20 mm, 25 mm and 30 mm, respectively. The different models {\it{Solid angle}}, {\it{Gauss}}, {\it{Gauss($\rho$)}}, {\it{Lorentzian($\rho$)}} and {\it{Keras}} are shown by the red, blue, green, pink and orange dot points, respectively.}
\label{fig:linearity_Diagram}
\end{figure*}

We define here the linearity diagram as the relationship between the mean value of the reconstructed-position distribution for the individual irradiations ($x_{r},y_{r}$) and the true position, which is delivered by our $xy$-scanning gantry with a micrometric accuracy (see Sec.~\ref{sec:Exp_setup}). Instead of representing the full ensemble of scan positions, the results for four representative directions across the crystal are shown: the central-horizontal line, the central-vertical line and the two crystal diagonals. This representation differs from the common linearity diagrams, as those shown in our previous work~\cite{BABIANO20191}. However, the new representation chosen here allows to display a larger amount of information (four axis and five methods) in a single diagram, still with an acceptable graphical clarity.
The ($x_{r},y_{r}$) values were calculated by means of the Robust covariance estimation~\cite{RobustCovarianceRousseeuw}. The Mahalanobis distance~\cite{Chi2Wilson} was used for detecting outliers, as implemented in the Python programming language using the sklearn package~\cite{scikit-learn,sklearn_api}. Based on the Minimum Covariance Determinant estimator, this is a robust and high-performance Gaussian fitting method, which allows one to realiably identify outlying events or random coincidences in the distribution. Once the outliers are filtered off via this technique, the location of the mean and covariance matrix are computed from the "pure" subset of observations, hereby minimizing the Covariance determinant as objective function. This methodology is especially interesting for the irradiations close to the edge of the crystal, where the signal to background ratio decreases.

The raw linearity diagrams are displayed Fig.~\ref{fig:linearity_Diagram}. The diagrams are ordered according to increasing crystal thickness: 10 mm, 15 mm, 20 mm, 25 mm and 30 mm, displayed in panels a), b), c), d), e), respectively. For each one of these diagrams, the results obtained from the different reconstruction algorithms use the same color code: {\it{Solid angle}}, {\it{Gauss}}, {\it{Gauss($\rho$)}}, {\it{Lorentzian($\rho$)}} and {\it{Keras}} are shown by the red, blue, green, pink and orange markers, respectively. The true positions of the individual axis are indicated by the solid black lines. 

The performance of the investigated models for irradiations close to the center of the crystal is pretty similar, with good linearity and small compression in all cases. For those positions the contribution from reflections in the crystal walls is still small compared to the direct light contribution. However, as the distance from the center increases, larger compression and lower linearity is observed in the diagrams. This is a known and expected effect due to the higher contribution of reflected light to the total registered light yield. This effect is specially relevant in the crystal corners, where the ratio reflected to direct light is even worse than in the central regions of the crystal walls.

An increasing compression trend with crystal thickness can be clearly observed in Fig~\ref{fig:linearity_Diagram}. This is ascribed to the larger contribution of scintillation photons reflected in the crystal walls with increasing crystal thickness.

\begin{figure}[htb!]
\begin{center}
\begin{tabular}{c c}
  \includegraphics[width=0.45\columnwidth]{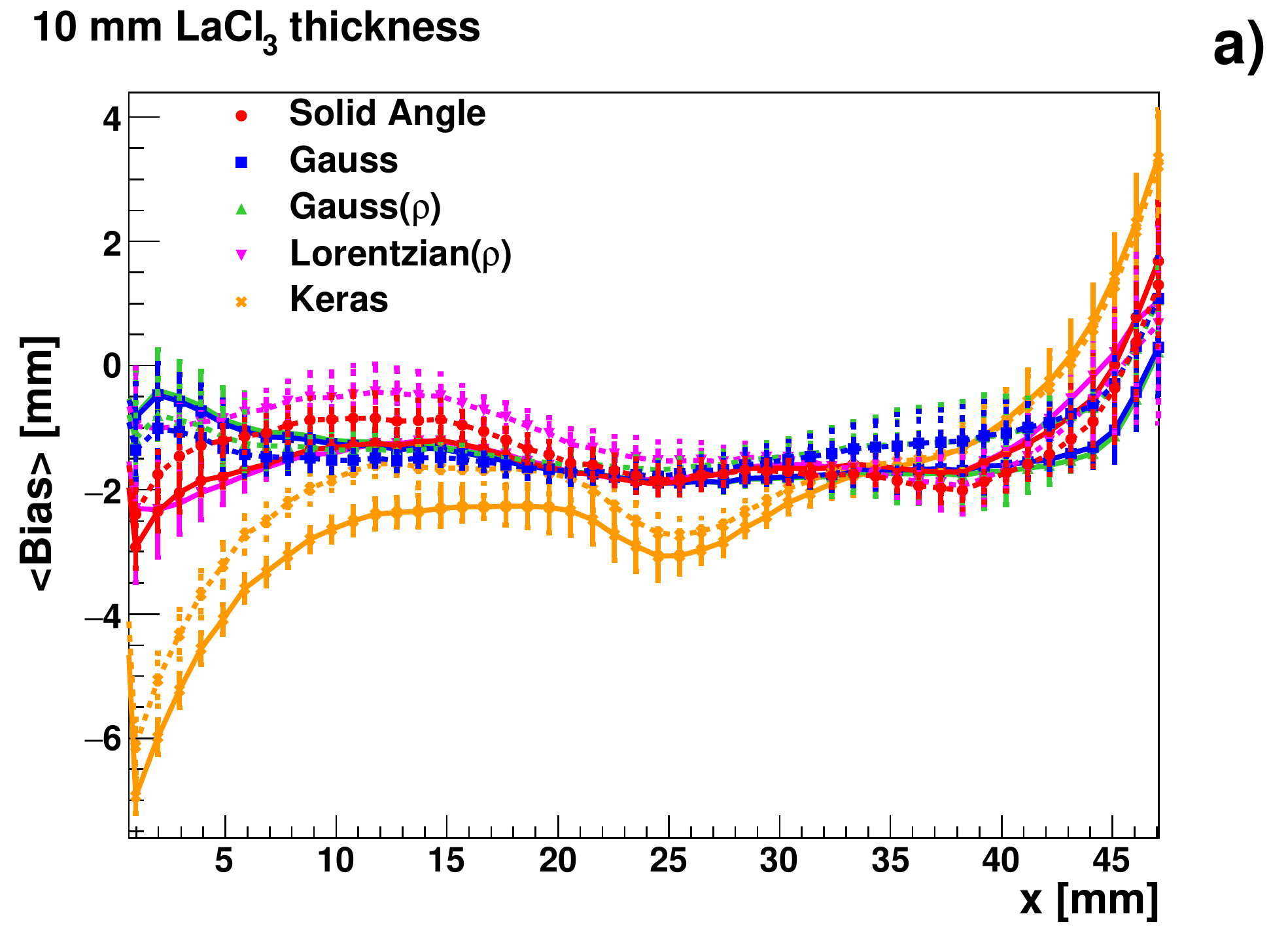} &
  \includegraphics[width=0.45\columnwidth]{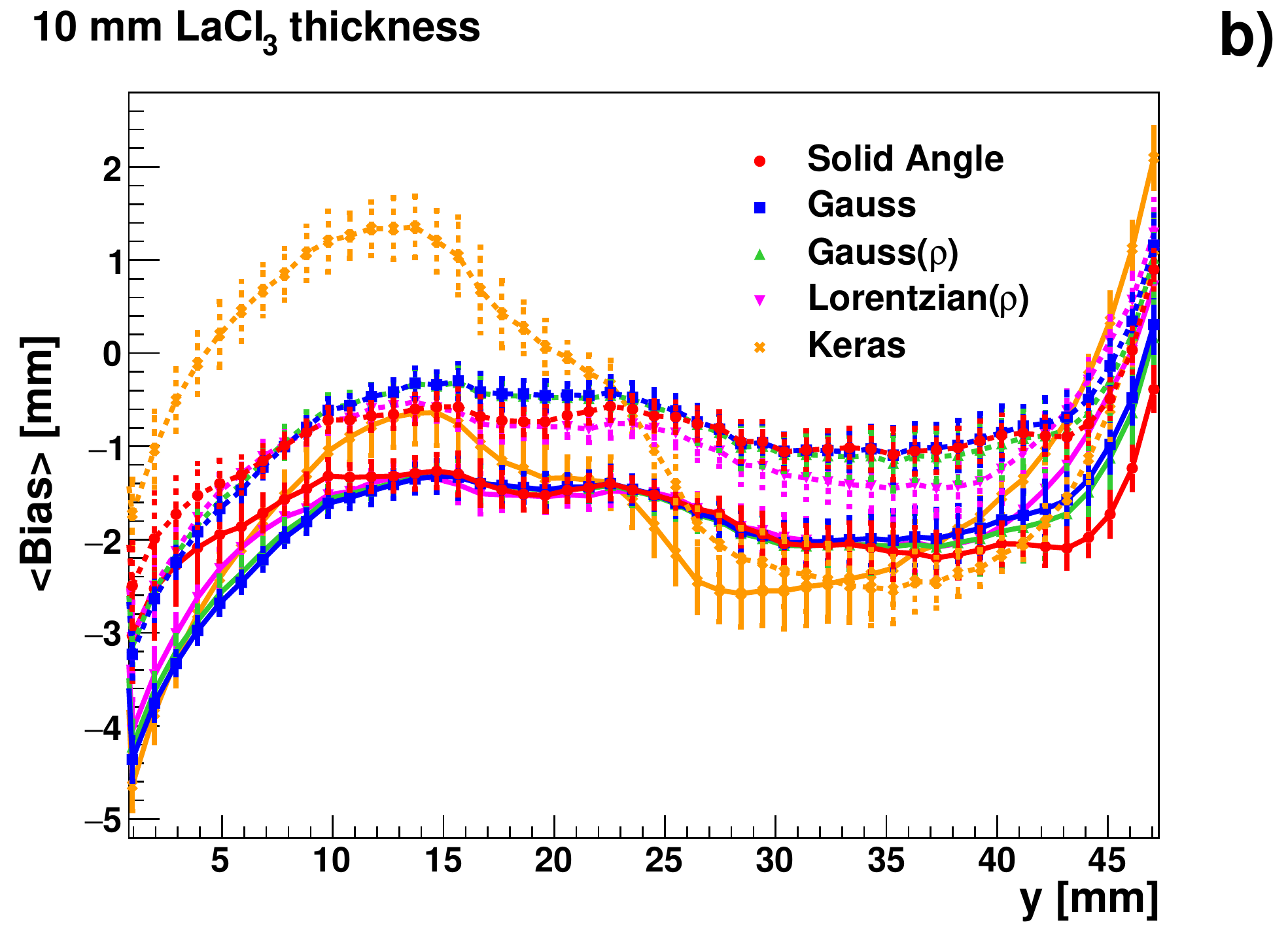} \\ 
  \includegraphics[width=0.45\columnwidth]{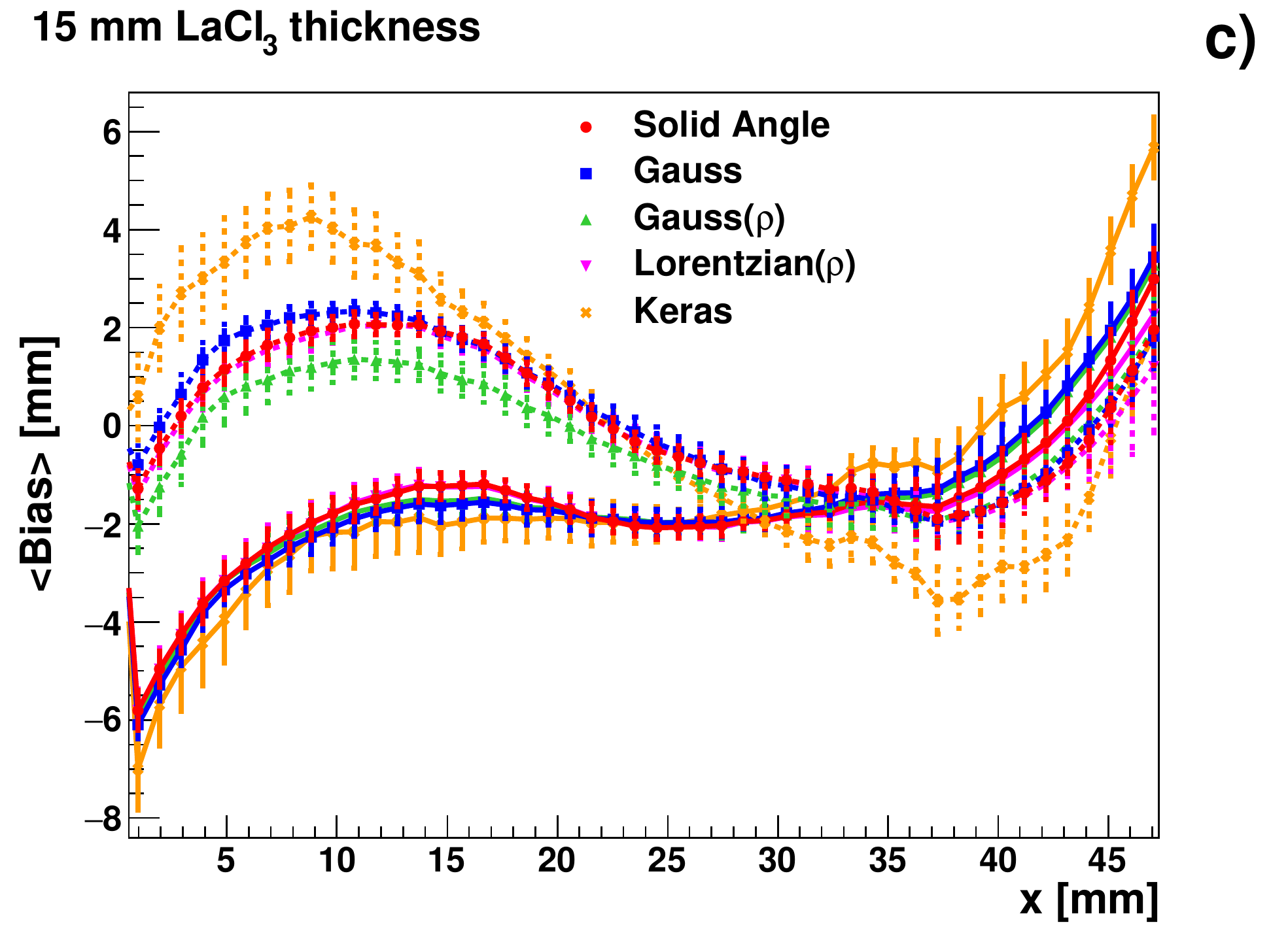} &
  \includegraphics[width=0.45\columnwidth]{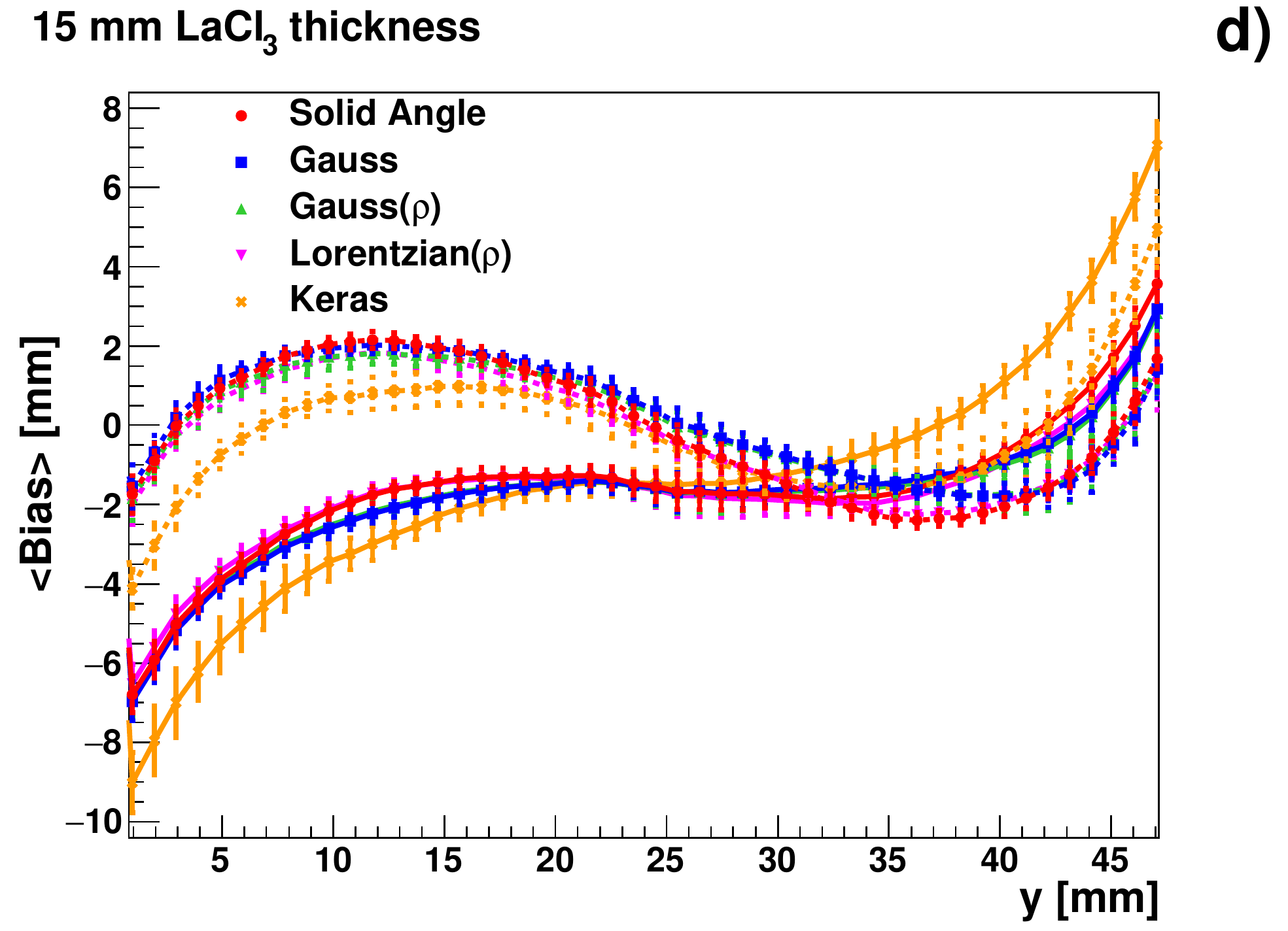} \\
  \includegraphics[width=0.45\columnwidth]{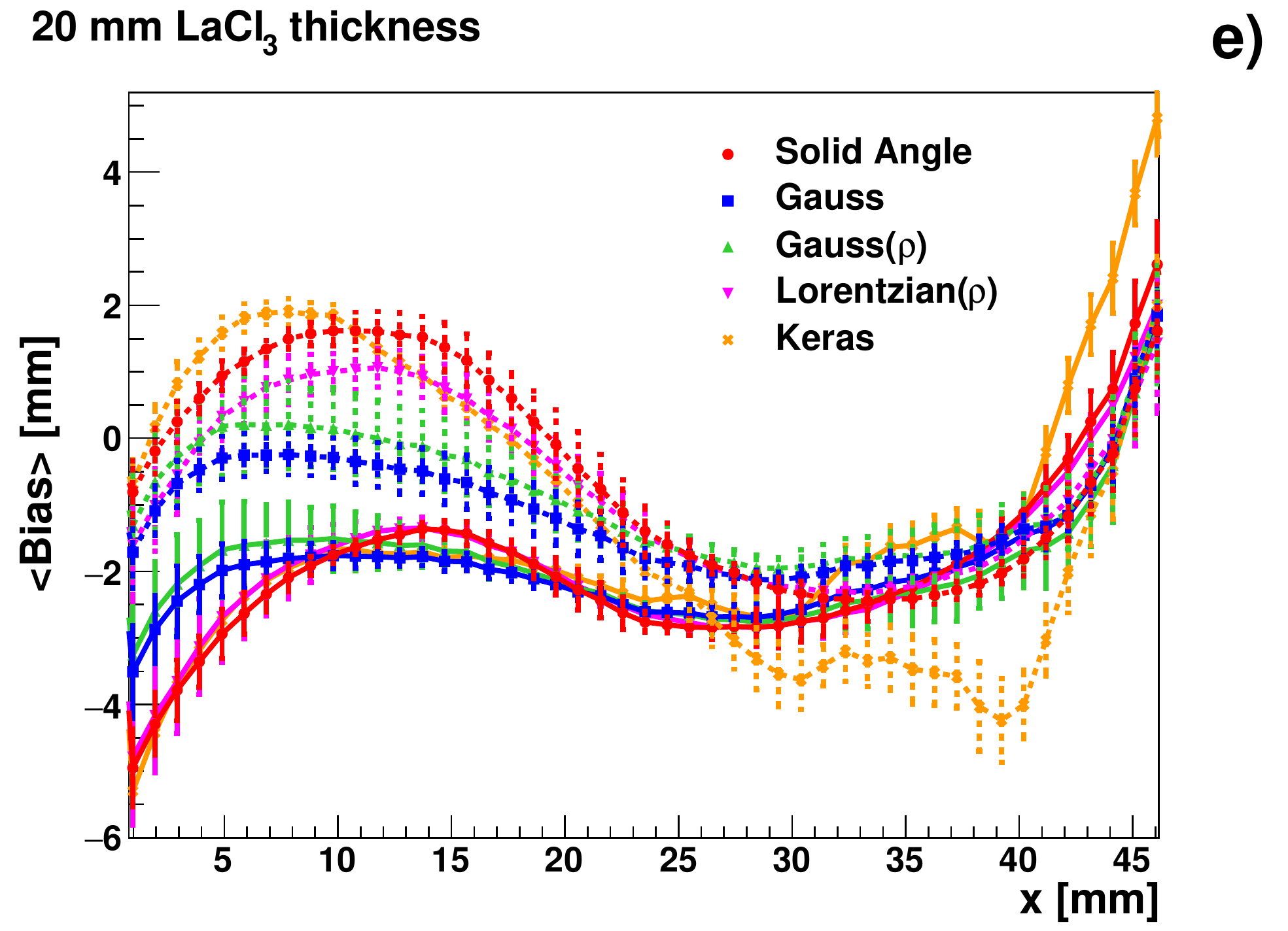} &
  \includegraphics[width=0.45\columnwidth]{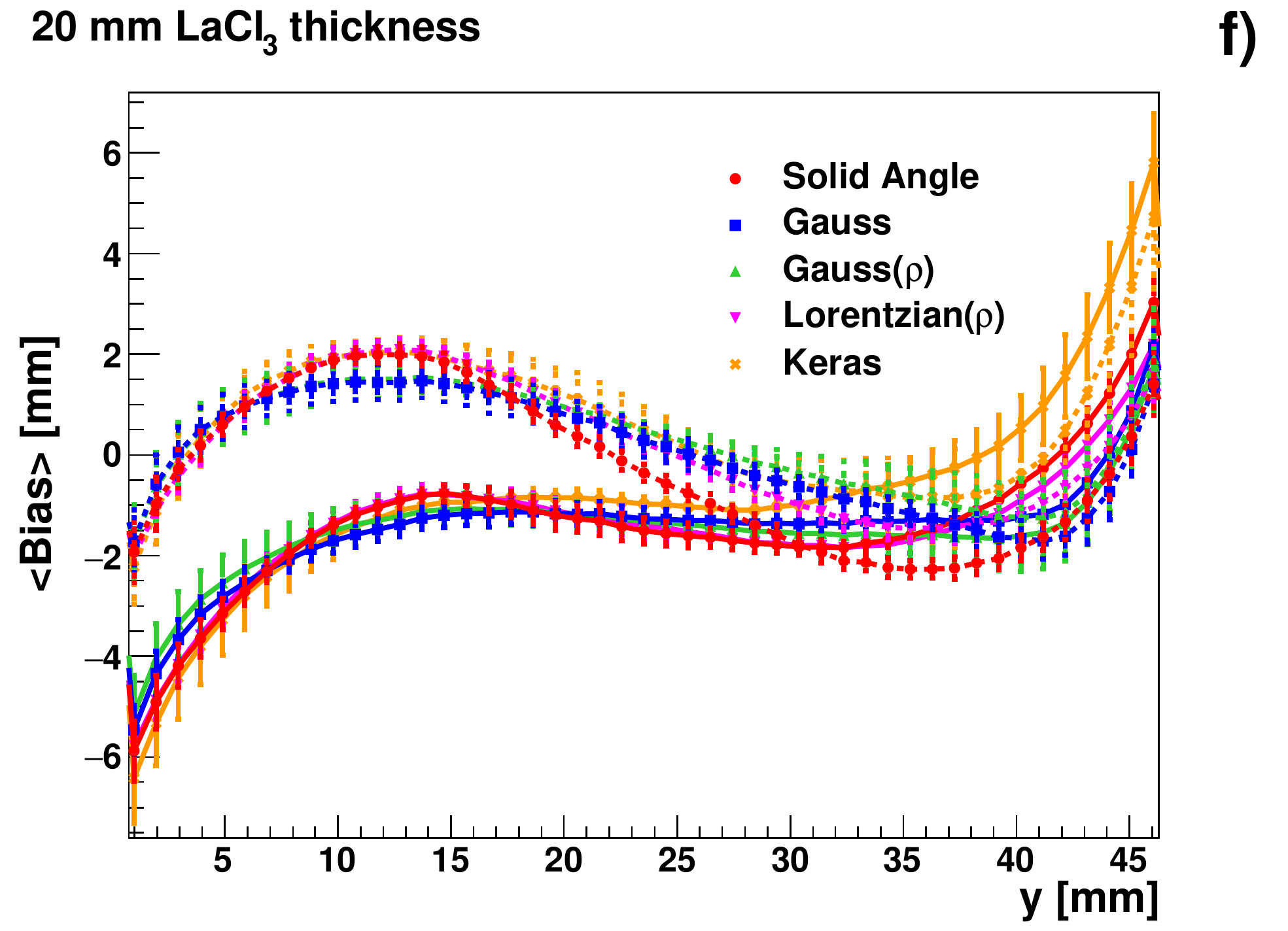} \\
  \includegraphics[width=0.45\columnwidth]{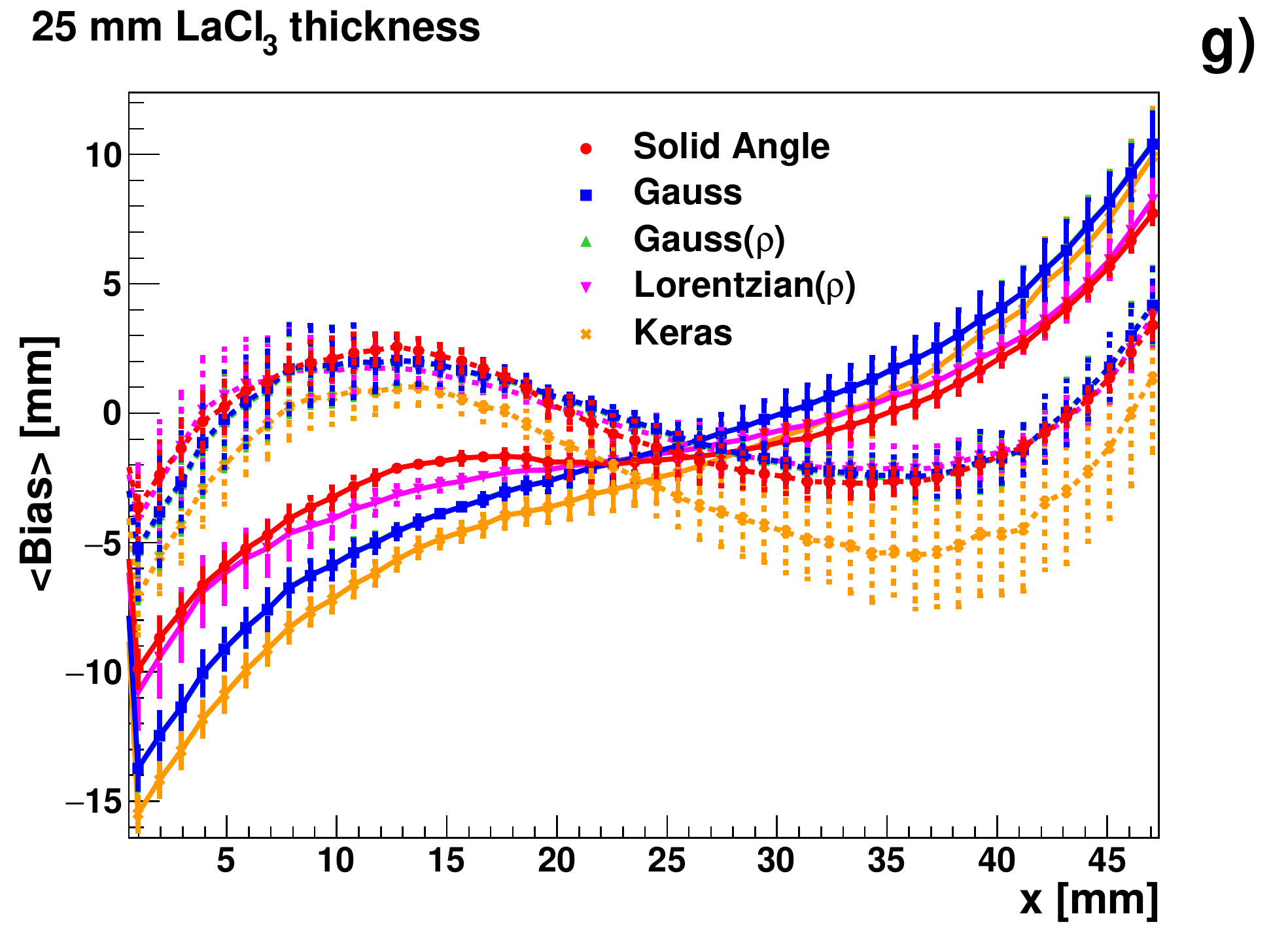} &
  \includegraphics[width=0.45\columnwidth]{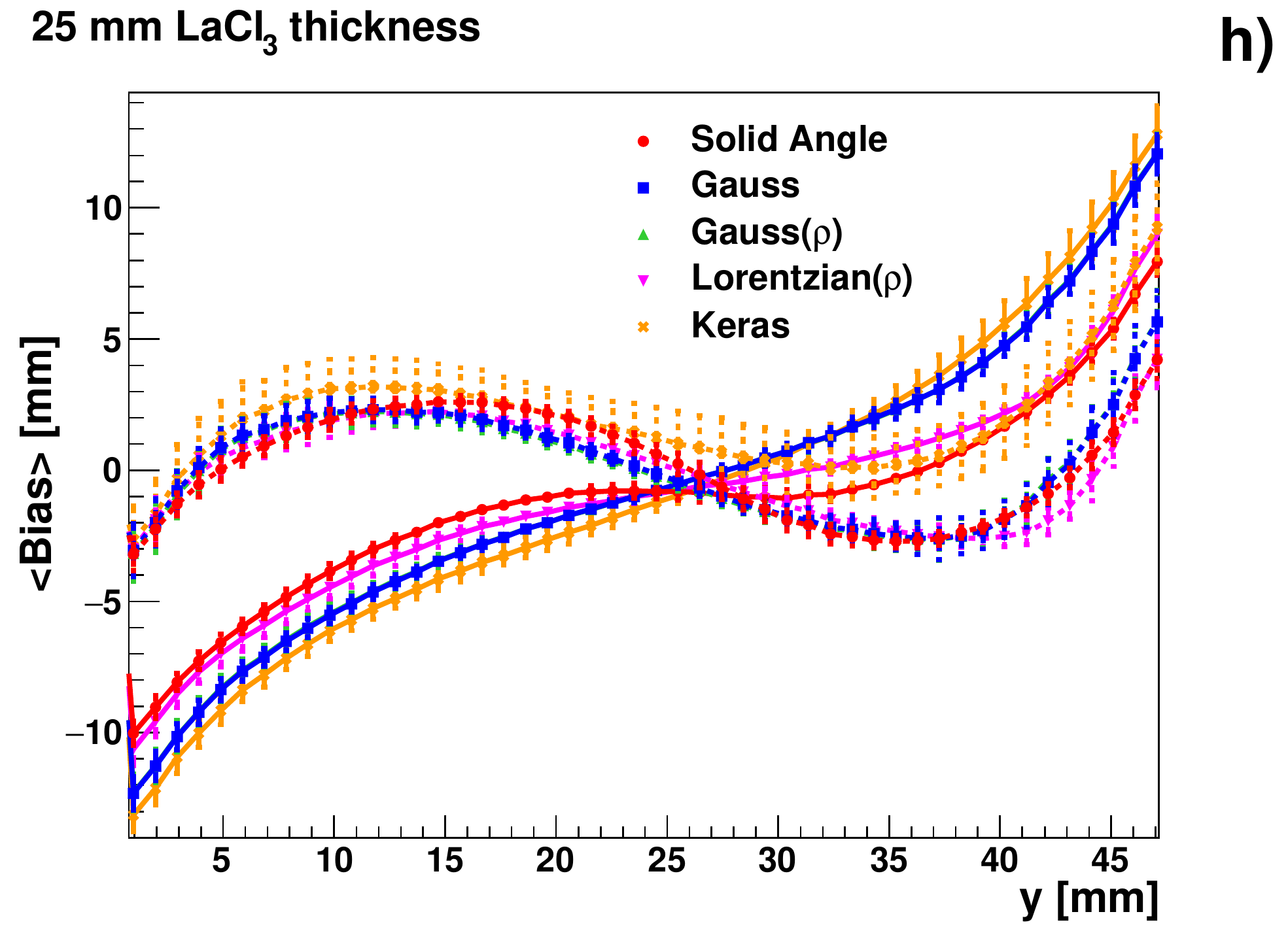} \\\ 
 \includegraphics[width=0.45\columnwidth]{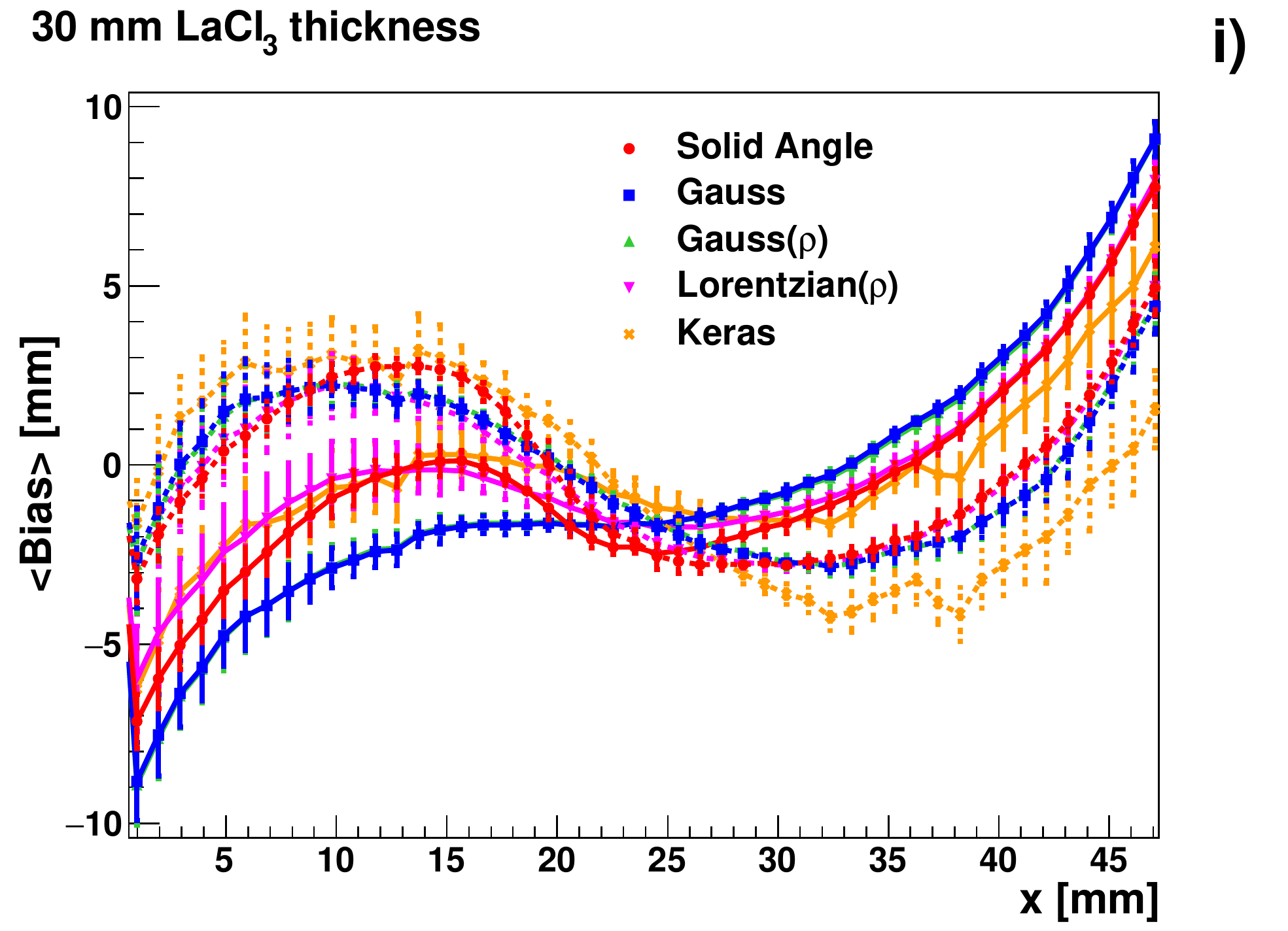} &
  \includegraphics[width=0.45\columnwidth]{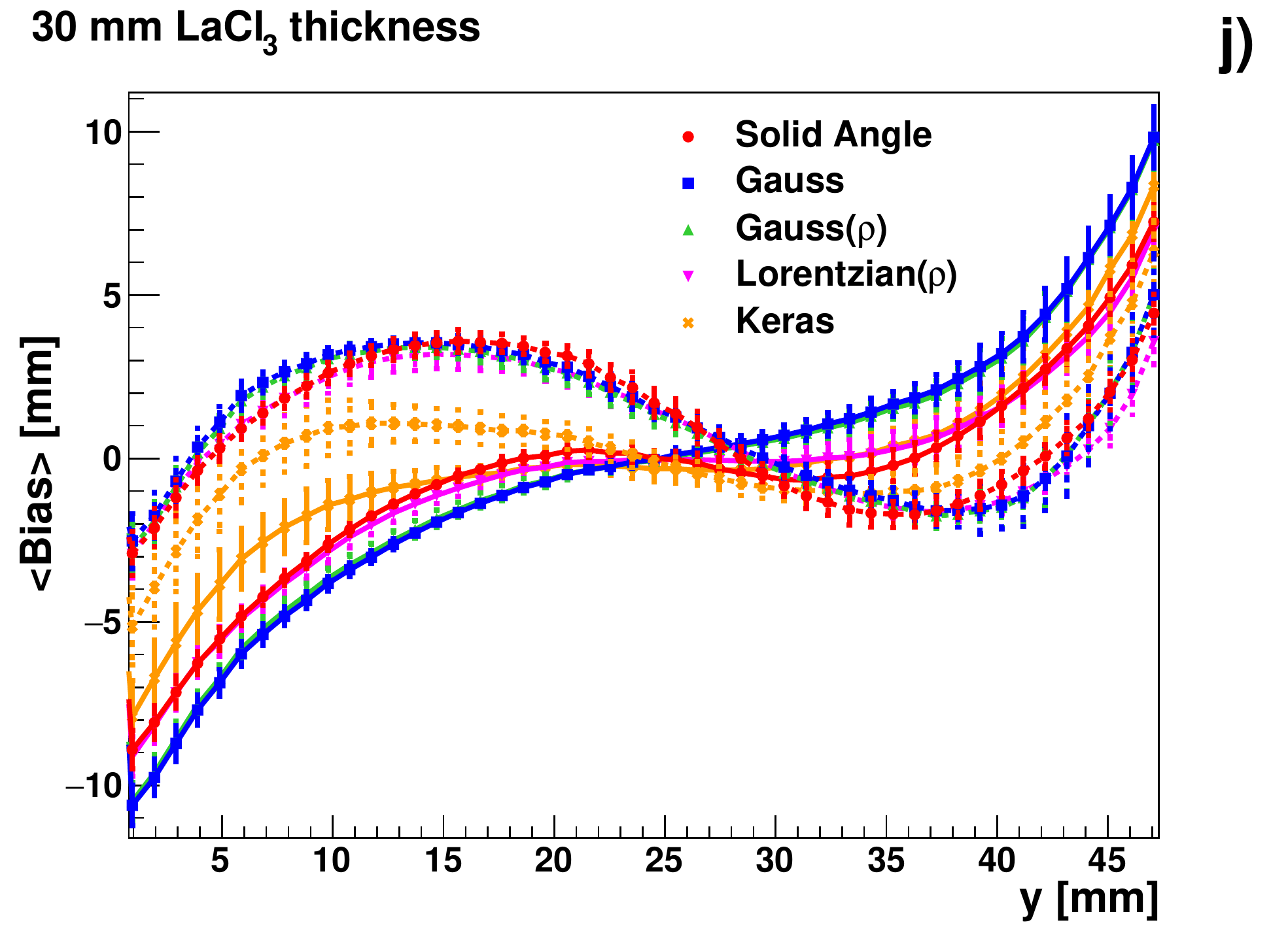} \\

\end{tabular}
 \end{center}
\caption{Average bias as a function of the true position measured for the five \lacls crystal thicknesses used in this work. Panels a), c), e), g), i) show the results for the horizontal line of the crystal. Panels a), c), e), g), i) show the results for the vertical line. The results obtained after the detector calibration discussed in Sec.~\ref{sec:Machine_Learning} are displayed by the same colors but in dashed lines (see text for details).}
\label{fig:Bias}
\end{figure}

The bias or average difference between true and reconstructed positions is shown in Fig.~\ref{fig:Bias} as a function of the true position. This quantity was determined for all the light yield models and \lacls crystal sizes used in this work. The average is computed from all available horizontal and vertical scan lines within the active area of the detector. In the latter figure, panels a), c), e), g) and i) present the results for the horizontal axis, while panels b), d), f), h) and j) correspond to the results for the vertical one. The colours used for the different thickness and methods are in concordance with the linearity diagrams presented in Fig.~\ref{fig:linearity_Diagram}. 

The results obtained with the analytical models for the 10~mm thick \lacls crystal are similar to the ones obtained in our previous work~\cite{BABIANO20191}. Also, they are comparable to similar studies that involve thin crystals~\cite{Li2010,Morrocchi_2016,CABELLO2013148}. However, the results obtained for {\it{Keras}} are significantly worse. In the case of the \lacls crystal with a thickness of 15~mm our results are similar to those reported in~\cite{8871159}.

In general, the bias plots in Fig.~\ref{fig:Bias} reflect the increasing non-linearity towards the edges of the crystal, which becomes more prominent with increasing crystal thickness. Comparable results in linearity and compression are obtained for all analytical models, {\it Solid angle}, {\it Gauss}, {\it Gauss($\rho$)} and {\it Lorentzian($\rho$)}. The limited performance observed for the CNN approach {\it{Keras}} can be most probably ascribed to the differences between the ideal detector response of the MC simulation and the true detector response, which includes pixel-gain fluctuations, pixel thresholds, crystal inhomegeneities and other experimental effects.

In summary, regarding linearity and pin-cushion effects, it can be concluded that all analytical models studied here perform very similar for all the crystal thickness. In some more detail, the {\it Solid angle} and {\it Lorentzian($\rho$)} models show slightly better overall performance compared to the rest of the models.

\begin{figure}
\begin{center}
 \includegraphics[width=\columnwidth]{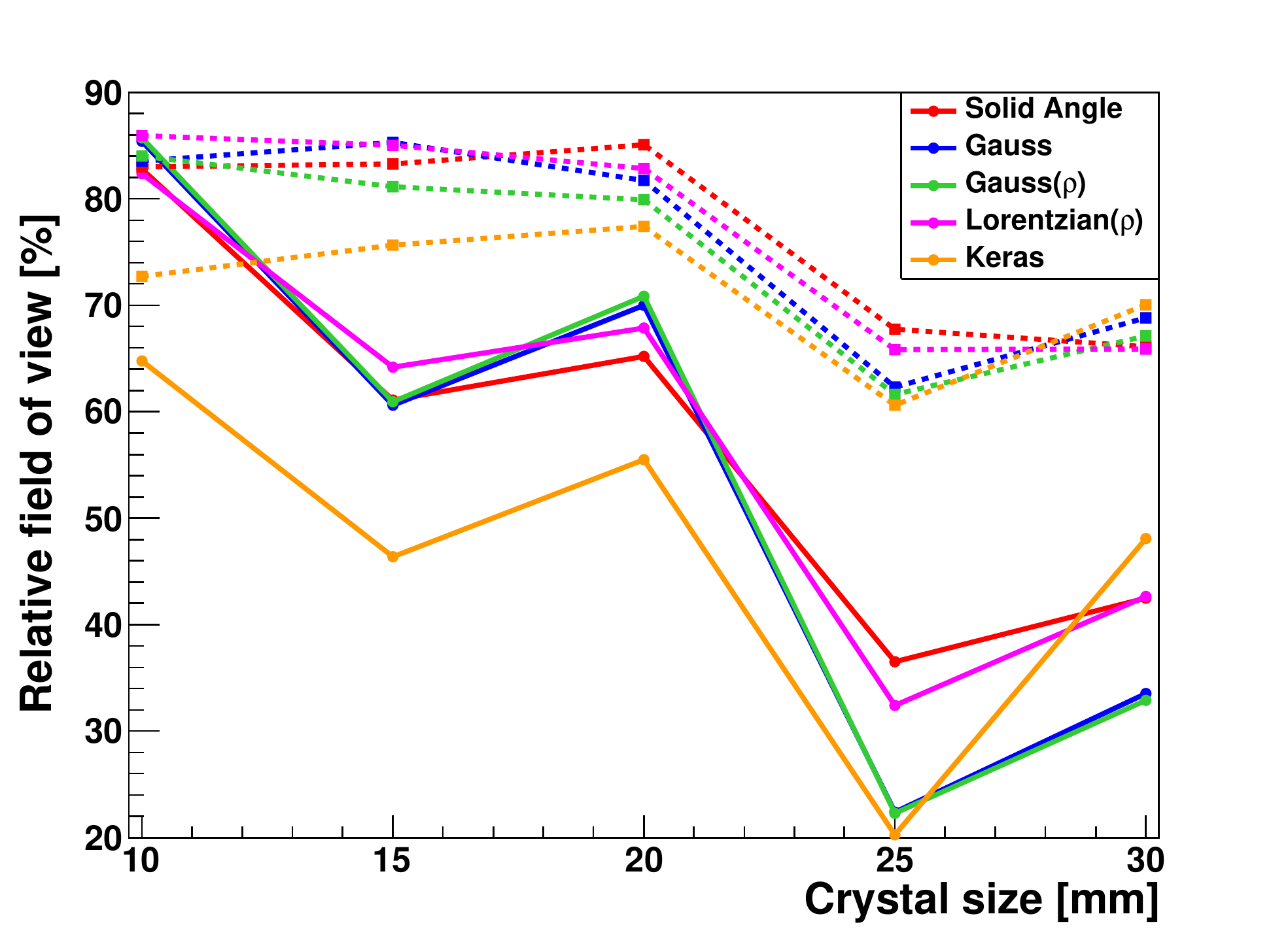}
 \end{center}
\caption{Relative FoV as a function of the \lacl thickness. The results for the different light models are shown by the solid coloured lines. The results obtained after the detector calibration are shown by the same colors but in dashed lines (see text for details).}
\label{fig:Useful_Field_of_View}
\end{figure}

In order to quantify and discuss the field-of-view (FoV) of every crystal we calculate the ratio between the reconstructed area calculated from the mean positions (x$_{r}$,y$_{r}$) and the effective area of the crystal of Tab.~\ref{tab:effective area}. This quantity, named as relative field-of-view, is displayed with solid lines in Fig.~\ref{fig:Useful_Field_of_View} for the different light-yield models as a function of the \lacls crystal thickness. The color code is the same as in previous figures. Again, the relative FoV shows a compression trend, which increases with the crystal thickness. It ranges from values of $\sim$80\% for the 10~mm thick crystal, down to $\sim$30\% for the thickest crystal (30~mm). This behavior is not very sensible to the reconstruction technique used. However, on average, a lower compression is obtained with the {\it{Solid angle}} and  {\it{Lorentzian($\rho$)}} methods, particularly for the 25~mm and 30~mm thick crystals. For the latter two, the results from {\it{Gauss}}, {\it{Gauss($\rho$)}} show, on the other hand, a poor performance. The {\it{Keras}} CNN technique shows the largest compression for all thicknesses, except for 30~mm thick crystal. This result probably indicates that CNN methods may become competitive for applications requiring position-sensitivity in very thick crystals (see for instance ~\cite{Agiaz2019IEEE}).

\begin{figure}
\begin{center}
  \includegraphics[width=\columnwidth]{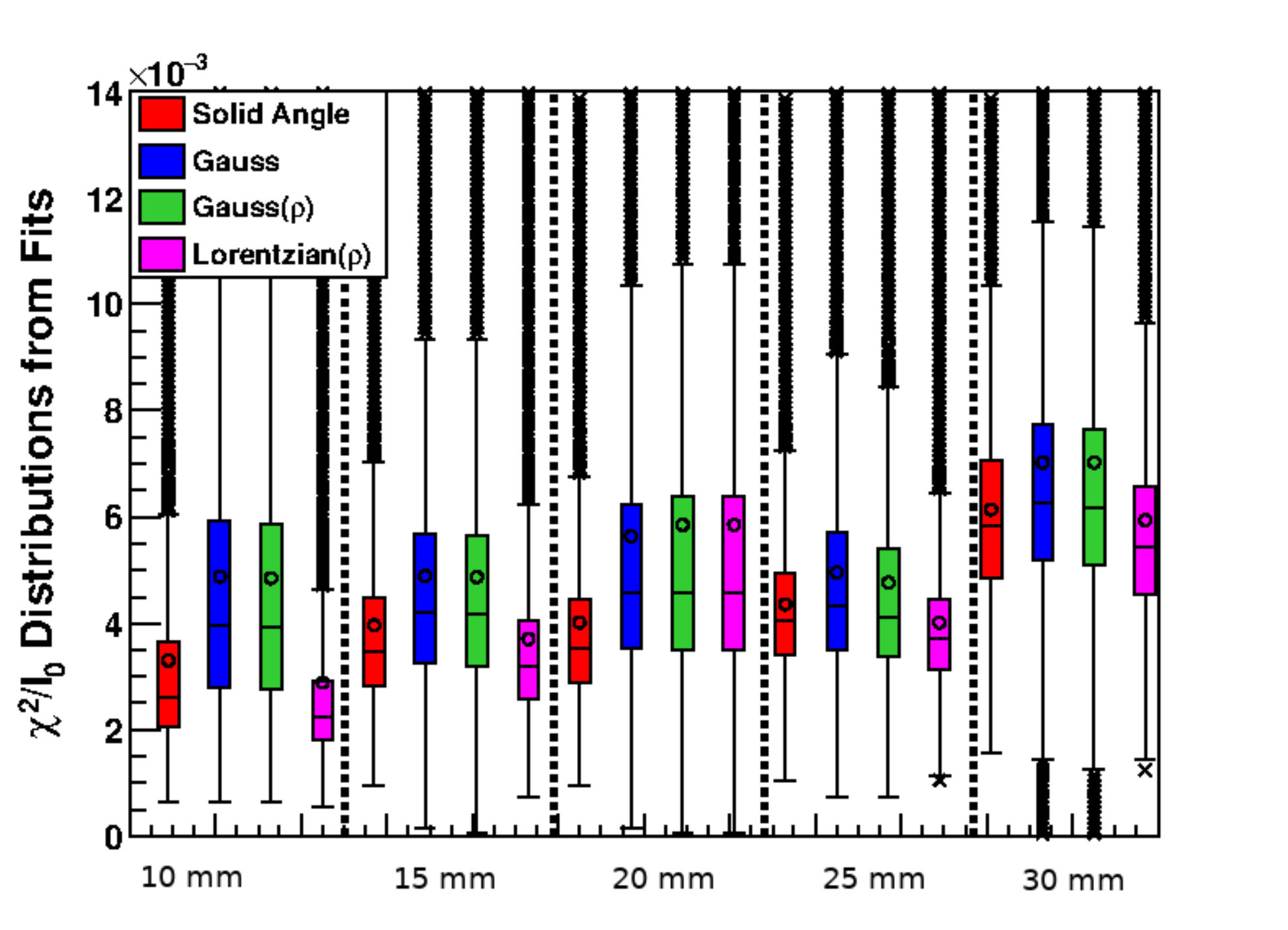}
 \end{center}
\caption{$\chi^{2}$/I$_{\circ}$ box plots obtained from the analytical model event position reconstruction for the various \lacls crystals thickness. The results for {\it{Solid angle}}, {\it{Gauss}}, {\it{Gauss($\rho$)}}, and {\it{Lorentzian($\rho$)}} are shown as red, blue, green and pink colors, respectively}
\label{fig:xi2-distributions}
\end{figure}

Finally, Fig.~\ref{fig:xi2-distributions} summarizes the main results for the analytical methods by means of a box plot of $\chi^{2}$ divided by the the QDC-integral of all the pixels fired in coincidence for the $\gamma$-ray event (I$_{\circ}$). This representation can be used to evaluate the goodness of each light-yield model for reproducing the light distribution as registered in the SiPM. This kind of plot is unusual in this type of studies, but it results very convenient to interpret and graphically categorize data distributions via the following six quantities:
\begin{itemize}
    \item The minimum value of the distribution is represented by the bottom whisker.
    \item The lower quartile of the data points is represented by the bottom of the box.
    \item The upper quartile of the data points is represented by the top of the box.
    \item The median of the distribution is represented by the line between the top and the bottom edge of the box.
     \item The mean of the distribution is represented by the circle.
    \item The maximum value of the distribution is given by the top whisker.
\end{itemize}

In addition, the outlier events of each distribution are displayed by black crosses outside the whisker markers. The $\chi^{2}$/I$_{\circ}$ box plots show that all the investigated analytical light-yield models are better adapted to thin crystals. More specifically, in terms of performance and thickness one can classify the different crystals in three groups. The best results are obtained for the 10~mm, 15~mm and 25~mm thick crystals, followed by the 20~mm thick crystal and last the 30~mm thick crystal. In terms of analytical models one can observe to distinct groups, the {\it{Solid Angle}} and {\it{Lorentzian ($\rho$)}} models on one side, and the {\it{Gauss}} and {\it{Gauss($\rho$)}} prescriptions on the other side.
 Only for the thickest crystal investigated all analytical models show a similar performance in terms of $\chi^{2}$/I$_{\circ}$. However, as it was shown in Fig.~\ref{fig:linearity_Diagram} and Fig.~\ref{fig:Bias}, the compression and non-linearity is larger for {\it{Gauss}} and {\it{Gauss($\rho$)}} for all the \lacl crystal thicknesses.

\section{Machine Learning aided linearity calibration}\label{sec:Machine_Learning}

Image compression effects and other non-linear distortions can be significantly improved when the dependency between the reconstructed and the true positions is still a monotonically increasing function. To this aim a Support Vector Machine (SVM) model with a linear kernel~\cite{Smola04atutorial,10.5555/1162264} was implemented in the present work. In short, the SVM algorithm uses an n-dimensional training data set to find a uni-dimensional function $f$($\xi$), so-called kernel, that has at most a deviation $\varepsilon$ from the actual known target values, with the restriction that the dependency $f$($\xi$) has to be flat.

This means that the algorithm neglects errors during the fitting procedure as long as they are lower than $\varepsilon$, but it will not accept deviations larger than that quantity~\cite{10.5555/1162264}. In this work, the training data set corresponds to the two-dimensional mean reconstructed positions in the mesh covering the transverse plane of the detector,
$X$=\{(x$_{1}$,y$_{1}$),(x$_{2}$,y$_{2}$),...(x$_{n}$,y$_{n}$)\}, and the target data are the known positions of the scanning gantry for one of the x or y axis, $Y$=\{x$
^{r}_{1}$,x$^{r}_{2}$...,x$^{r}_{n}$\} or $Y$=\{y$^{r}_{1}$,y$^{r}_{2}$,...,y$^{r}_{n}$\}. 

In the case of a linear kernel, the function $f$($\xi$) is described by:

\begin{equation}
    f(\xi)=\langle w,\xi\rangle + b \quad with \quad w \in \Xi , b \in R
\end{equation}

where $\xi$ is the position coordinate, either $x$ or $y$, $\langle w,\xi \rangle$ denotes the dot product in the $\Xi$ vectorial space and $b$ is a constant term. A flat $f(\xi)$ dependency means that the algorithm looks for the smallest possible value of $w$. For such conditions, the mathematical problem can be written as a convex optimization problem described by:

\begin{equation*}
    \text{minimize} \quad \frac{1}{2}\|w\|^{2}
\end{equation*}
\begin{equation}
    \text{Subject to} \begin{dcases} \xi_{i}-\langle w, \xi_{i} \rangle -b \leq \varepsilon \\
                        \langle w, \xi_{i} \rangle +b - \xi_{i} \leq \varepsilon
                     \end{dcases}
\end{equation}

A similar approach, but using more complex functions can be applied to the calculation with non-linear kernels~\cite{Smola04atutorial,10.5555/1162264}.

In this work, the linear kernel was preferred to other non-linear kernels such as Gaussian or polynomial because of its better overall performance. To implement the model we have used the SVM algorithms available in the Python {\it{sklearn-learn}} package~\cite{scikit-learn}. For the simultaneous fitting of both $x-$ and $y-$axis, we construct two SVM models, one for each axis. They were fitted together to the experimental data using a {\it{MultiOutputRegressor}} provided by the same Python package. This function grants the synchronous minimization of $n$ models to $n$-dimensional known target data during the fitting procedure. Low sensitivity to the SVM parameters was found during the fitting procedure. For this reason, all of them except for the $\varepsilon$ parameter, which was set to a value of 0.2, were fixed to the default values.

\begin{figure*}
\begin{center}
\begin{tabular}{c c}
  \includegraphics[width=\columnwidth]{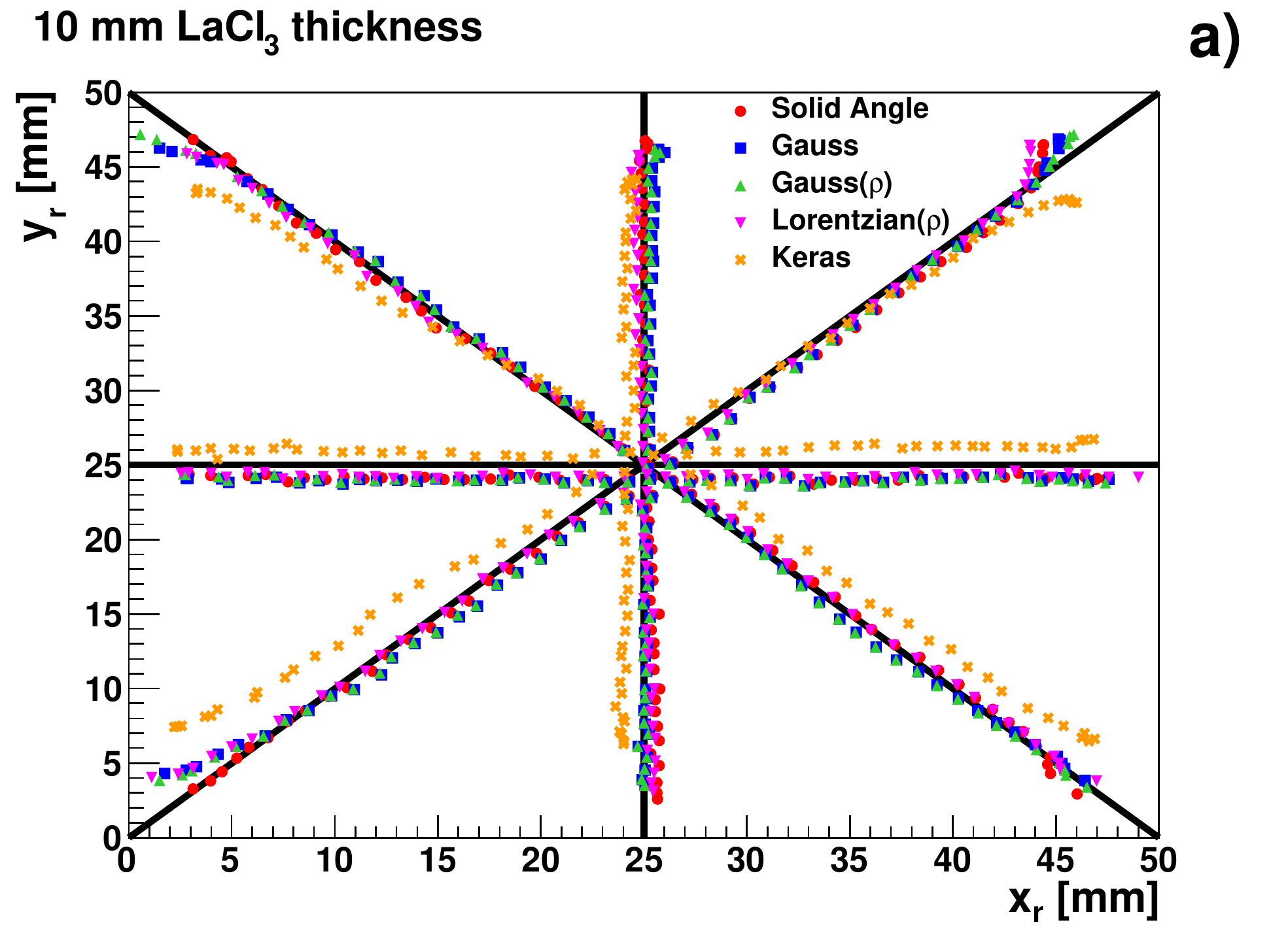} &
  \includegraphics[width=\columnwidth]{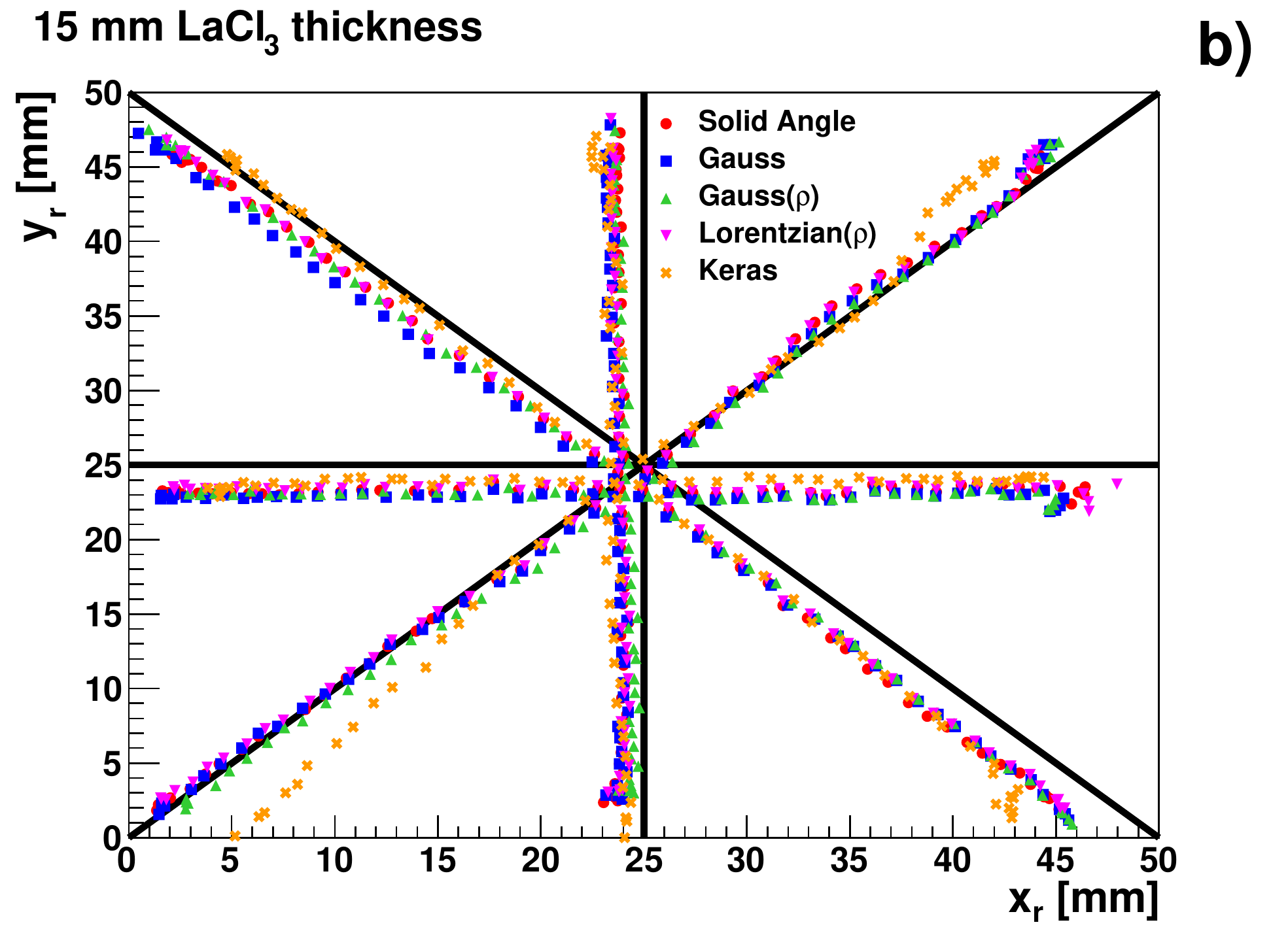} \\
  \includegraphics[width=\columnwidth]{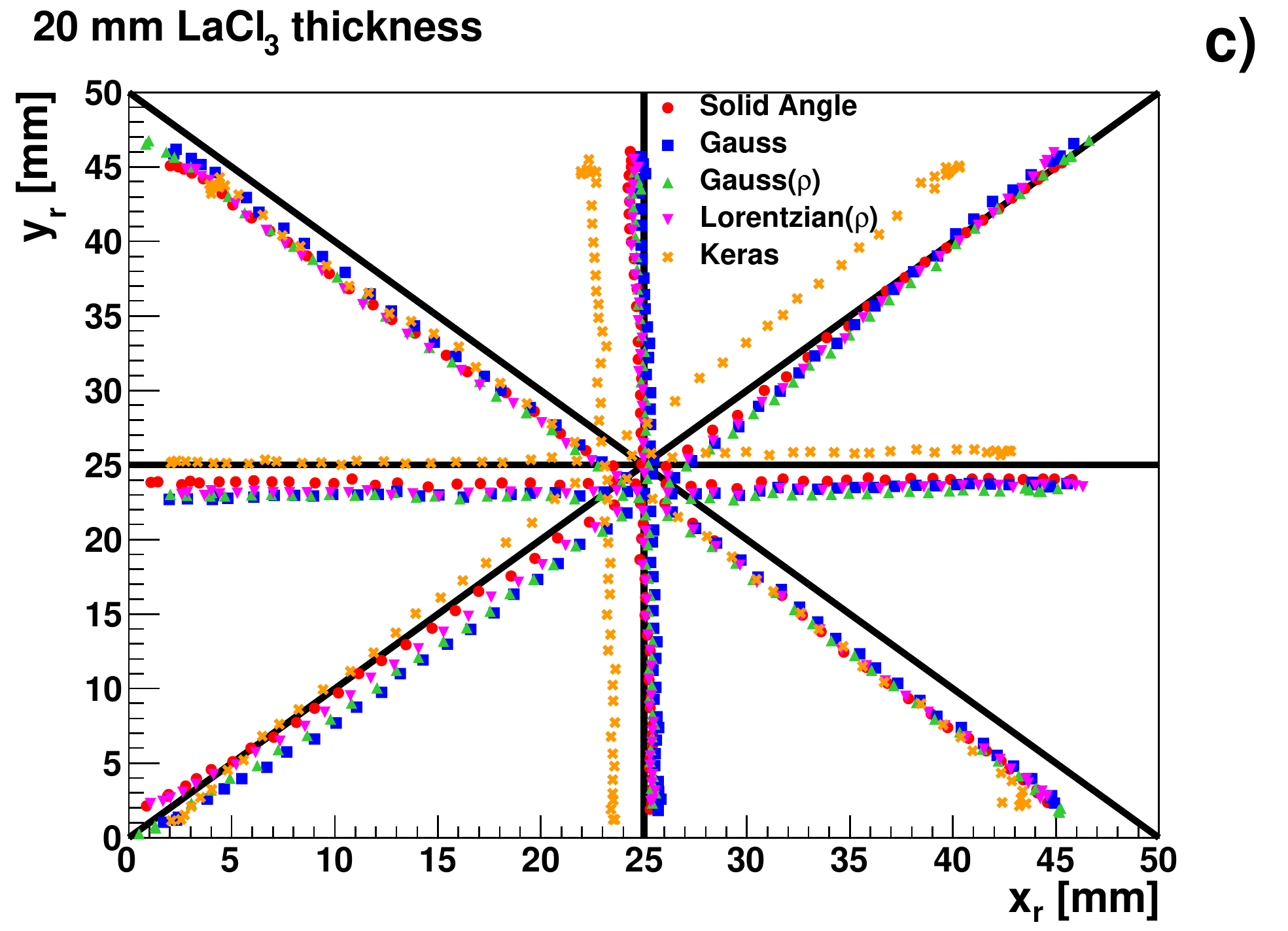} &
  \includegraphics[width=\columnwidth]{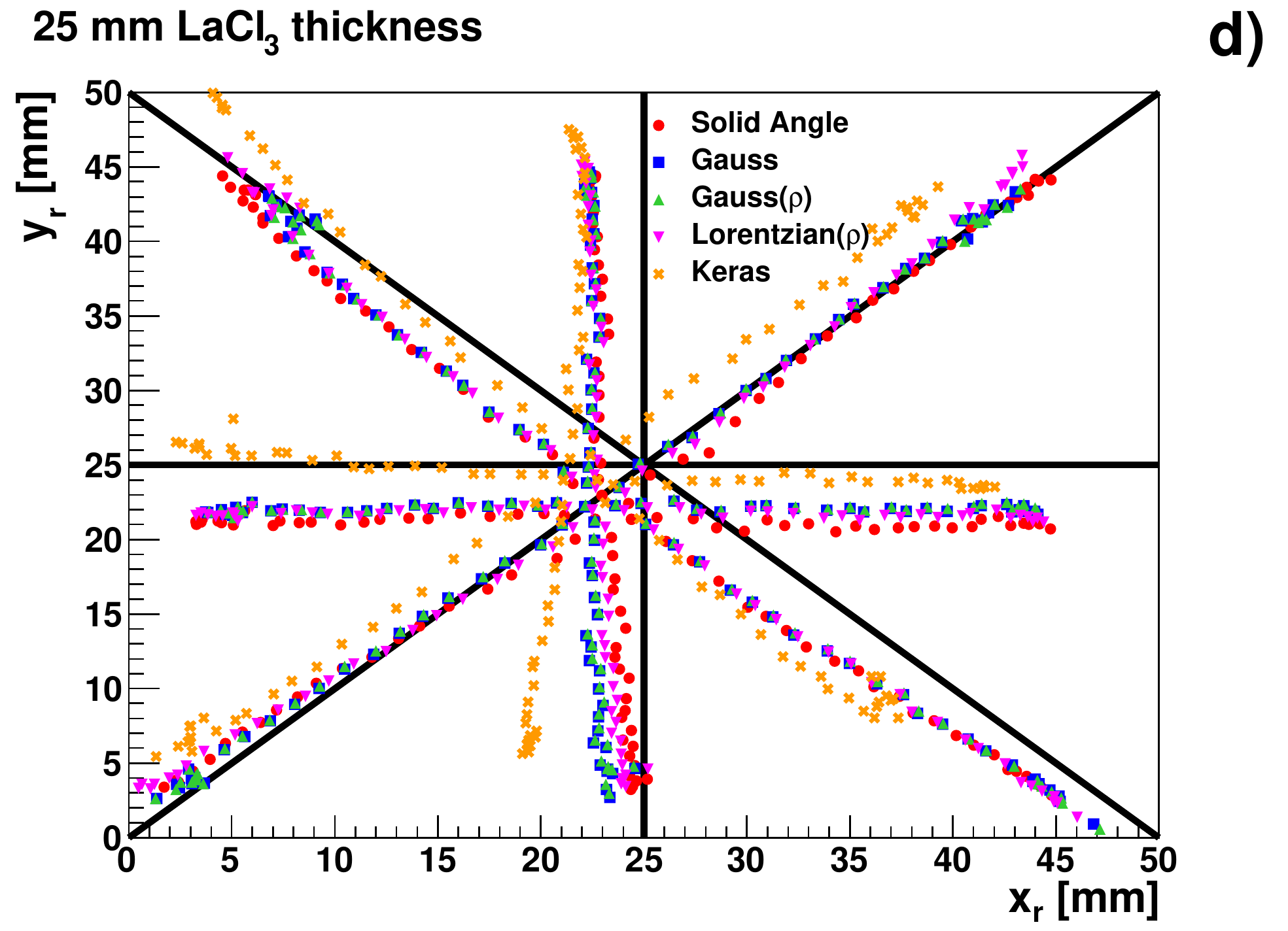} \\ 
  \includegraphics[width=\columnwidth]{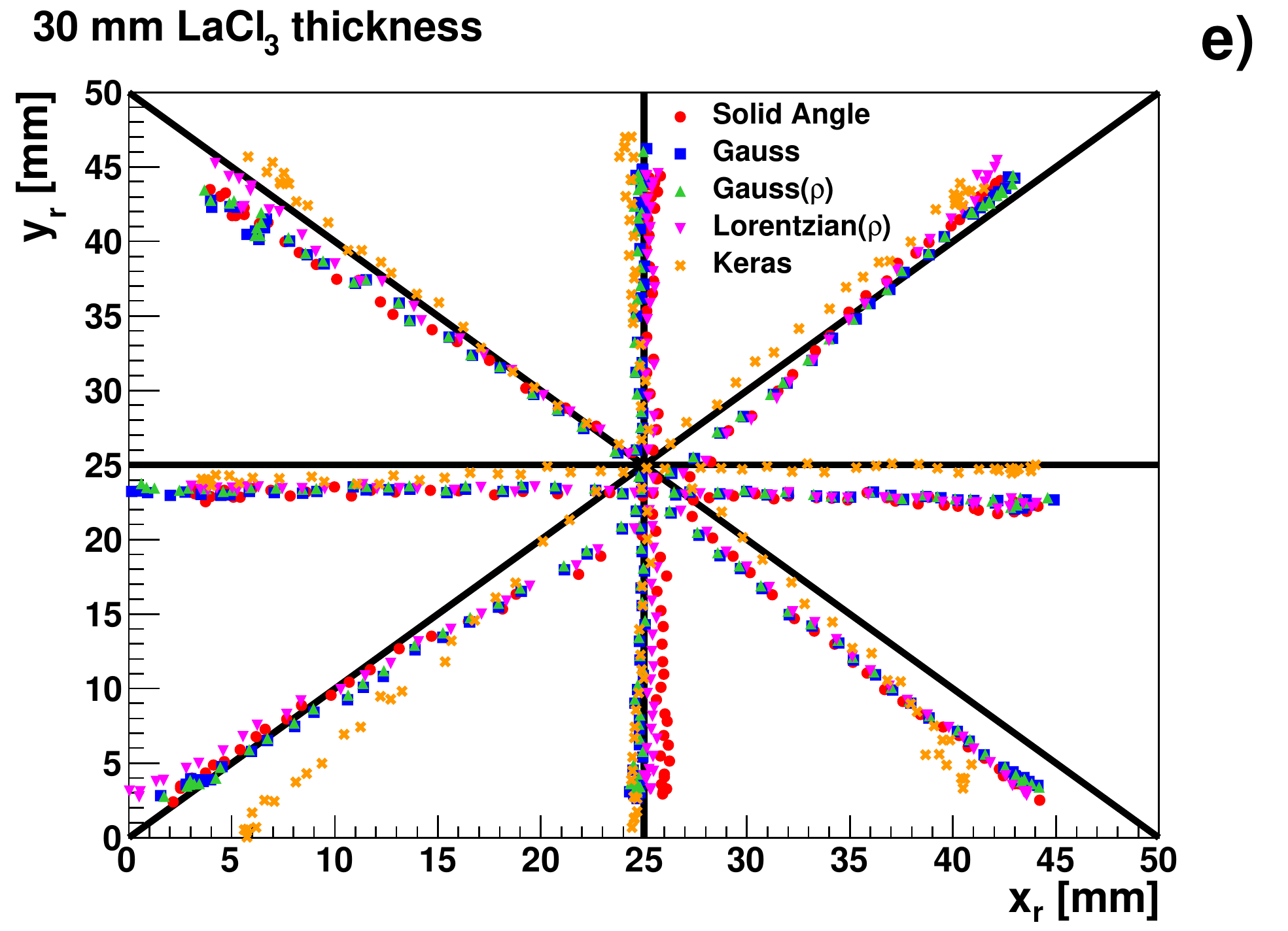} &
\end{tabular}
 \end{center}
\caption{SVM corrected linearity diagram  for the different \lacls crystal thickness. From the top left to the bottom right, the results are ordered as follow: 10~mm, 15~mm, 20~mm, 25~mm and 30~mm. The results for {\it{Solid angle}}, {\it{Gauss}}, {\it{Gauss($\rho$)}}, {\it{Lorentzian($\rho$)}} and {\it{Keras}} are shown by the red, blue, green, pink and orange dot points, respectively.}
\label{fig:Corrected_linearity_Diagram}
\end{figure*}
After the fitting procedure, the SVM correction model was applied to the individual events, transforming the reconstructed $\gamma$-ray position into a corrected reconstructed position. In a similar way as before, the Robust Covariance method was applied to get the mean of the distribution. New linearity diagrams for the SVM-corrected positions are produced for the different crystal thicknesses. The latter are shown in  Fig.~\ref{fig:Corrected_linearity_Diagram}. As it can be appreciated by comparing to the raw linearity diagrams shown in Fig.~\ref{fig:linearity_Diagram}, the compression effects are now remarkably reduced, thereby achieving a similar linearity for every \lacls crystal thickness and methodology. This is especially true in the case of the 30~mm thick crystal. A special case is the 25~mm thick \lacls crystal, for which the large deviations already present in the raw data (Fig.~\ref{fig:linearity_Diagram}) become significantly magnified after applying the SVM linearity corrections.

In some cases, small differences between the determination of the edge of the crystal and the $xy$-gantry positioning introduce an additional bias or mismatch between the reconstructed and ideal positions in the linearity diagrams.The comparison of bias as a function of the known position after the SVM correction is also displayed as dashed lines in the panels of Fig.~\ref{fig:Bias}. For the \lacls crystal with a thickness of 15~mm the results obtained in this work are comparable to other similar works~\cite{8871159,5783323}. In particular, the values obtained in the central region of the crystal are comparable to the results reported using Voronoi diagrams~\cite{8871159}.

In order to quantify the remaining compression observed in the corrected positions with respect to the true positions, the ratio between the corrected FoV and effective crystal area is calculated again as a function of the crystal thickness in a similar fashion as before. For comparison purposes, the results are displayed, together with the results without the SVM linearity correction, in Fig.~\ref{fig:Useful_Field_of_View}. The figure shows that the relative FoV, after the SVM calibration, becomes significantly larger for all crystal thicknesses above 10~mm. For the 25 mm and 30 mm thick \lacls crystals the improvement in FoV ranges between a factor 2 and 3.

In summary, the SVM correction for all models and \lacls crystals yields similar results in terms of spatial resolution. However, the main impact of the SVM calibration is on the detector FoV, where it helps to significantly mitigate pin-cushion effects particularly towards increasing crystal thickness. It is worth to note that the {\it{Solid Angle}} model shows an overall performance which is slightly better than the rest of approaches in terms of linearity and bias. From the five models investigated, the second best option seems to be the {\it{Lorentzian($\rho$)}} model, followed by the two Gaussian models. The CNN technique based on {\it{Keras}} yields a rather limited performance, both in terms of resolution and linearity. However, the results found here with CNNs are comparable to other works using MC data as input for the training of the model~\cite{4545078,1344371,Wang_2013,Iborra_2019}.

\section{Resolution in the transverse crystal plane}\label{sec:resolution}
\begin{figure}
\begin{center}
 \includegraphics[width=\columnwidth]{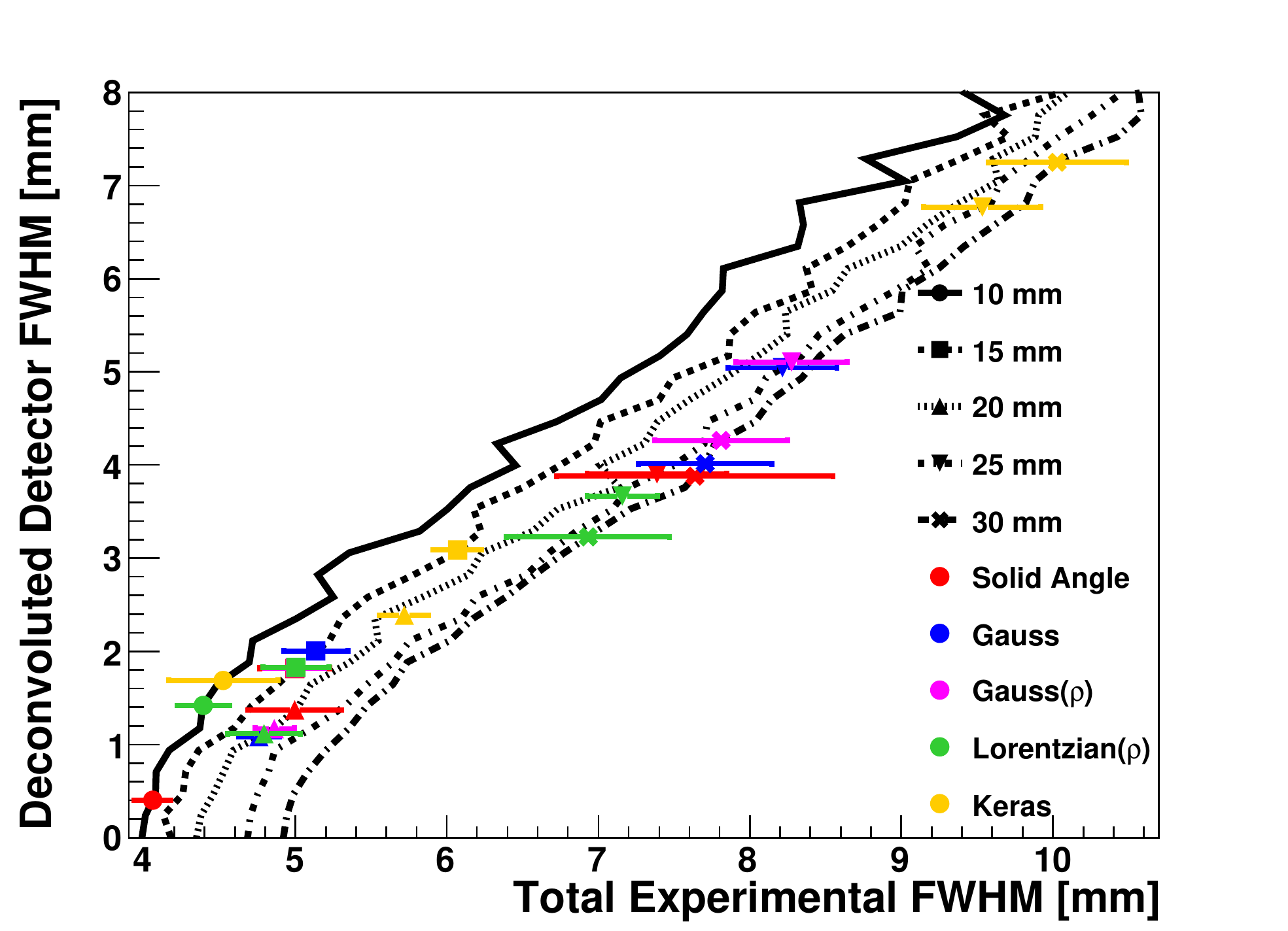}
 \end{center}
\caption{Intrinsic detector resolution as a function of the total measured width, as derived from MC simulations. The functions for the different crystal thickness are represented by the several lines. The results for the models investigated are shown with different markers (see labels and text for more details).}
\label{fig:resolution_curve}
\end{figure}

It is not possible to directly measure the intrinsic detector spatial resolution because the width of the reconstructed position distribution is affected by the divergence of the collimated $\gamma$-ray beam. In order to account for this spread and infer the intrinsic detector resolution MC simulations were performed using the \textsc{Geant4} toolkit~\cite{ALLISON2016186}. To this aim, a realistic realization of the experimental setup was included in the simulation, paying special attention to the sensitive distances and the hole diameter of the different collimators as it shown in Fig.~\ref{fig:MC_setup} of Sec.\ref{sec:Exp_setup}. As described in Ref.~\cite{BABIANO20191}, the relationship between the intrinsic spatial resolution and the total measured experimental width can be graphically represented in order to deconvolve the value of interest. This is shown in Fig.~\ref{fig:resolution_curve}, where the lines represent the MC simulated trends for each crystal thickness and the different symbols indicate, on top of each line, the resolution obtained with each one of the models. The total experimental \fwhms resolutions shown in that figure and reported in Tab.~\ref{tab:resolutions} were calculated as the average of all the available data-points from the 60 $\times$ 60~mm$^2$ scan, excluding the first and last 10~mm of irradiations on each crystal side.

\begin{table*}
    \centering
    \begin{tabular}{c|c c c c c} \hline
   \lacls thickness & & Average \fwhms (mm)& & \\
   (mm) & {\it{Solid Angle}} & {\it{Gauss}} & {\it{Gauss($\rho$)}} & {\it{Lorentzian($\rho$)}} & {\it{Keras}}\\ \hline \hline
    10 & 0.30(22) & 0.25(3) & 0.25(3) & 0.7(6) & 1.4(12) \\
    15 & 1.7(7) & 2.0(7) & 1.8(6) & 1.8(7) & 3.1(4) \\
    20 & 1.4(8) & 1.0(5) & 1.2(4) & 1.1(6) & 2.4(4) \\
    25 & 3.9(11) & 5.0(9) & 5.0(9) & 3.6(6) & 6(1) \\
    30 & 4(2) & 4(1) & 4(1) & 3(1) & 7(1) \\ \hline
    \end{tabular}
    \caption{Average deconvoluted detector \fwhms for all the detector sizes and methods applied in this work, excluding the first and last 5 mm of crystal irradiations.}
    \label{tab:resolutions}
\end{table*}

%In order to have a determination of the best attainable intrinsic spatial resolution, with negligible statistical and systematic uncertainty, a dedicated scan was carried out. The latter consisted of a 600s scan in the central part of the crystal using a stepsize of 0.2 mm in both axis. An a total of 961 irradiations covering a total area of 6.2x6.2 mm$^{2}$.

%In order to have a determination of the best attainable intrinsic spatial resolution, with negligible statistical and systematic uncertainty, the average total experimental \fwhm, excluding the first and last 5 mm of the crystal, was calculated. For this value the deconvolutedThe deconvoluted values are summa 

\begin{figure}
\begin{center}
\begin{tabular}{c c}
  \includegraphics[width=0.5\columnwidth]{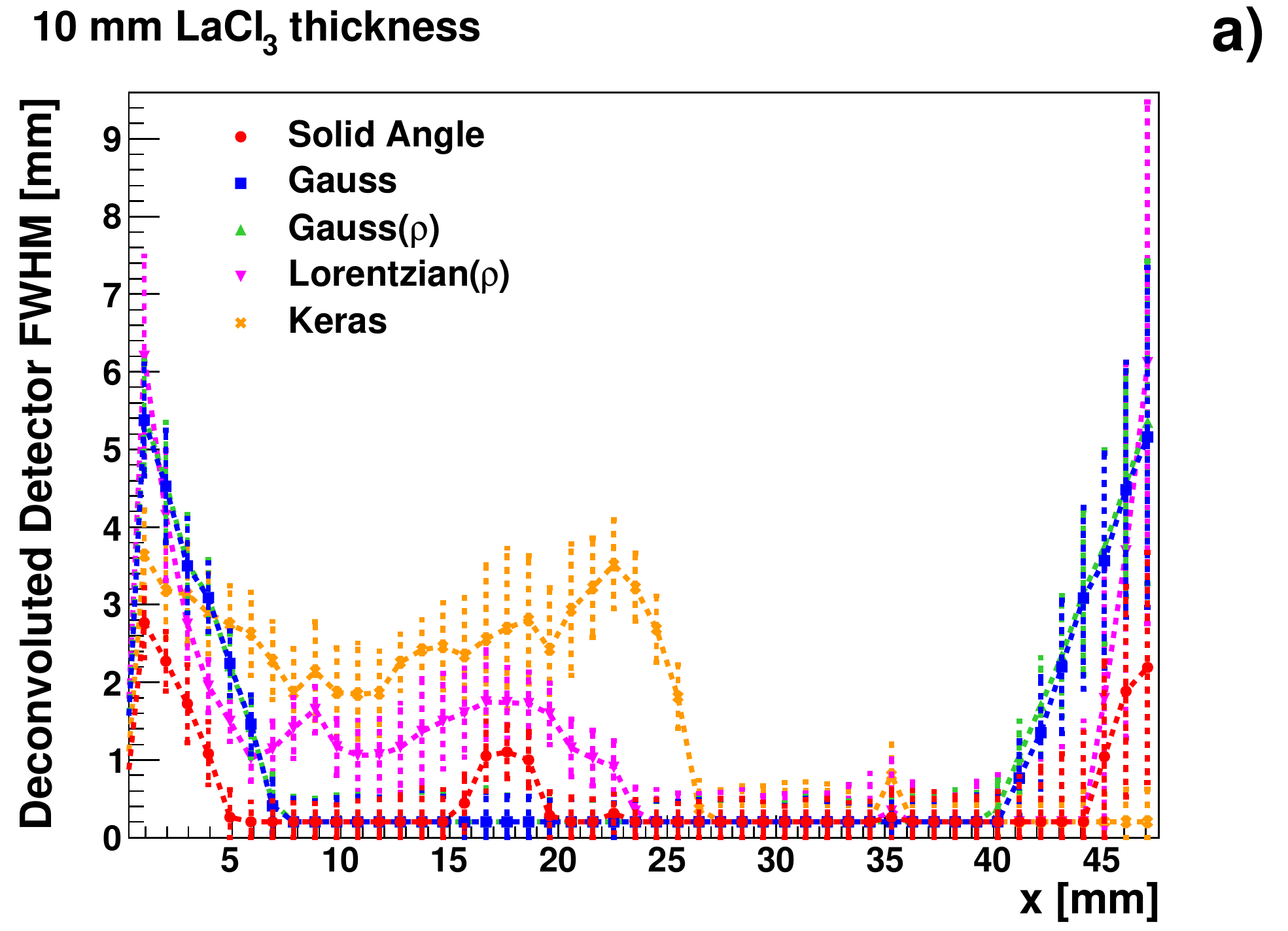} &
  \includegraphics[width=0.5\columnwidth]{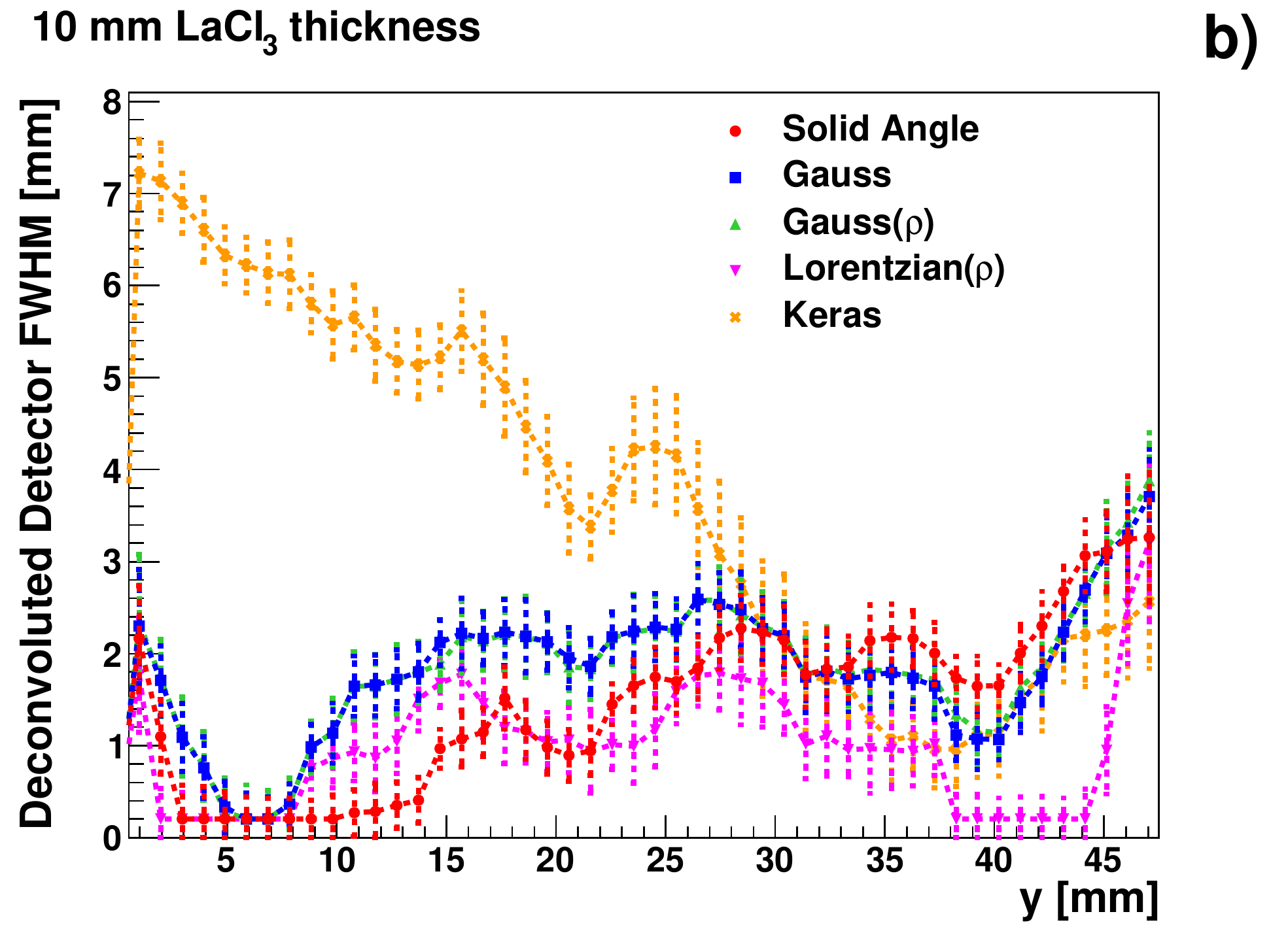} \\ 
  \includegraphics[width=0.5\columnwidth]{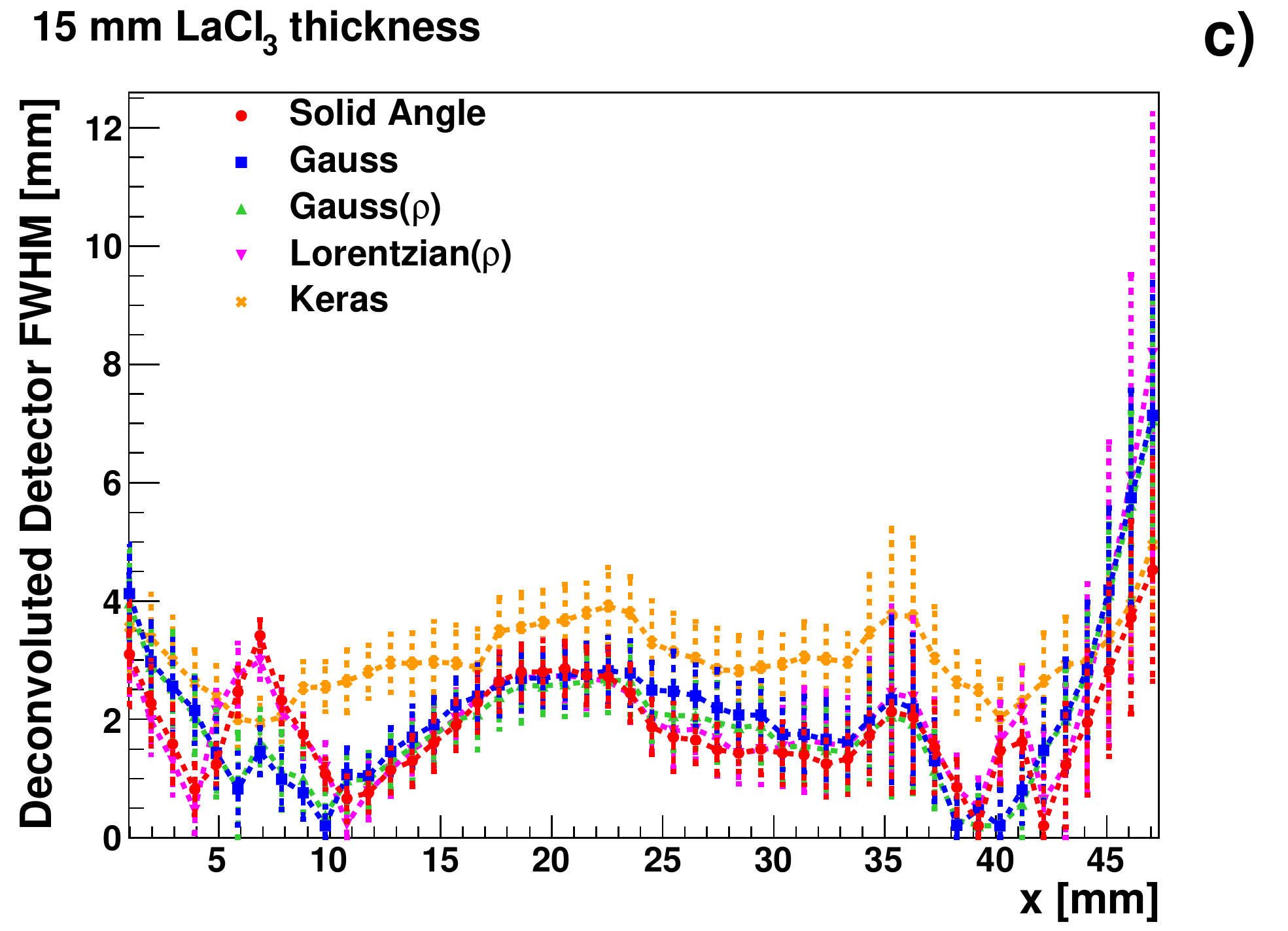} &
  \includegraphics[width=0.5\columnwidth]{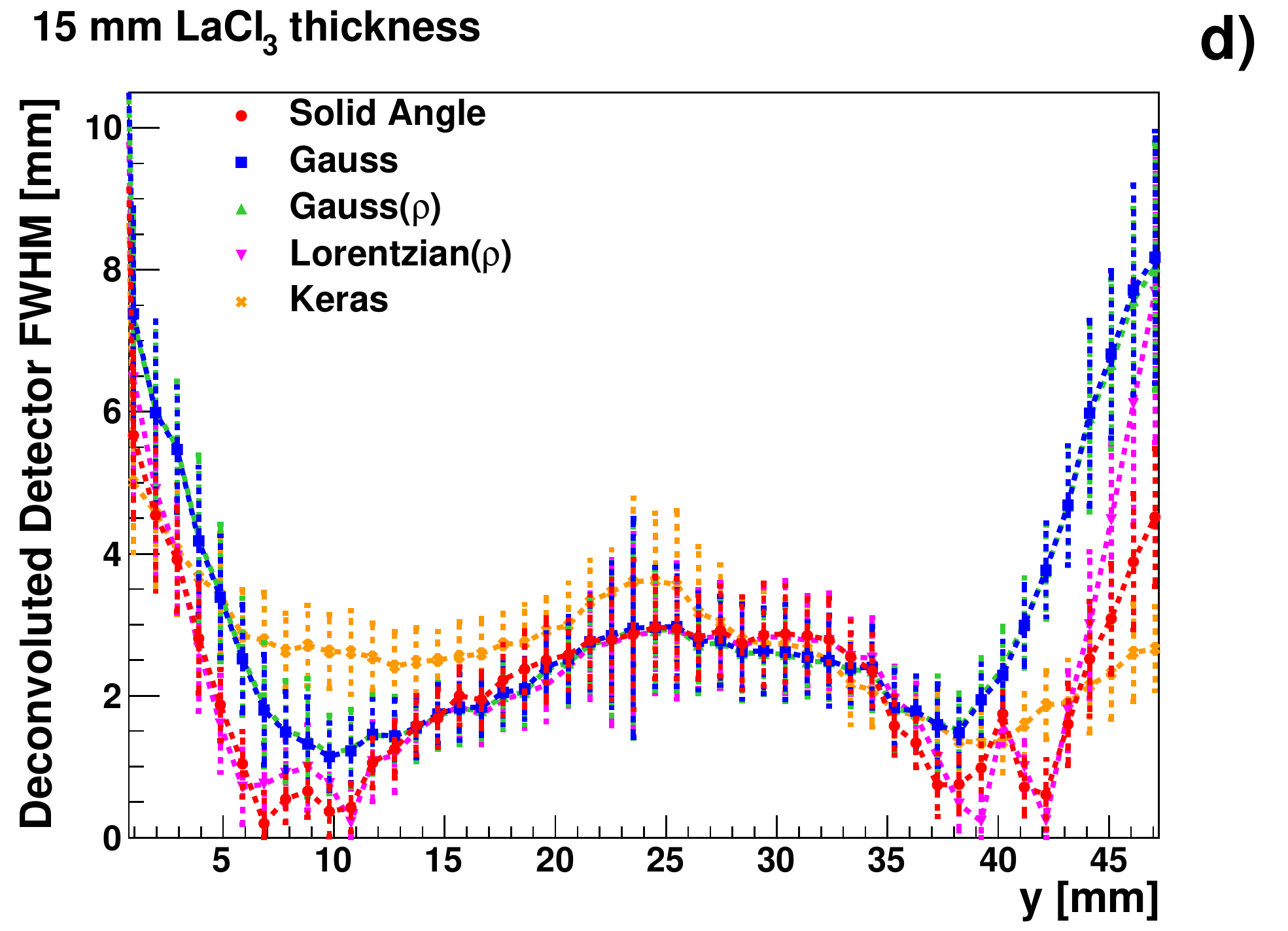} \\ 
  \includegraphics[width=0.5\columnwidth]{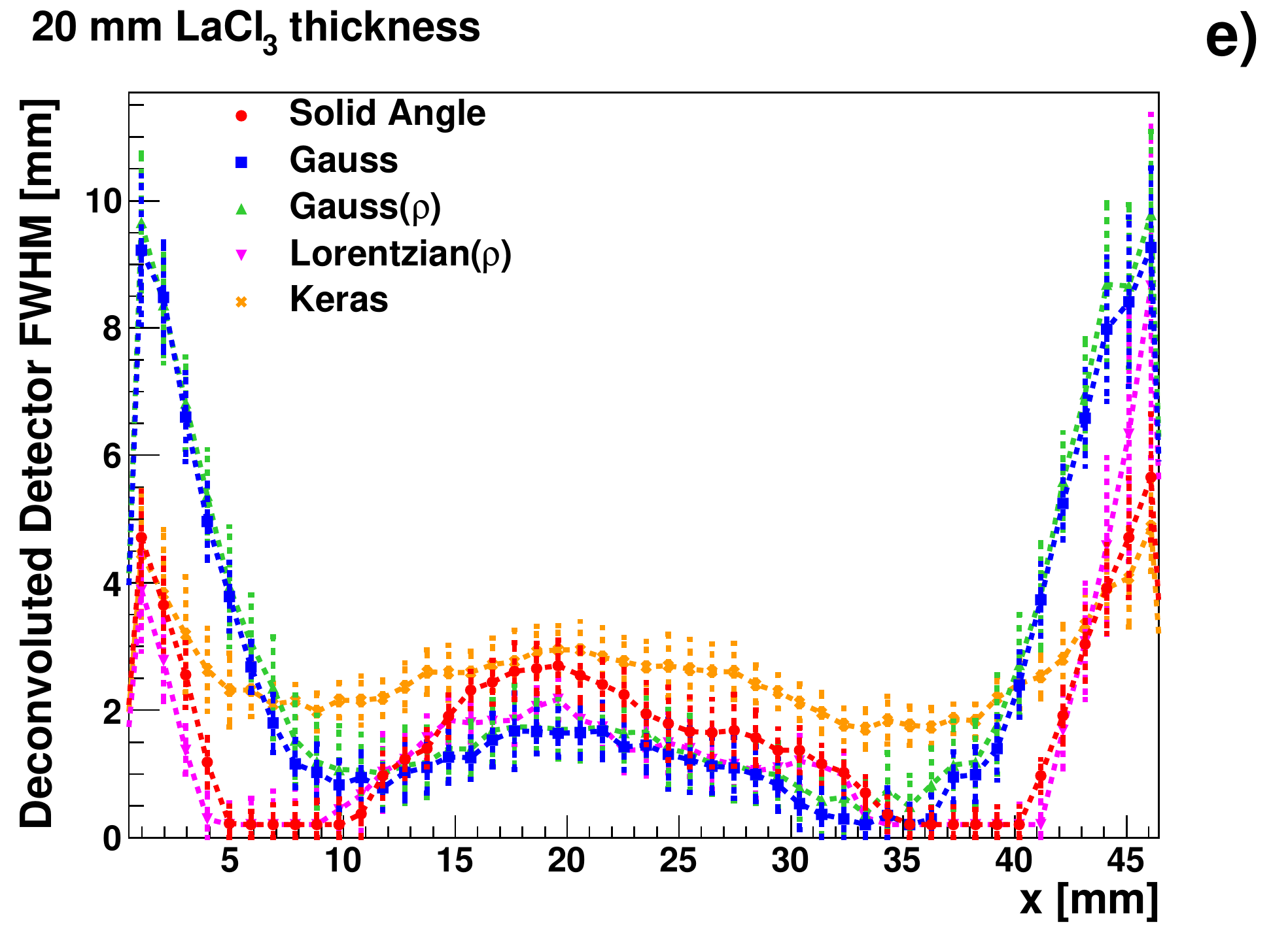} &
  \includegraphics[width=0.5\columnwidth]{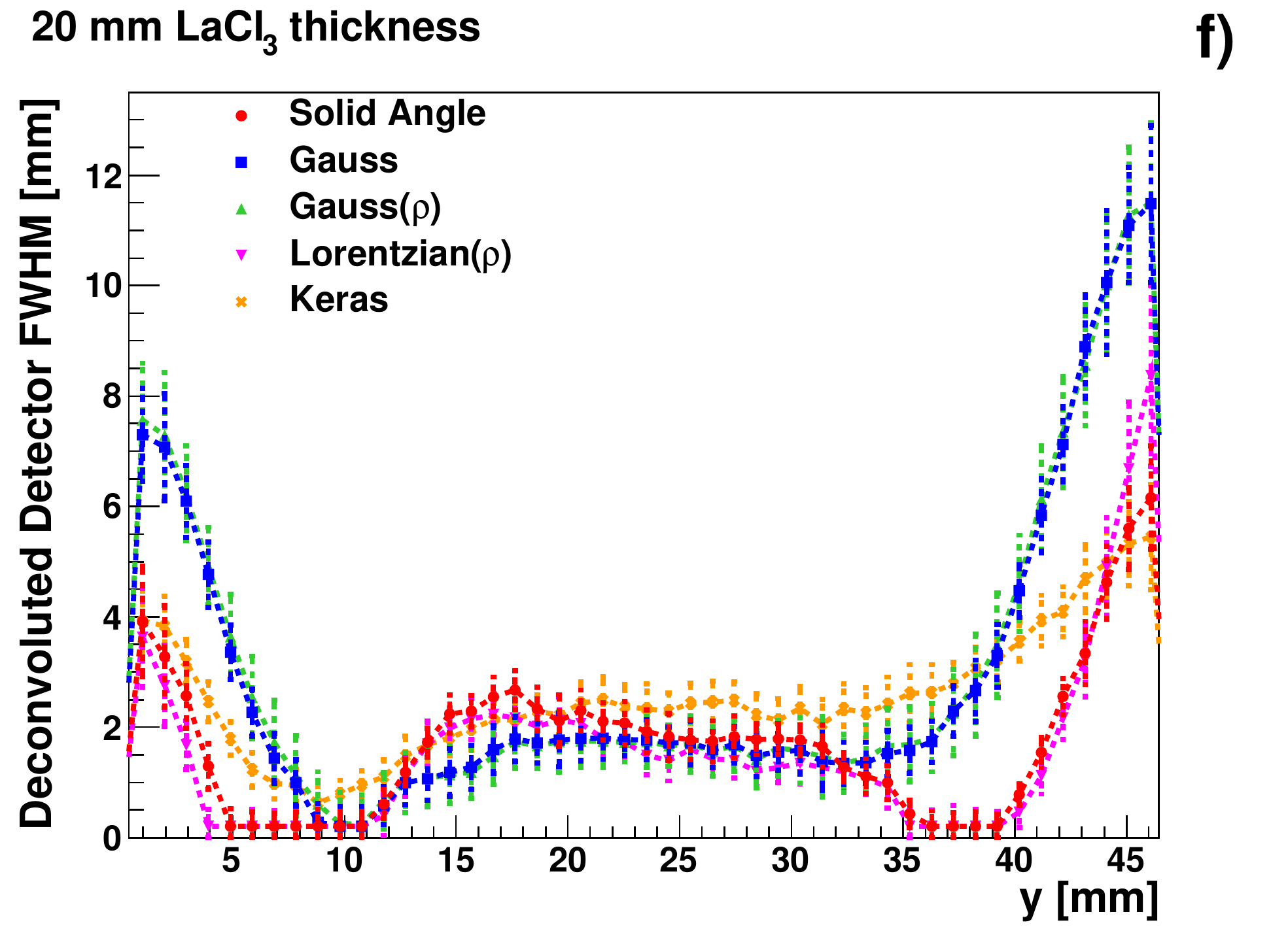} \\ 
  \includegraphics[width=0.5\columnwidth]{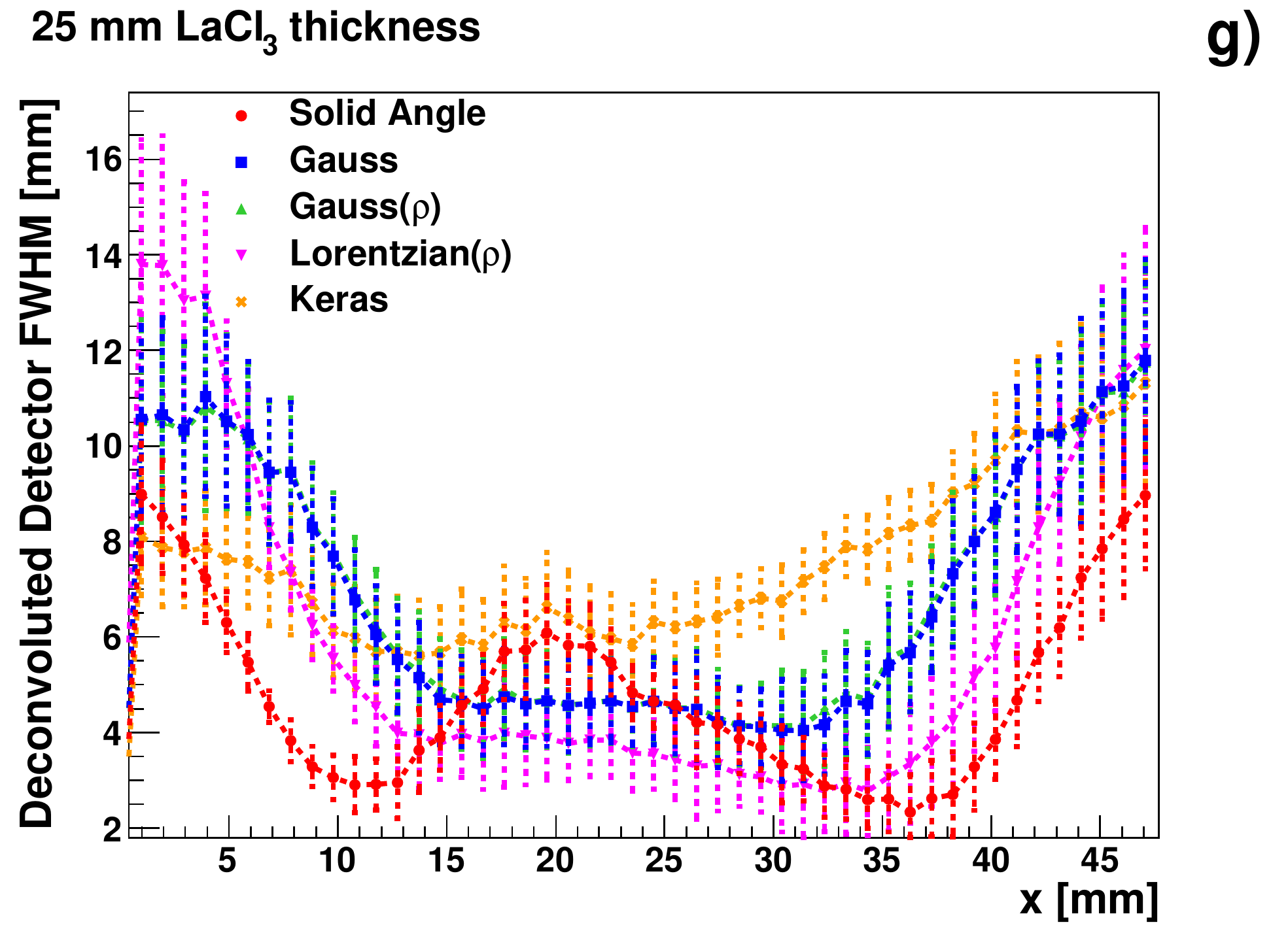} &
   \includegraphics[width=0.5\columnwidth]{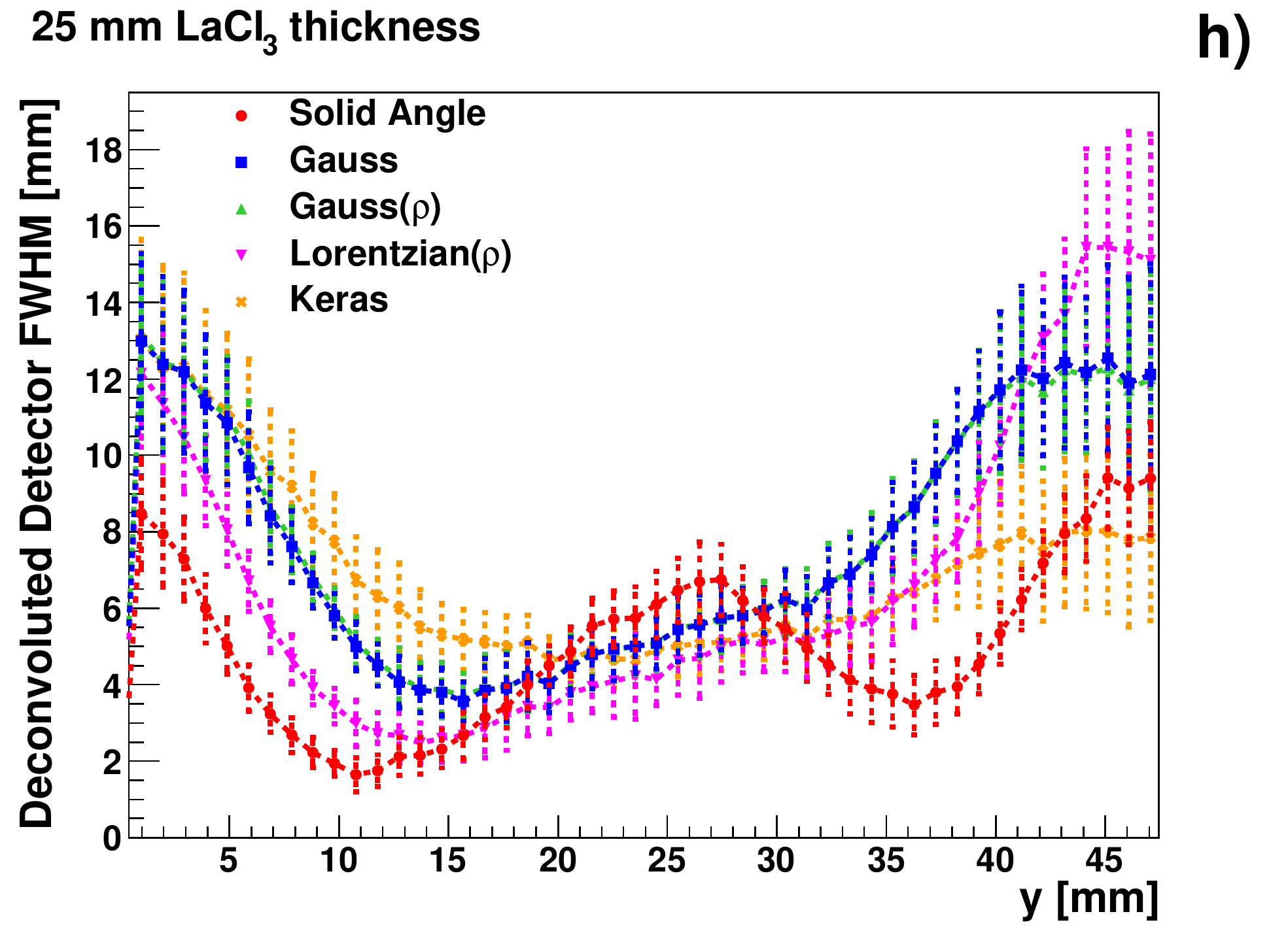} \\ 
  \includegraphics[width=0.5\columnwidth]{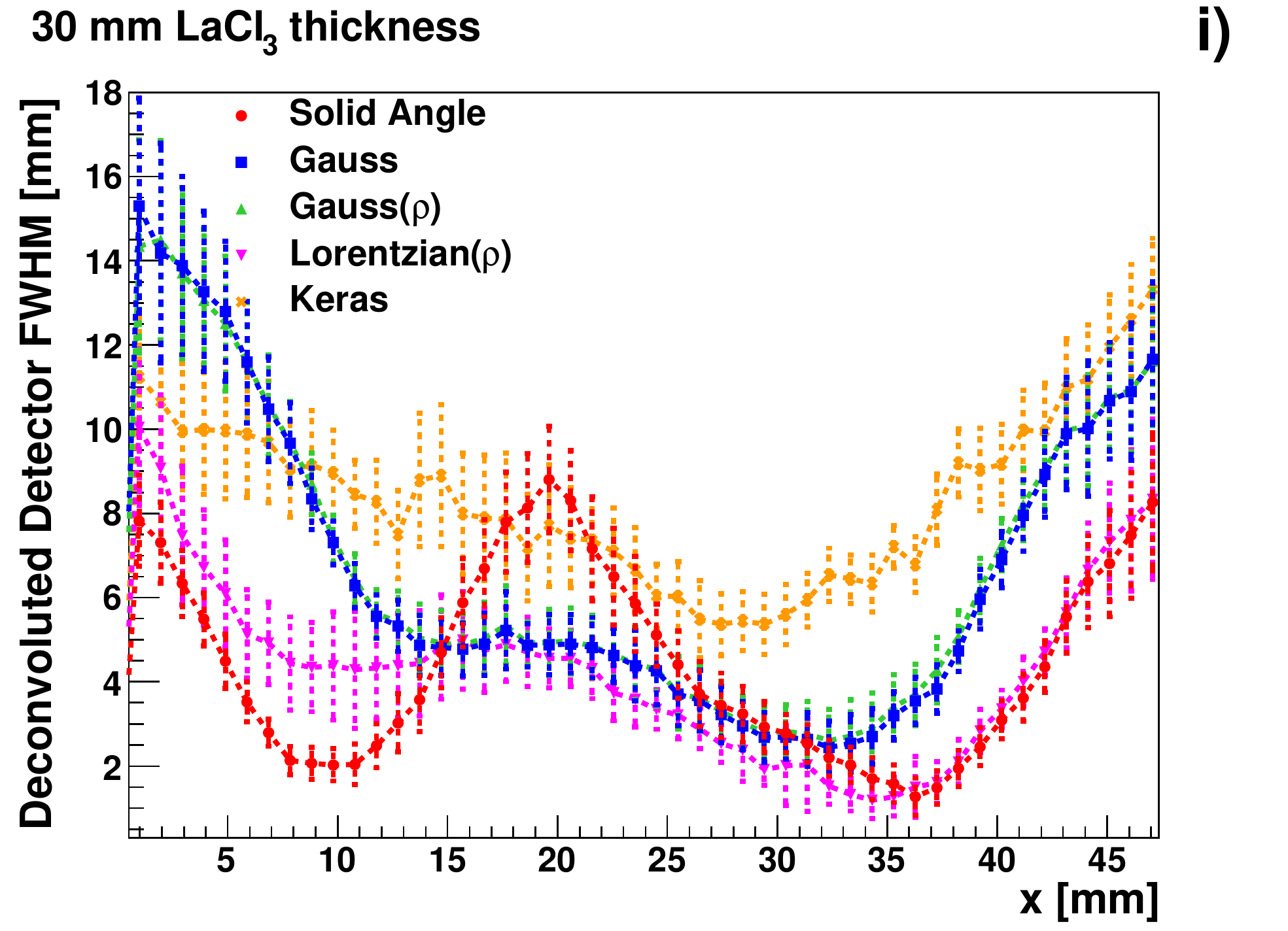} &
  \includegraphics[width=0.5\columnwidth]{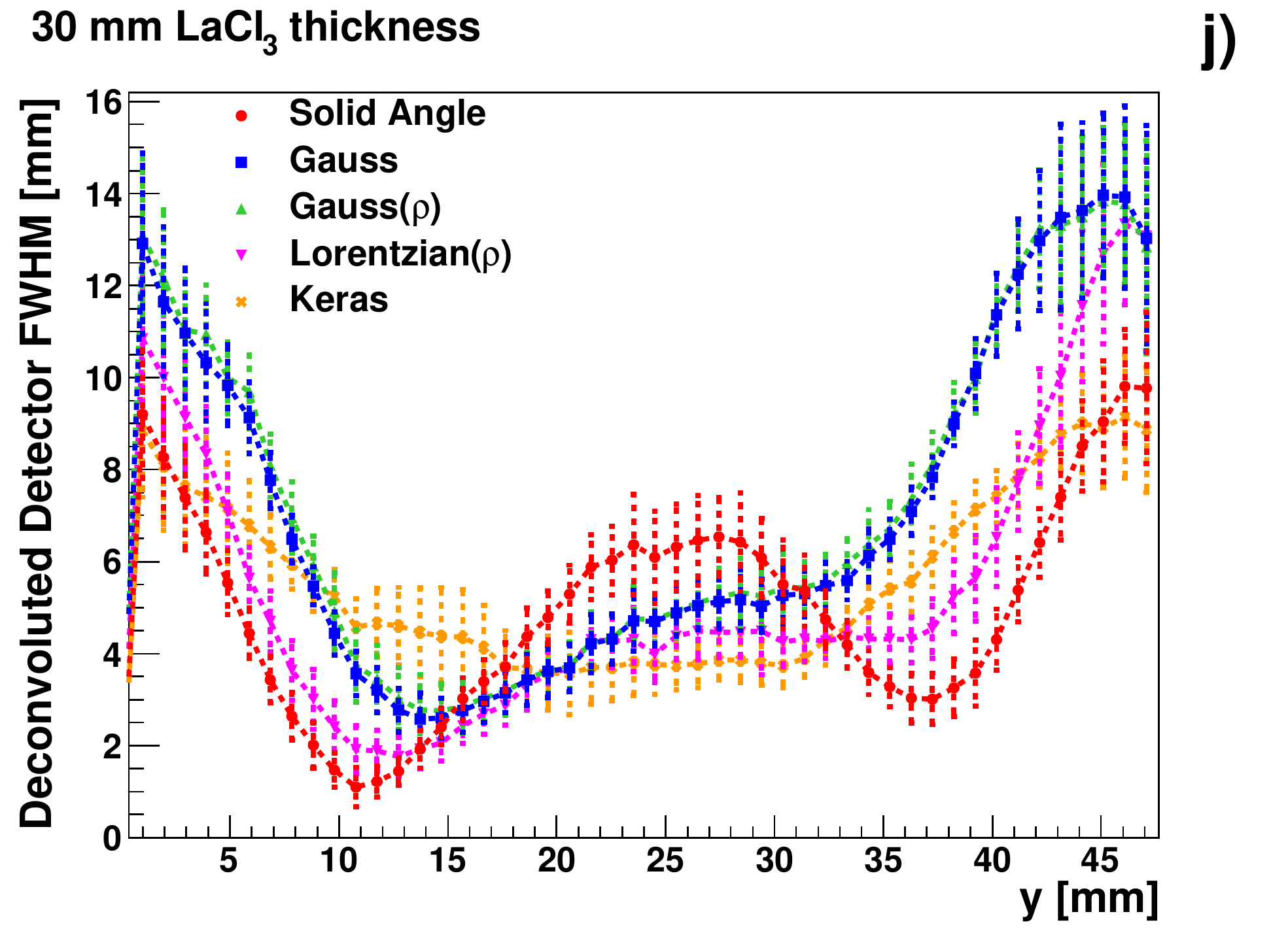} \\ 
\end{tabular}
 \end{center}
\caption{Average and deconvoluted detector resolution (\fwhm) after the detector calibration as a function of the known position for the five \lacls-crystal sizes. Panels a), c), e), g), and i) show the results for the horizontal lines. Panels b), d), f), h), and j) show the results for the vertical lines.}
\label{fig:FWHM}
\end{figure}

The average intrinsic or deconvoluted resolution (\fwhm) as a function of the reconstructed positions for the vertical and horizontal lines after the detector SVM calibration are shown by the different lines in Fig.~\ref{fig:FWHM}. The resolution, as in the case of the bias (Fig.~\ref{fig:Bias}), becomes larger in the peripheral region of the crystal. Overall, the results obtained with the {\it{Solid angle}} model for the different thicknesses show the best performance. In the case of the 10~mm thick crystal, the \fwhms is comparable to other works using smaller crystal sizes~\cite{Seifert_2013,Borghi_2016} and the one reported in our previous work~\cite{BABIANO20191}. For the 15~mm thickness, the results obtained using the {\it{Solid angle}} are comparable with the results reported using Voronoi diagrams~\cite{8871159}. For thicker \lacls crystals of 20~mm, 25~mm and 30 mm the results reported here seem to be slightly worse than those presented in our previous work~\cite{BABIANO20191}. This is due to the fact that in the present study we did not apply any restriction on the $\chi^{2}$-values of the fitting procedure. Avoiding such event rejection based on the $\chi^{2}$-values will enhance the final efficiency of i-TED, a key aspect for its final purpose.

As expected, the best results for the average resolution are obtained with the 10~mm thick crystal in the $x,y$-plane. On average, \fwhms resolutions of 1~mm or less can be attained by means of the {\it Solid Angle}, {\it Gauss} , {\it Gauss($\rho$)} and {\it Lorentzian ($\rho$)}. Slightly worse values, of 1.4~mm \fwhm, are obtained for the {\it Keras} approach. It is worth to mention that the average value quoted for the {\it Solid Angle}, {\it Gauss} and {\it Gauss($\rho$)}, which  approaches values of $\sim$0.3~mm \fwhms has to be taken with caution and represents rather a bottom-limit. Indeed, given the steepness of the divergence-deconvolution function (Fig.~\ref{fig:resolution_curve}) in that region the uncertainty to be ascribed to that value is rather large, of about 50\%.

For the 15~mm thick crystal the spatial resolution obtained with all analytical models is similar, with average values of about 1.8~mm \fwhms. A factor of two worse resolution was found, on average, for the CNN approach based on {\it Keras}. Surprisingly, for the 20~mm thick crystal better overall results are obtained, when compared to the 15~mm thick crystal. In this case resolutions between 1~mm and 1.4~mm \fwhms are obtained with all analytical models, a performance in between the results for the 10~mm and 15~mm thick crystals. There is no clear explanation for this result, and most probably can be ascribed to particular variations or imperfections between both crystals.

For thicker crystals a clear trend appears on the average resolution values. For crystal thicknesses between 25~mm and 30~mm one can observe, in general, a better performance by {\it Solid Angle} and {\it Lorentzian($\rho$)}, when compared to {\it Gauss} and {\it Gauss($\rho$)}. In this thickness range, the average resolution is of $\sim$4-5~mm for all the analytical models investigated. Therefore, a thickness of 20~mm seems to represent the limit where average resolutions below 2~mm \fwhms are attainable. Increasing the crystal thickness by 5~mm or 10~mm seems to have a significant impact on the average spatial resolutions in the transverse crystal plane. In view of these results, it would be certainly interesting to research crystals thicker than 30~mm in future studies with the aim of studying the evolution of this behavior and eventually determine the largest crystal thickness, which still enables a certain level of spatial sensitivity in the transverse plane. At present, at least for our application and considering also the preceding results, the use of crystals thicker than 30~mm is limited by the non-linearity and pin-cushion effects, rather than by the intrinsic position resolution in the $x,y$ plane.\newline
Finally, it is worth to mention that in terms of average resolution CNN approaches seem only competitive for thin crystals (10~mm). For larger thicknesses the average resolution obtained with Keras is significantly worse than that obtained with any analytical approach. This result is also in contrast with the conclusion obtained in the preceding section, where we found that in terms of linearity the CNN approach seems to become competitive only for very thick crystals ($\gtrsim$30~mm).

\section{Depth of interaction}\label{sec:Z_axis}

\begin{figure}
\begin{center}
  \includegraphics[width=\columnwidth]{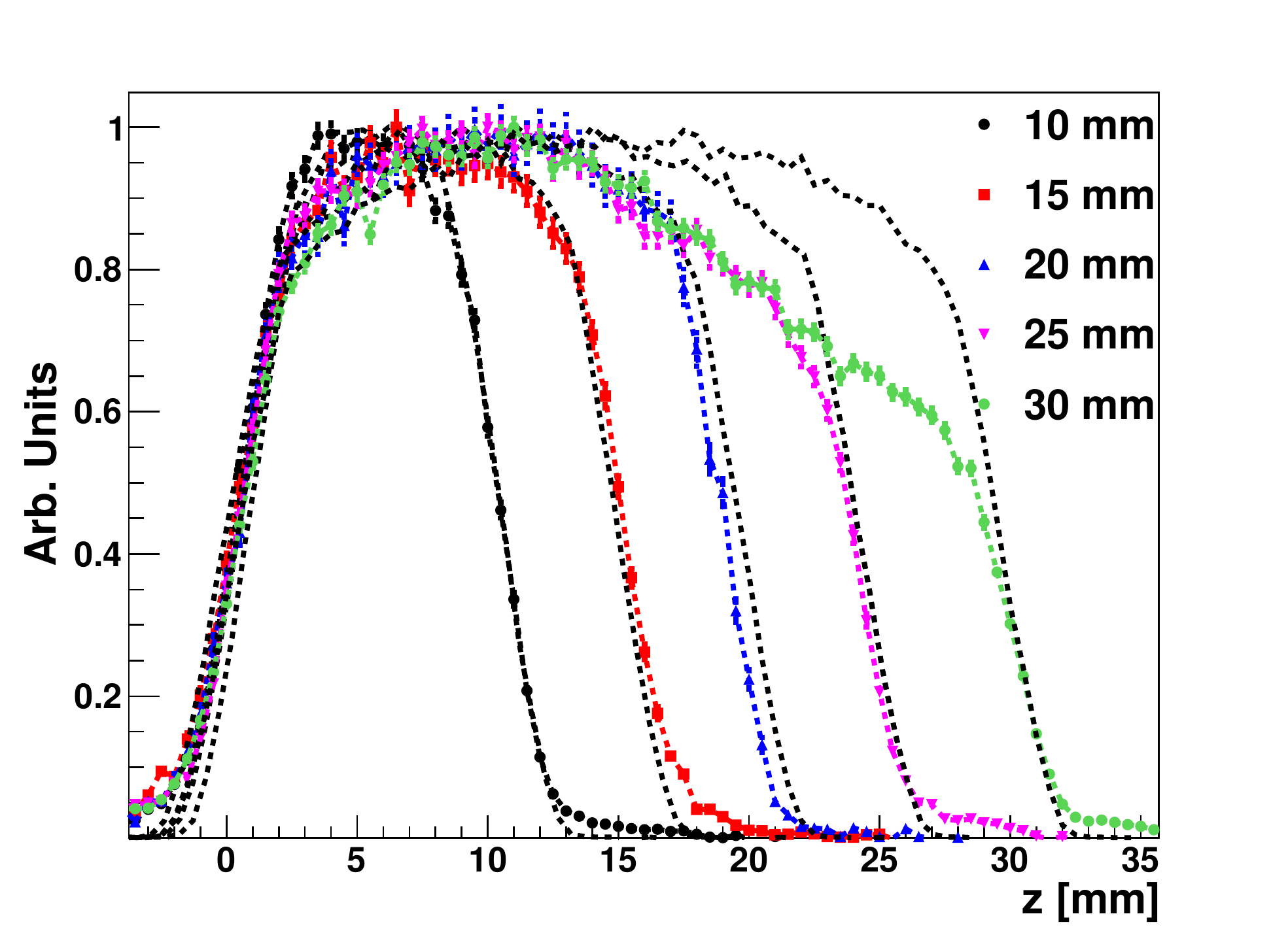}
 \end{center}
\caption{Detected 511~keV time-coincidence rate profile for the several \lacls crystals. The black dashed lines is the expected behaviour calculated from the MC simulations.}
\label{fig:doi_coincidence_rate}
\end{figure}

The calibration in the $z$-axis of the crystal was made using the experimental data from the lateral irradiations of each detector, which was described in section~\ref{sec:Exp_setup}. The high granularity of the irradiation mesh (0.5~mm steps) used for these measurements helps to perform an accurate calibration along this specific axis. 
The registered 511~keV time-coincidence rate $z$-profile is displayed in Fig.~\ref{fig:doi_coincidence_rate} for all \lacls crystal thicknesses. 
The behavior expected from MC-simulations of the experimental setup as it was described in section~\ref{sec:Exp_setup} is also shown in the same figure. For the sake of clarity, the $z$-profiles are normalized to the maximum value. For convenience, the starting point of the \lacls crystals is defined here as the 30\% of the maximum detected counting rate. The $z$ coordinate is defined as the perpendicular direction from the SiPM surface to the $\gamma$-ray interaction position in the crystal.

In Fig.~\ref{fig:doi_coincidence_rate} a clear decrease in detection efficiency with increasing $z$-distance is observed for the two thickest crystals. At the moment we do not have an explanation for such effect. The fact that it appears always for $z$ values beyond $\sim$20~mm might indicate that it may be related to difficulties or variations along the growing process of the \lacls crystals.

\begin{figure}[b!]
\begin{center}
  \includegraphics[width=\columnwidth]{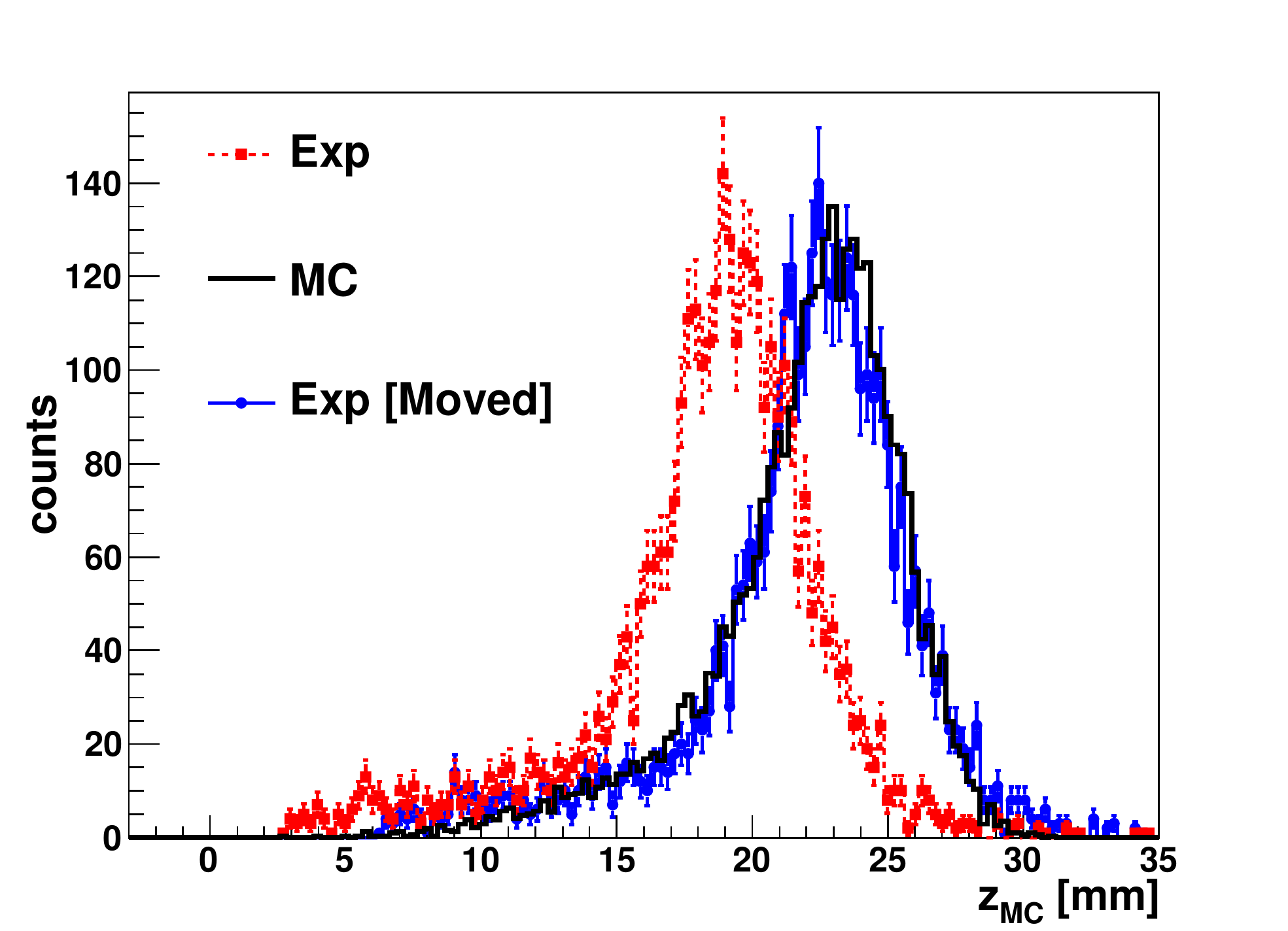}
 \end{center}
\caption{Experimental and MC distributions obtained from a single lateral irradiation using the {\it{Solid angle}} model. The dashed red curve corresponds to the raw experimental distribution. The MC distribution, convoluted with an arbitrary resolution $\sigma_{R}$ is shown by the solid black line. The shifted experimental distribution is displayed by the solid blue line. See text for details.}
\label{fig:doi_calibration_distribution}
\end{figure}

The $z$-axis calibration procedure for {\it{Solid angle}} and {\it{Keras}} models is carried out by comparing the experimental distribution for the reconstructed $z$-coordinate with the MC distribution, $z_{MC}$. Aiming at matching both distributions, a calibration resolution $\sigma_{R}$ is introduced in $z_{MC}$ and the experimental distribution is shifted by a calibration $z$-offset ($z,z_{MC}$). An example of the calibration is shown in Fig.~\ref{fig:doi_calibration_distribution} for the \lacls crystal with 25~mm thickness. 

\begin{figure*}
\begin{center}
\begin{tabular}{c c}
  \includegraphics[width=\columnwidth]{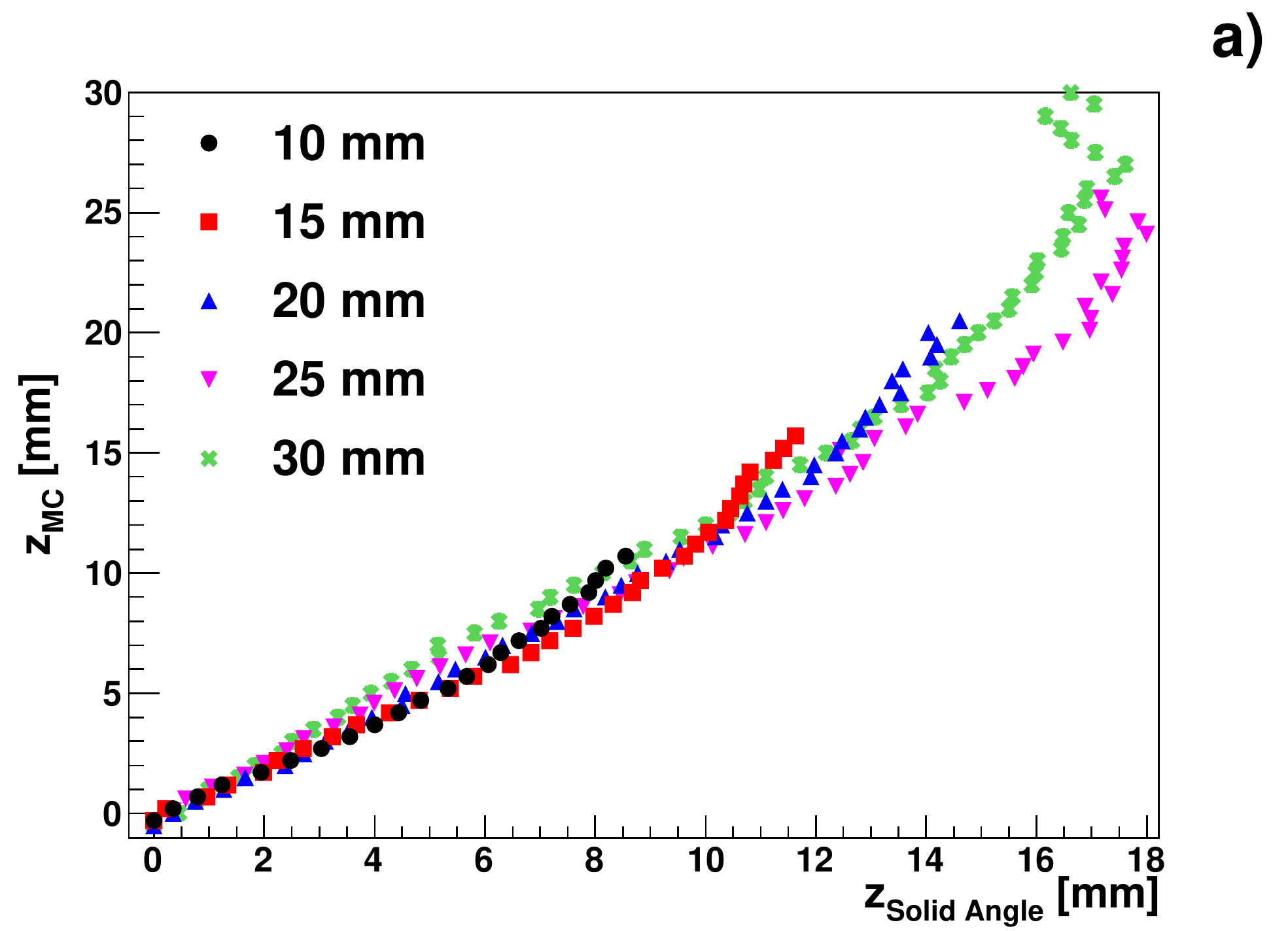} &
  \includegraphics[width=\columnwidth]{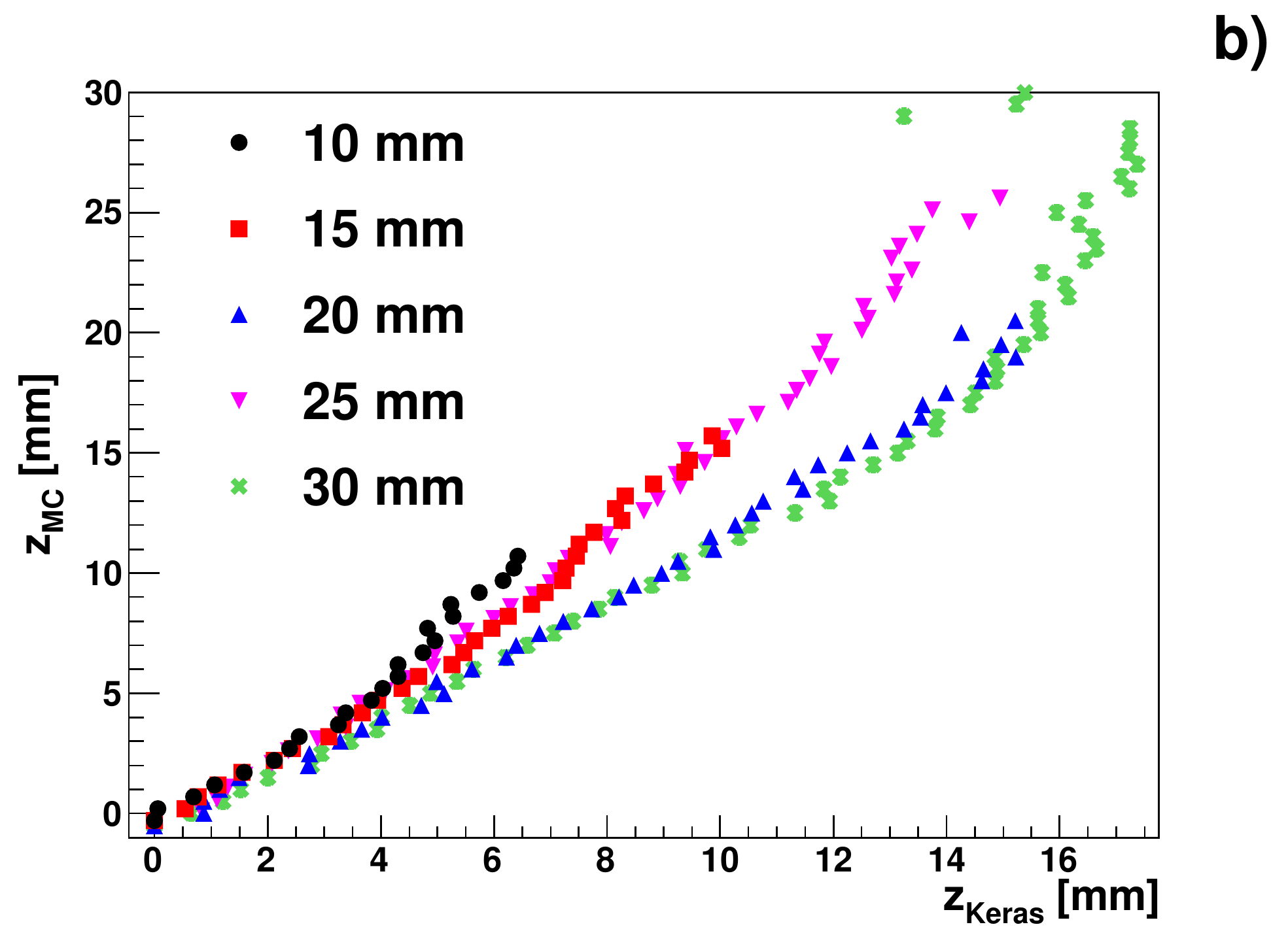} \\
  \includegraphics[width=\columnwidth]{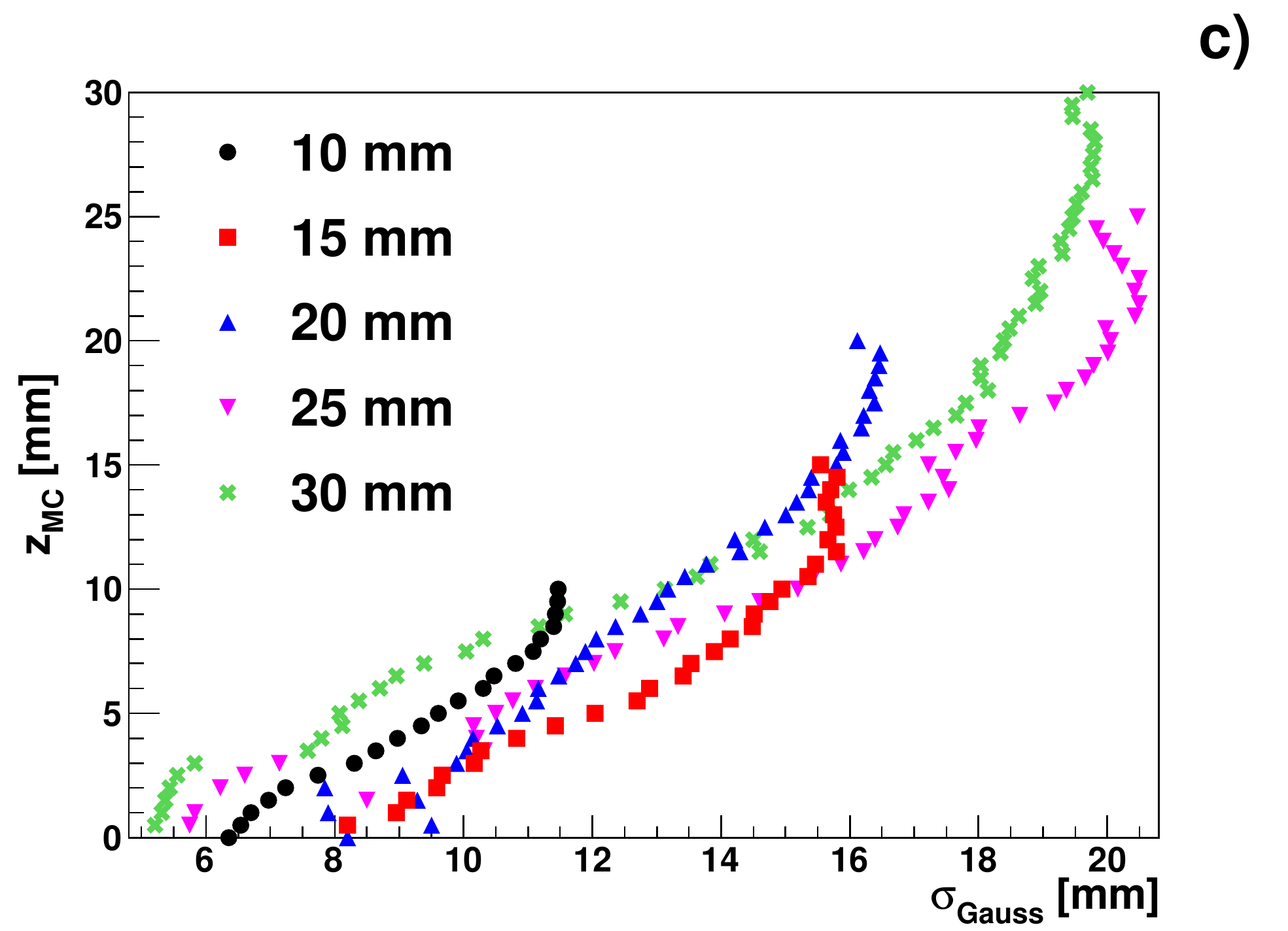} &
  \includegraphics[width=\columnwidth]{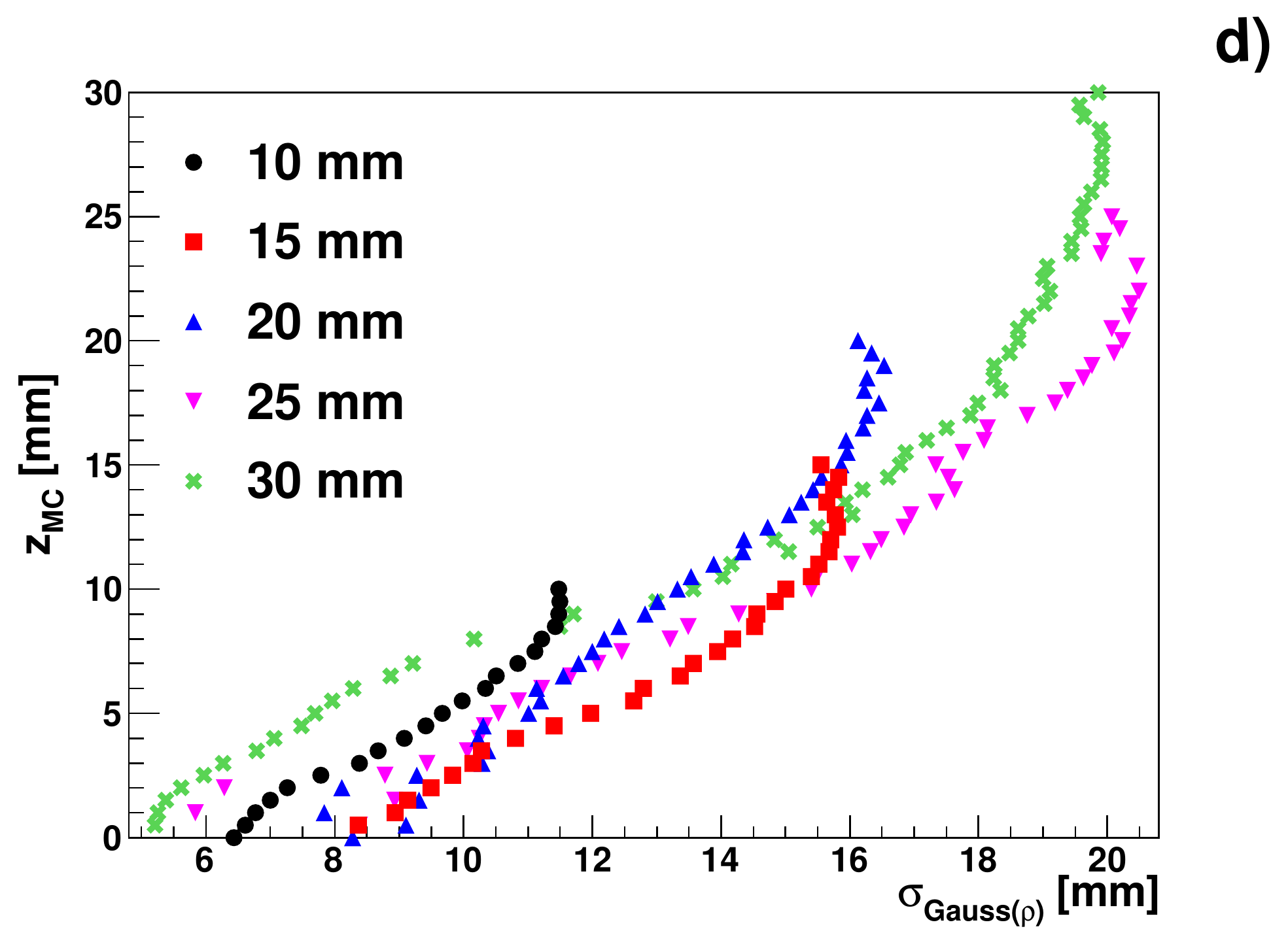}\\
  \includegraphics[width=\columnwidth]{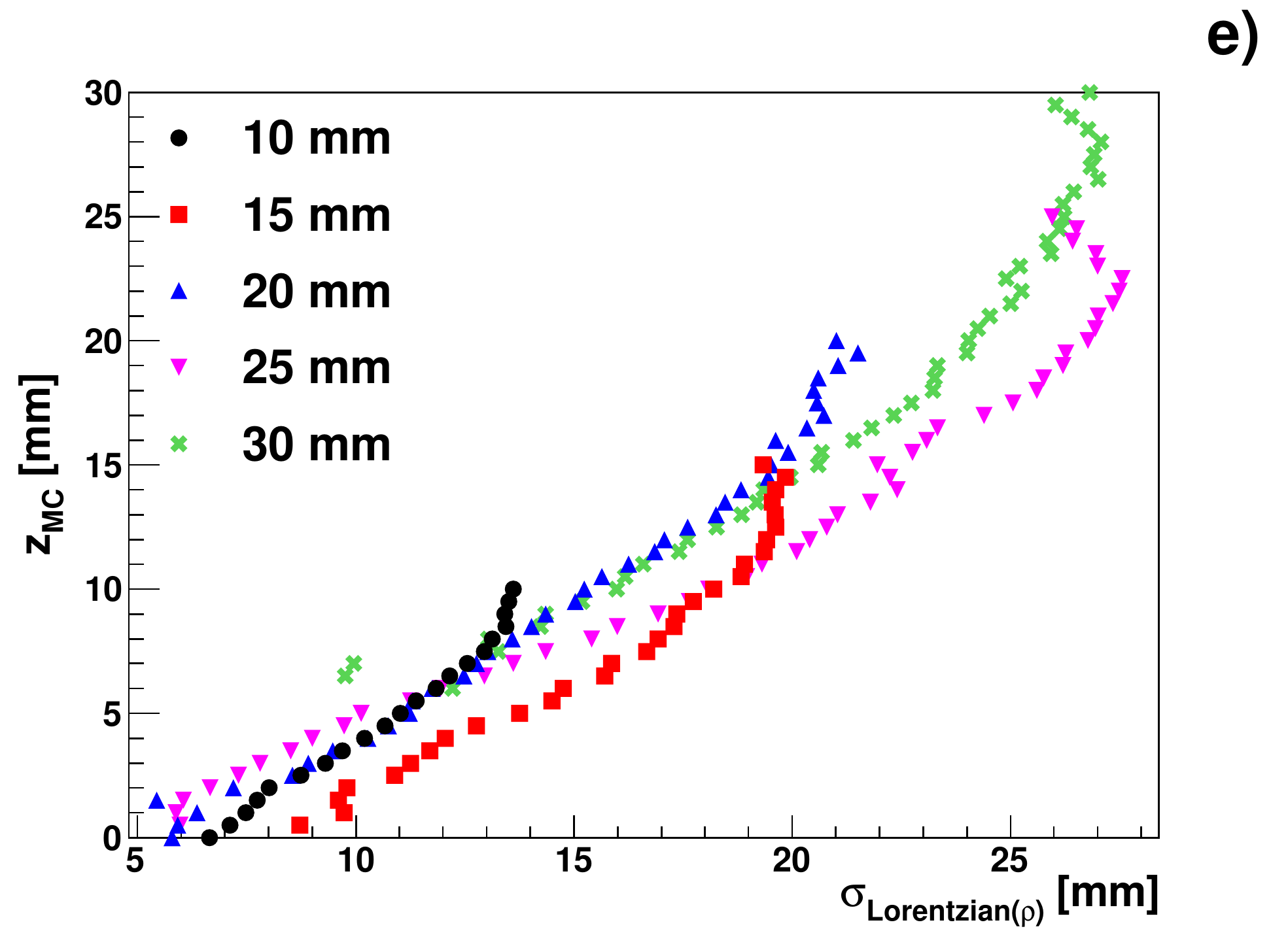} &
\end{tabular}
 \end{center}
\caption{Calibration curves for the different methodologies. From the top-left to the bottom right, the calibration curves correspond to {\it{Solid angle}}, {\it{Keras}}, {\it{Gauss}}, {\it{Gauss($\rho$)}}, and {\it{Lorentzian($\rho$)}}. The results for the several crystal thicknesses are labeled as follow: black, red, blue, pink and green for the 10~mm, 15~mm, 20~mm, 25~mm and 30~mm, respectively.}
\label{fig:doi_calibration}
\end{figure*}

Compared to the MC distribution (solid black line), the experimental distribution (dashed red line) is shifted toward low $z$ values. This effect can be interpreted as a compression in this axis, in a similar way as it happens in the other two crystal axis. In order to obtain the calibration $z$-offset for this particular irradiation, the experimental distribution is shifted toward the position of the MC distribution (solid blue line). The resolution is adjusted by the convolution of the MC distribution with a Gaussian function of width $\sigma_{R}$ until a good agreement is obtained between both distributions.

The calibration for the {\it{Gauss}}, {\it{Gauss($\rho$)}}, and {\it{Lorentzian($\rho$)}} models is based on the width of the detected light distribution~\cite{PANI2016,LERCHE2009624}. For the calibration, the width of the detected light distribution, $\sigma$, is calculated as:

\begin{equation}
    \sigma=\sqrt{\sigma^{2}_{x}+\sigma^{2}_{y}-2\rho\sigma_{x}\sigma_{y}}
\end{equation}

wherein the case of the {\it{Gauss}} model the correlation factor, $\rho$, is zero. For these models, the calibration curve in the z-axis, ($\sigma$,z$_{MC}$), is calculated fitting the experimentally reconstructed $\sigma$ distribution for each particular irradiation to a Gaussian function plus a constant value. From the mean value obtained in the fit, the corresponding $\sigma$ is associated to the z$_{MC}$ data-point for the $z$-axis calibration curve.

The calibration curves calculated for the different light-yield models and \lacls crystal thicknesses are displayed in the several panels of Fig.~\ref{fig:doi_calibration}. Panels a), b), c), d), and e) show the calibration curves of {\it{Solid angle}}, {\it{Keras}}, {\it{Gauss}}, {\it{Gauss($\rho$)}} and {\it{Lorentz($\rho$)}} models, respectively. All the calibration curves obtained in this work are not linear, increasing the non-linearity with the $z$-coordinate or the distance from the SiPM. In the case of {\it{Solid Angle}} model, the calibration of the full crystal distance is possible for \lacls crystal thicknesses up to 20~mm. For larger thicknesses, the results indicate that the last 15-20\% of the $z$-distances (first DoI values), cannot be calibrated because of the non-monotonically increasing calibration function. A similar situation is found for the {\it{Keras}} model in panel b). However, in this case the calibration is possible up to a crystal thickness of 25~mm.

%Panel b) The calculated curve presents a compressed but linear dependency for positions close to the SiPM, i. e. for low $z$ values. However, as the $z$ value increases the calibration becomes non-linear. This is particularly true in the last 15-20\% of crystal thickness. For the 25~mm and 30~mm thick crystals the calibration curve becomes a non monotonically increasing function, which makes a z-determination not possible within that z-range of small DOI-values.

%The b) panel displays the calibration curve for the {\it{Keras}} model. The calibration curve is similar to the {\it{Solid angle}} model. In this case, an slightly better performance can be observed for the two thickest crystals in the small DOI range or large $z$-values.

\begin{figure}[htb!]
\begin{center}
\begin{tabular}{c}
  \includegraphics[width=\columnwidth]{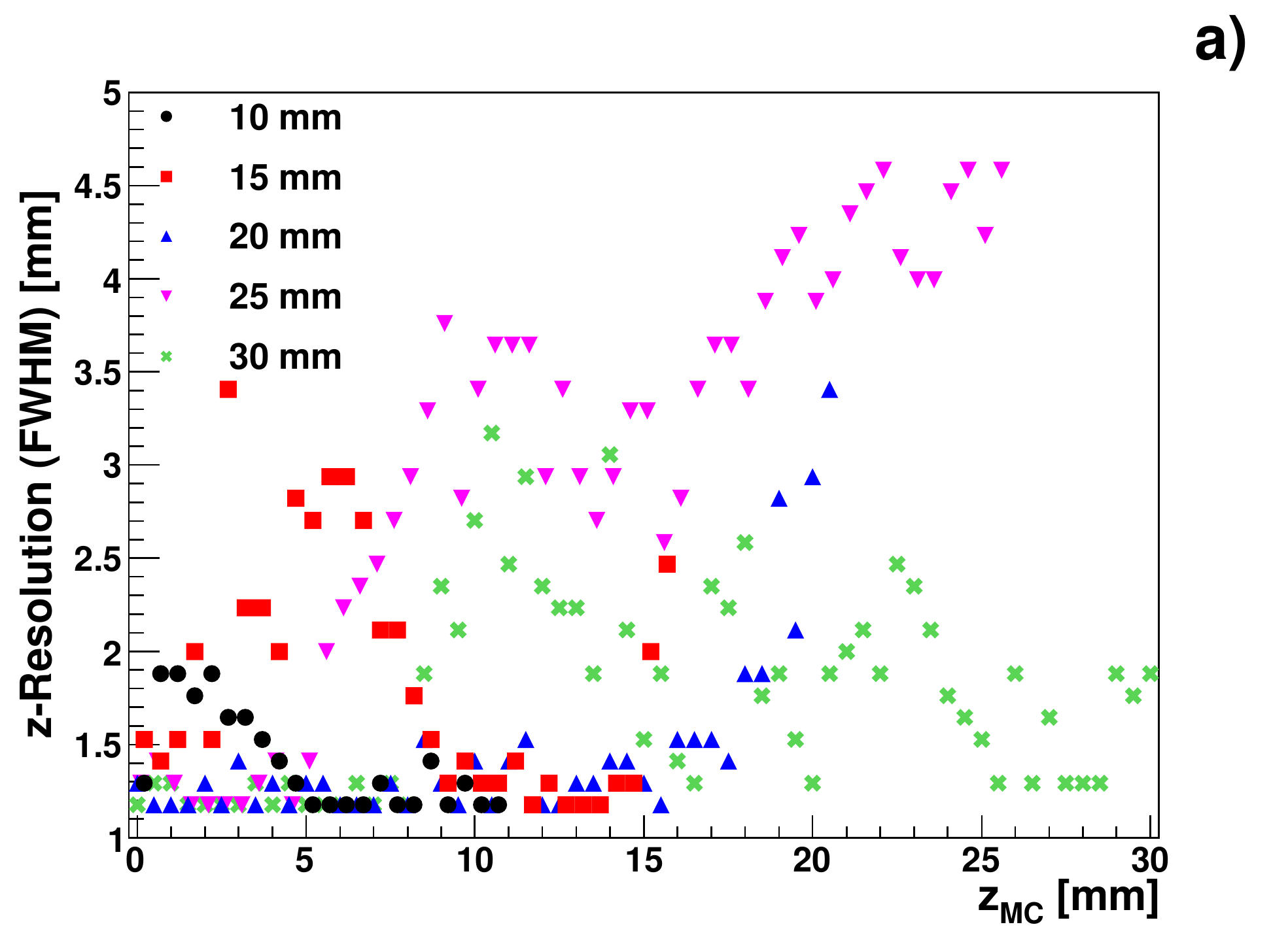} \\
  \includegraphics[width=\columnwidth]{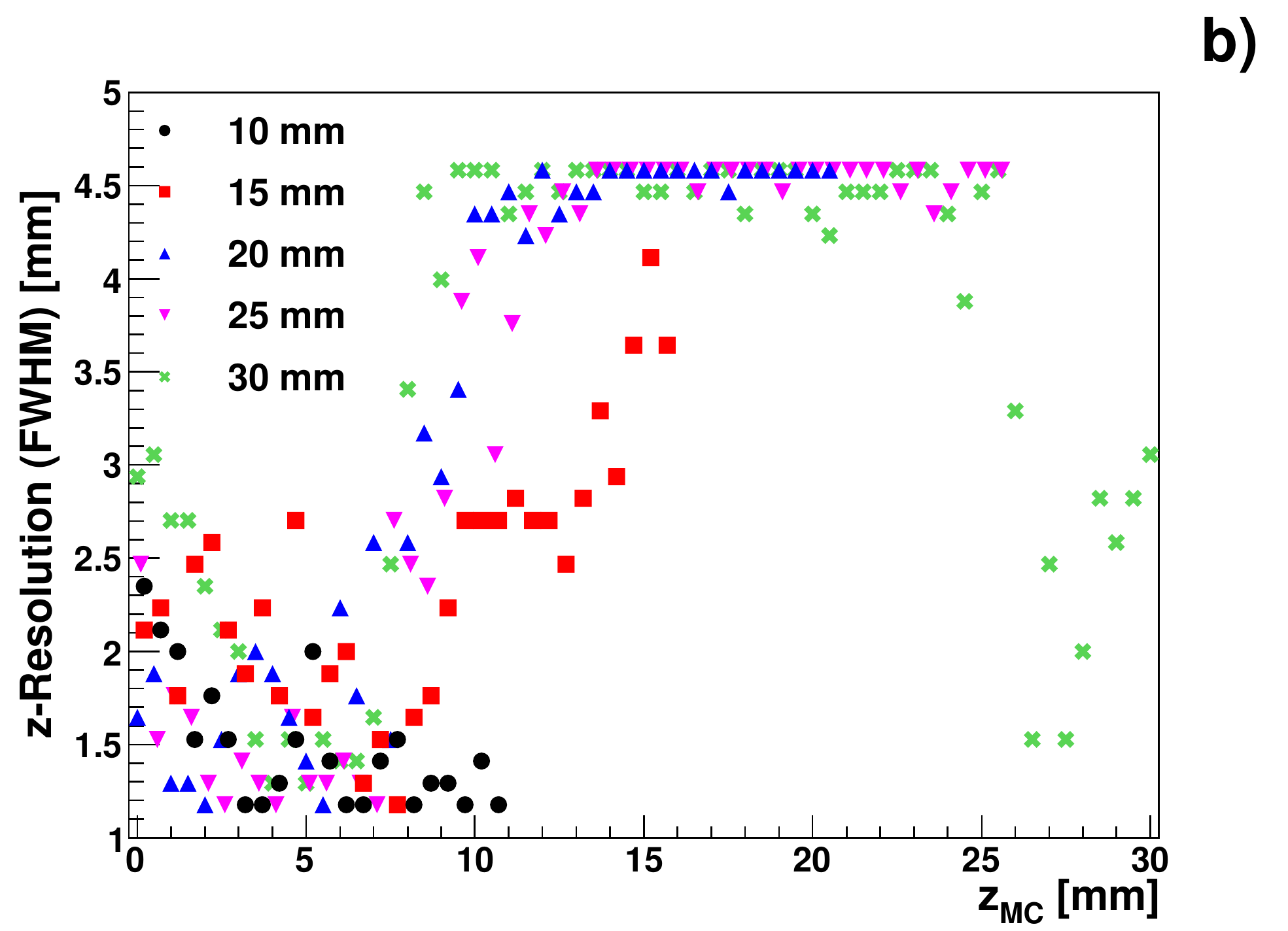} \\
\end{tabular}
 \end{center}
\caption{Spatial resolution (\fwhm) along the $z$-axis for {\it{Solid angle}} model in panel a) and for {\it{Keras}} model in panel b).}
\label{fig:doi_calibration_resolution}
\end{figure}

The calibration curves for the {\it{Gauss}}, {\it{Gauss($\rho$)}} and {\it{Lorentzian($\rho$)}} models are displayed in panels c), d) and e), respectively. For these models, the calibration is less reliable than {\it{Solid Angle}} and {\it{Keras}} models because of the large non-linearity obtained. In fact, all models based on the light-distribution width show the saturation effect in the last 15\% to 20 \% of the crystal thickness, observed in the previous models only for the thickest crystals. This is in agreement with other works using large monolithic crystals, where the small DoI values cannot be calibrated reliably~\cite{PANI2016,Bettiol_2016}. 

The spatial resolution on the third space coordinate $z$ for the {\it{Solid angle}} and {\it{Keras}} models is displayed in Fig.~\ref{fig:doi_calibration_resolution}. It was calculated as the \fwhms of the Gaussian broadening $\sigma_{R}$ adjusted during the $z$-offset fitting procedure. Even with the complex response shown by both curves a trend can be appreciated. The resolution values in the region close to the SiPM (large DoI values) are better, than those obtained at large $z$-value (small DoIs). In general, as the $z$ value increases a worse resolution is obtained for all crystal thicknesses and models. This might be related to the large spread of the scintillation light beyond a certain $z$-value, where the charge distributions are so broad that there is little sensitivity to further variations in the $z$-interaction coordinate.

The values obtained for the {\it{Solid angle}} model are better than those of the CNN-technique based on {\it{Keras}}. As it happens in the other two axis, experimental effects such as pixel thresholds and inhomogeneities in light collection may affect the performance of the {\it{Keras}} model, that is trained with ideal MC responses.

The resolution obtained in this work for the 10~mm thick \lacls crystal is comparable to the results reported in other previous works~\cite{LERCHE2009624,PANI2011324} for similar crystal thicknesses. For the 20~mm thick \lacls crystal the resolution values obtained in this work are better than the results reported in Ref.~\cite{PANI2016} using a 12 $\times$ 12 SiPM array. The results obtained for larger crystal thicknesses are more difficult to compare directly with previous works because scarce studies have been reported thus far for monolithic crystals thicker than 20~mm. Additionally, there are differences in experimental set-up (pixelation, crystal type, finishing and size).

\section{Conclusions and outlook}\label{sec:Conclusions}

In this work, the 3D-position reconstruction in five 50 $\times$ 50 mm$^{2}$ \lacls crystals of different thicknesses, from 10 to 30~mm, have been studied using five different algorithms. Four models were based on analytical prescriptions, and one additional model was based on a convolutional neural network technique. 
Such a large systematic study was possible thanks to the GPU computing implemented in all cases, which allowed us to speed up their performance in a factor 30. This technique can be of interest for applications, such as medical imaging were similar approaches have been implemented~\cite{LERCHE2009359}, or wherever an online real-time position monitoring becomes relevant.

The main results obtained for the $x,y$ transverse crystal plane regarding linearity are graphically displayed in Fig.~\ref{fig:Corrected_linearity_Diagram}. The results found here are comparable to other previous works when similar crystal thicknesses were utilized~\cite{8871159,Li2010,BABIANO20191}. The relative field-of-view of the crystal, relative to the effective crystal surface, is represented in Fig.~\ref{fig:Useful_Field_of_View} as a function of the \lacls crystal thickness. Initially, as expected, a strong compression effect was found with increasing crystal thickness. To overcome this issue in this work we have implemented a Machine Learning solution, SVM-linear, whose results are displayed also in Fig.~\ref{fig:Useful_Field_of_View}. The performance achieved with this method, where comparable in terms of crystal thickness, is similar to other techniques such as Voronoi diagrams~\cite{8871159}, maximum likelihood algorithms~\cite{4782175} and \knns algorithms~\cite{5783323}. The SVM method has allowed us to achieve a relative FoV of about 85\% of the effective crystal surface for crystal thicknesses between 10~mm and 20~m, and of about 70\% for crystal thicknesses of 25~mm and 30~mm.
Finally, this ML-technique allowed us to significantly improve the linearity for all crystal thicknesses, as it can be observed by comparing Fig.~\ref{fig:Corrected_linearity_Diagram} and Fig.~\ref{fig:linearity_Diagram}.

In terms of spatial resolution in the transverse $x,y$ crystal plane, the best overall performance is obtained with the {\it Solid Angle} and {\it Lorentzian ($\rho$)} analytical models (see Fig.~\ref{fig:FWHM}). After implementing the aforementioned linearity corrections, intrinsic average spatial resolutions of $\sim$0.3~mm, 1.7~mm, 1.4~mm, 3.9~mm and 4~mm are obtained across the full detector FoV using the {\it Solid Angle} technique for crystal thicknesses of 10~mm, 15~mm, 20~mm, 25~mm and 30~mm, respectively. 

The {\it Solid Angle} approach was also the best performing in terms of determining the third space coordinate $z$ of the $\gamma$-ray interaction position in the monolithic crystal. Depending on crystal thickness, we estimated $z$-resolutions between 1.5~mm and 4.5~mm \textsc{fwhm}, becoming the uncertainties larger for small DoI values. Fig.~\ref{fig:doi_calibration} shows that the last 15-20\% of the crystal thickness (large $z$-values or equivalently small DoIs) is not accessible by some models (in agreement with other works~\cite{PANI2016,Bettiol_2016}) while with {\it{Solid angle}} and {\it{Keras}} it becomes possible to calibrate in those regions, at least for crystal thicknesses of 10~mm to 20~mm.

In summary, in terms of performance, the different crystal thicknesses can be classified into two groups. Resolutions of about 1-2~mm \fwhms in the transverse plane and relative FoV of about 80\% of the effective crystal base surface can be obtained for crystal thicknesses of up to 20~mm. On the other hand, for thicknesses between 25~mm and 30~mm spatial resolutions of $\sim$4~mm and relative FoV of 70\% can be achieved. In future studies it would be interesting to extend this study to crystals thicker than 30~mm, in order to see where is the limit in terms of reasonable performances. This latter aspect is of interest for nuclear applications like the one pursued in this work, where the $\gamma$-ray energy range commonly extends up to energies of 6-8~MeV and high detection efficiency is required.

\begin{figure}[htb!]
\begin{center}
\begin{tabular}{c}
  \includegraphics[width=\columnwidth]{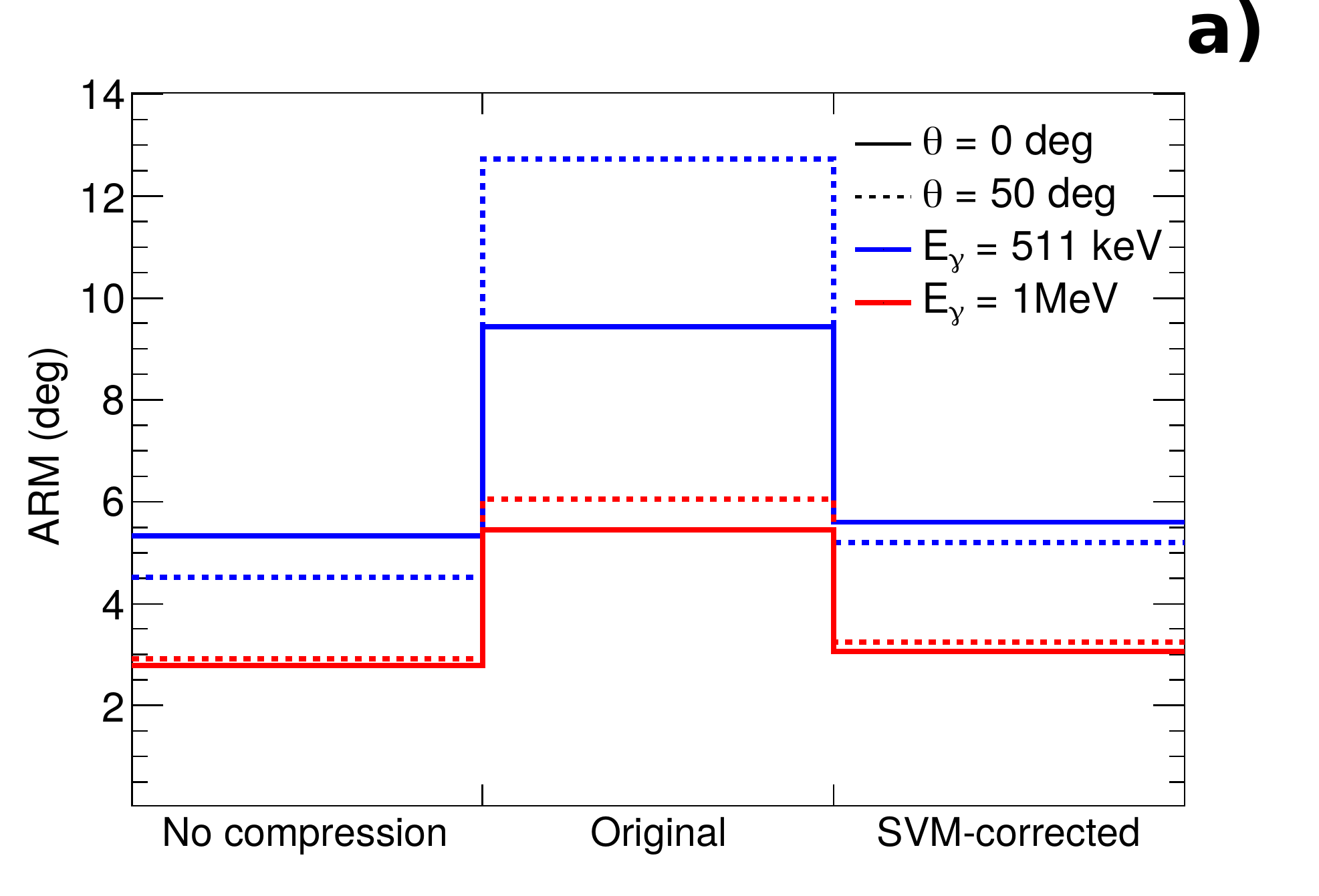} \\
  \includegraphics[width=\columnwidth]{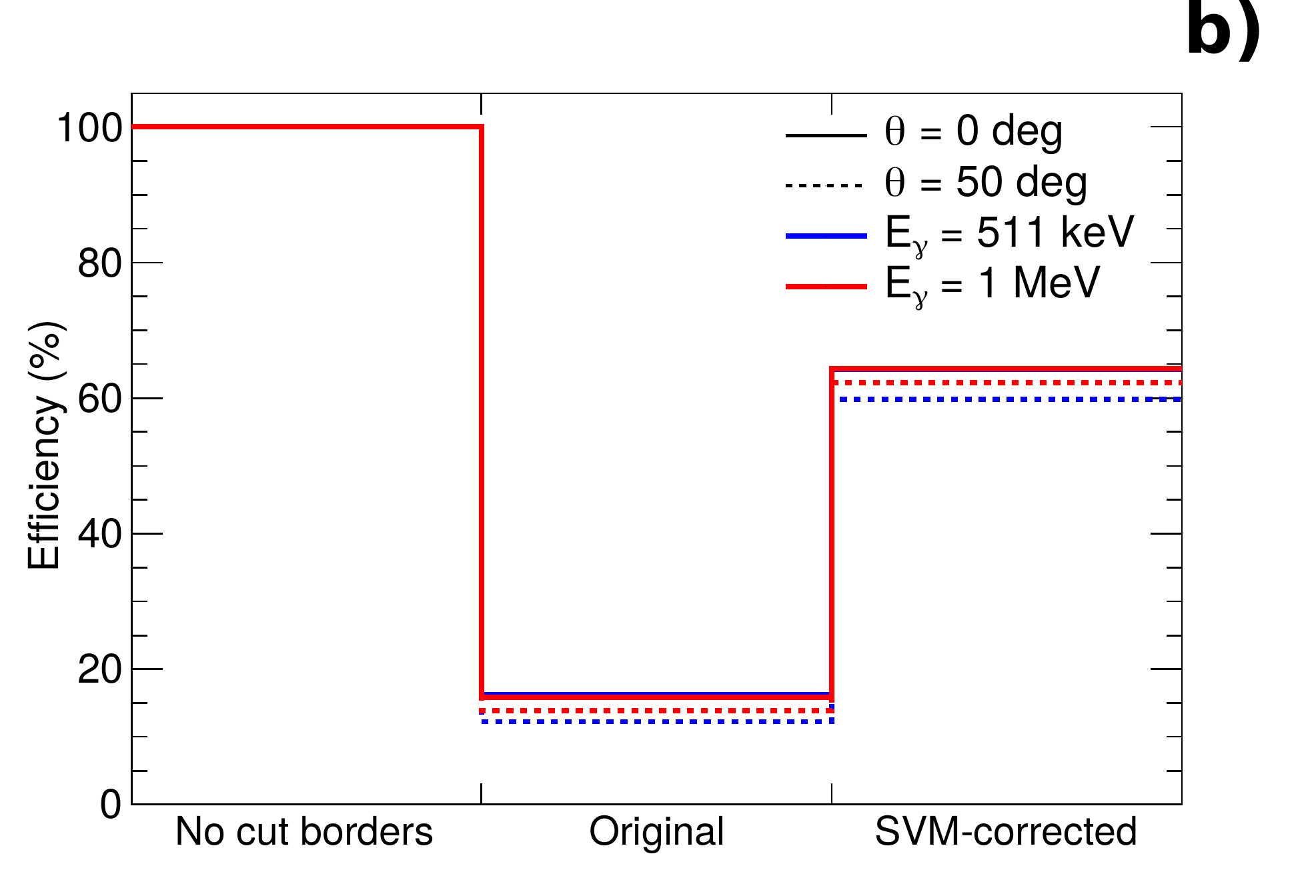} \\
\end{tabular}
 \end{center}
\caption{Impact of the compression in the performance of iTED in terms of angular resolution (panel a)) and efficiency (panel b)). The performance of ideally linear PSDs is compared to the results with the original compression and the upgraded situation after the SVM-linear correction.}
\label{fig:MC_Compression}
\end{figure}

Finally, the impact of these results on the performance of i-TED has been estimated by means of the MC model described in Sec.~\ref{sec:Motivation}. The angular resolution (\fwhms of the ARM distribution) and efficiency for one i-TED Compton module are depicted in Fig.~\ref{fig:MC_Compression} for the ideal case (No compression), before SVM-linearity correction (Original) and after implementing the SVM-correction (SVM-corrected). A significant worsening of the angular resolution is found for the original pin-cushion effects of about 10~mm on each edge of the crystal, which becomes again almost perfect after the SVM correction. The improvement is especially remarkable for low $\gamma$-ray energies and sources placed in peripheral locations ($\theta = 50^{\circ}$). Prior to the application of this ML-aided solution, the only solution to avoid the compression was rejecting the events hitting the compressed area, hence reducing the effective area of the crystals down to 30-40\% for thick crystals (see Fig~\ref{fig:Useful_Field_of_View}). This had an unaffordable cost in the relative imaging efficiency of the i-TED detection system, which dropped down to below 20\%, as shown in Fig.~\ref{fig:MC_Compression}. After the SVM solution, the enhanced FoV increases the imaging efficiency in a factor 3. This will be a key aspect for the feasibility of neutron capture measurements with i-TED at neutron-capture time-of-flight facilities, where the use of beam time for the experiment has to be kept under reasonable limits.

\section*{Acknowledgments}
This work has received funding from the European Research Council (ERC) under the European Union's Horizon 2020 research and innovation programme (ERC Consolidator Grant project HYMNS, with grant agreement nr. 681740). The authors acknowledge support from the Spanish Ministerio de Ciencia e Innovaci\'on under grants FIS2015-71688-ERC, FPA2017-83946-C2-1-P, PID2019-104714GB-C21 and CSIC for funding PIE-201750I26.

\section*{Declaration of competing interest}
The authors declare that they have no known competing financial interests or personal relationships that could have appeared to influence the work reported in this paper.
% https://www.elsevier.com/authors/journal-authors/policies-and-ethics/credit-author-statement

\section*{CRediT authorship contribution statement}
\textbf{J. Balibrea-Correa:} Conceptualization, Investigation, Methodology, Validation, Formal analysis, Data curation, Visualization, Software, Writing - original draft.
\textbf{J. Lerendegui-Marco:} Investigation, Methodology, Formal analysis, Data curation, Visualization, Software, Writing - original draft. 
\textbf{V.~Babiano:} Investigation, Software.  \textbf{L.~Caballero:} Methodology, Investigation. \textbf{D.~Calvo:} Methodology, Investigation. \textbf{I. Ladarescu:} Methodology, Software. \textbf{P. Olleros:} Investigation, Software. \textbf{C. Domingo-Pardo:} Investigation, Software, Methodology, Supervision, Writing -review \& editing, Project administration, Funding acquisition. 

\bibliography{bibliography}

\begin{thebibliography}{10}

\bibitem{RevModPhys.29.547}
E.~M. Burbidge et~al.
\newblock {Synthesis of the Elements in Stars}.
\newblock {\em Rev. Mod. Phys.}, 29:547--650, Oct 1957.

\bibitem{RevModPhys.83.157}
F.~K\"appeler et~al.
\newblock {The $s$-process: Nuclear physics, stellar models, and observations}.
\newblock {\em Rev. Mod. Phys.}, 83:157--193, Apr 2011.

\bibitem{CAMERON1957AJ629C}
A.~G.~W. {Cameron}.
\newblock {On the origin of the heavy elements.}
\newblock {\em aj}, 62:9--10, February 1957.

\bibitem{PhysRev.159.1007}
R.~L. Macklin and J.~H. Gibbons.
\newblock {Capture-Cross-Section Studies for 30---220-keV Neutrons Using a New
  Technique}.
\newblock {\em Phys. Rev.}, 159:1007--1012, Jul 1967.

\bibitem{ABBONDANNO2004454}
U.~Abbondanno et~al.
\newblock {New experimental validation of the pulse height weighting technique
  for capture cross-section measurements}.
\newblock {\em Nuclear Instruments and Methods in Physics Research Section A:
  Accelerators, Spectrometers, Detectors and Associated Equipment}, 521(2):454
  -- 467, 2004.

\bibitem{BORELLA2007626}
A.~Borella et~al.
\newblock {The use of C6D6 detectors for neutron induced capture cross-section
  measurements in the resonance region}.
\newblock {\em Nuclear Instruments and Methods in Physics Research Section A:
  Accelerators, Spectrometers, Detectors and Associated Equipment}, 577(3):626
  -- 640, 2007.

\bibitem{HYMNS}
C.~Domingo-Pardo.
\newblock {High-sensitivitY measurements of key stellar nucleo-synthesis
  reactions (HYMNS), ERC-consolidator grant agreement no. 681740}.

\bibitem{DOMINGOPARDO201678}
C.~Domingo-Pardo.
\newblock {i-TED: A novel concept for high-sensitivity (n,$\gamma$)
  cross-section measurements}.
\newblock {\em Nuclear Instruments and Methods in Physics Research Section A:
  Accelerators, Spectrometers, Detectors and Associated Equipment}, 825:78 --
  86, 2016.

\bibitem{BABIANO2020163228}
V.~Babiano et~al.
\newblock {First i-TED demonstrator: A Compton imager with Dynamic Electronic
  Collimation}.
\newblock {\em Nuclear Instruments and Methods in Physics Research Section A:
  Accelerators, Spectrometers, Detectors and Associated Equipment}, 953:163228,
  2020.

\bibitem{Olleros2018}
P.~Olleros et~al.
\newblock {On the performance of large monolithic {LaCl}3(Ce) crystals coupled
  to pixelated silicon photosensors}.
\newblock {\em Journal of Instrumentation}, 13(03):P03014--P03014, mar 2018.

\bibitem{8871159}
M.~{Freire} et~al.
\newblock {Calibration of Gamma Ray Impacts in Monolithic-Based Detectors Using
  Voronoi Diagrams}.
\newblock {\em IEEE Transactions on Radiation and Plasma Medical Sciences},
  4(3):350--360, 2020.

\bibitem{4782175}
W.~C.~J. {Hunter} et~al.
\newblock {Calibration Method for ML Estimation of 3D Interaction Position in a
  Thick Gamma-Ray Detector}.
\newblock {\em IEEE Transactions on Nuclear Science}, 56(1):189--196, 2009.

\bibitem{5783323}
H.~T. {van Dam} et~al.
\newblock {Improved Nearest Neighbor Methods for Gamma Photon Interaction
  Position Determination in Monolithic Scintillator PET Detectors}.
\newblock {\em IEEE Transactions on Nuclear Science}, 58(5):2139--2147, 2011.

\bibitem{Schaart_2009}
D.~R. Schaart et~al.
\newblock {A novel, {SiPM}-array-based, monolithic scintillator detector for
  {PET}}.
\newblock {\em Physics in Medicine and Biology}, 54(11):3501--3512, may 2009.

\bibitem{7012118}
G.~{Borghi} et~al.
\newblock {Experimental Validation of an Efficient Fan-Beam Calibration
  Procedure for $k$-Nearest Neighbor Position Estimation in Monolithic
  Scintillator Detectors}.
\newblock {\em IEEE Transactions on Nuclear Science}, 62(1):57--67, 2015.

\bibitem{Liprandi2017}
S.~Liprandi et~al.
\newblock {Sub-3mm spatial resolution from a large monolithic LaBr3 (Ce)
  scintillator}.
\newblock {\em Current Directions in Biomedical Engineering}, 3(2):655 -- 659,
  01 Sep. 2017.

\bibitem{Li2010}
Z.~Li et~al.
\newblock {Nonlinear least-squares modeling of 3D interaction position in a
  monolithic scintillator block}.
\newblock {\em Physics in Medicine and Biology}, 55(21):6515--6532, oct 2010.

\bibitem{BABIANO20191}
V.~Babiano et~al.
\newblock {$\gamma$-Ray position reconstruction in large monolithic LaCl3(Ce)
  crystals with SiPM readout}.
\newblock {\em Nuclear Instruments and Methods in Physics Research Section A:
  Accelerators, Spectrometers, Detectors and Associated Equipment}, 931:1 --
  22, 2019.

\bibitem{PANI2016}
R.~{Pani} et~al.
\newblock {A Novel Method for $\gamma - \text{photons}$ Depth-of-Interaction
  Detection in Monolithic Scintillation Crystals}.
\newblock {\em IEEE Transactions on Nuclear Science}, 63(5):2487--2495, 2016.

\bibitem{LERCHE2009624}
C.~Lerche et~al.
\newblock {Depth of interaction detection for $\gamma$-ray imaging}.
\newblock {\em Nuclear Instruments and Methods in Physics Research Section A:
  Accelerators, Spectrometers, Detectors and Associated Equipment}, 600(3):624
  -- 634, 2009.

\bibitem{8069405}
M.~{Occhipinti} et~al.
\newblock {Light response estimation and $\gamma$ events reconstruction in
  gamma-detectors based on continuous scintillators and SiPMs}.
\newblock In {\em 2016 IEEE Nuclear Science Symposium, Medical Imaging
  Conference and Room-Temperature Semiconductor Detector Workshop
  (NSS/MIC/RTSD)}, pp. 1--4, 2016.

\bibitem{4774303}
M.~{Mikeli} et~al.
\newblock {A new position reconstruction method for position sensitive
  photomultipliers}.
\newblock In {\em 2008 IEEE Nuclear Science Symposium Conference Record}, pp.
  4736--4741, 2008.

\bibitem{PANI2011324}
R.~Pani et~al.
\newblock {DoI position resolution in a continuous LaBr3(Ce) scintillation
  crystal for $\gamma$-ray imaging}.
\newblock {\em Nuclear Physics B - Proceedings Supplements}, 215(1):324 -- 327,
  2011.
\newblock Proceedings of the 12th Topical Seminar on Innovative Particle and
  Radiation Detectors (IPRD10).

\bibitem{LERCHE1487684}
C.~W. {Lerche} et~al.
\newblock {Depth of $\gamma$-ray interaction within continuous crystals from
  the width of its scintillation light-distribution}.
\newblock {\em IEEE Transactions on Nuclear Science}, 52(3):560--572, 2005.

\bibitem{LERCHE2005326}
C.~Lerche et~al.
\newblock {Depth of interaction detection with enhanced position-sensitive
  proportional resistor network}.
\newblock {\em Nuclear Instruments and Methods in Physics Research Section A:
  Accelerators, Spectrometers, Detectors and Associated Equipment}, 537(1):326
  -- 330, 2005.
\newblock Proceedings of the 7th International Conference on Inorganic
  Scintillators and their Use in Scientific adn Industrial Applications.

\bibitem{Bettiol_2016}
M.~Bettiol et~al.
\newblock {A Depth-of-Interaction encoding method for {SPECT} monolithic
  scintillation detectors}.
\newblock {\em Journal of Instrumentation}, 11(12):C12054--C12054, dec 2016.

\bibitem{6152614}
R.~{Pani} et~al.
\newblock {Continuous DoI determination by gaussian modelling of linear and
  non-linear scintillation light distributions}.
\newblock In {\em 2011 IEEE Nuclear Science Symposium Conference Record}, pp.
  3386--3389, 2011.

\bibitem{Scrimger_1967}
J.~W. Scrimger and R.~G. Baker.
\newblock {Investigation of Light Distribution from Scintillations in a Gamma
  Camera Crystal}.
\newblock {\em Physics in Medicine and Biology}, 12(1):101--103, jan 1967.

\bibitem{8360486}
F.~{Müller} et~al.
\newblock {Gradient Tree Boosting-Based Positioning Method for Monolithic
  Scintillator Crystals in Positron Emission Tomography}.
\newblock {\em IEEE Transactions on Radiation and Plasma Medical Sciences},
  2(5):411--421, 2018.

\bibitem{8554136}
F.~{Müller} et~al.
\newblock {A Novel DOI Positioning Algorithm for Monolithic Scintillator
  Crystals in PET Based on Gradient Tree Boosting}.
\newblock {\em IEEE Transactions on Radiation and Plasma Medical Sciences},
  3(4):465--474, 2019.

\bibitem{4545078}
P.~{Bruyndonckx} et~al.
\newblock {Evaluation of Machine Learning Algorithms for Localization of
  Photons in Undivided Scintillator Blocks for PET Detectors}.
\newblock {\em IEEE Transactions on Nuclear Science}, 55(3):918--924, 2008.

\bibitem{1344371}
P.~{Bruyndonckx} et~al.
\newblock {Neural network-based position estimators for PET detectors using
  monolithic LSO blocks}.
\newblock {\em IEEE Transactions on Nuclear Science}, 51(5):2520--2525, 2004.

\bibitem{Wang_2013}
Y.~Wang et~al.
\newblock {3D position estimation using an artificial neural network for a
  continuous scintillator {PET} detector}.
\newblock {\em Physics in Medicine and Biology}, 58(5):1375--1390, feb 2013.

\bibitem{Iborra_2019}
A.~Iborra et~al.
\newblock {Ensemble of neural networks for 3D position estimation in monolithic
  {PET} detectors}.
\newblock {\em Physics in Medicine {\&} Biology}, 64(19):195010, oct 2019.

\bibitem{9036979}
A.~{LaBella} et~al.
\newblock {Convolutional Neural Network for Crystal Identification and Gamma
  Ray Localization in PET}.
\newblock {\em IEEE Transactions on Radiation and Plasma Medical Sciences},
  4(4):461--469, 2020.

\bibitem{NIPS1989_293}
Y.~LeCun et~al.
\newblock {Handwritten Digit Recognition with a Back-Propagation Network}.
\newblock In D.~S. Touretzky, editor, {\em Advances in Neural Information
  Processing Systems 2}, pp. 396--404. Morgan-Kaufmann, 1990.

\bibitem{ProofProspectsiTED2020}
V.~Babiano et~al.
\newblock {i-TED: First experimental proof-of-concept and future prospects
  based on Machine-Learning techniques}.
\newblock {\em Nuclear Instruments and Methods in Physics Research Section A:
  Accelerators, Spectrometers, Detectors and Associated Equipment},
  (submitted).

\bibitem{ALLISON2016186}
J.~Allison et~al.
\newblock {Recent developments in Geant4}.
\newblock {\em Nuclear Instruments and Methods in Physics Research Section A:
  Accelerators, Spectrometers, Detectors and Associated Equipment}, 835:186 --
  225, 2016.

\bibitem{WILDERMAN1998}
S.~J. {Wilderman} et~al.
\newblock {Fast algorithm for list mode back-projection of Compton scatter
  camera data}.
\newblock {\em IEEE Transactions on Nuclear Science}, 45(3):957--962, June
  1998.

\bibitem{Hosokoshi2019}
H.~Hosokoshi et~al.
\newblock {Development and performance verification of a 3-D position-sensitive
  Compton camera for imaging MeV gamma rays}.
\newblock {\em Scientific Reports}, 9(1):18551, Dec 2019.

\bibitem{SENSL2020}
OnSemiconductor.
\newblock {Sensl documentation}.
\newblock \url{http://sensl.com/documentation}, 2008.
\newblock [Online; accessed 15-July-2020].

\bibitem{DIFRANCESCO2016194}
A.~{Di Francesco} et~al.
\newblock {TOFPET 2: A high-performance circuit for PET time-of-flight}.
\newblock {\em Nuclear Instruments and Methods in Physics Research Section A:
  Accelerators, Spectrometers, Detectors and Associated Equipment}, 824:194 --
  195, 2016.
\newblock Frontier Detectors for Frontier Physics: Proceedings of the 13th Pisa
  Meeting on Advanced Detectors.

\bibitem{Zaber_Gantry}
Zaber.
\newblock T-g-lsm200a200a.
\newblock
  \url{https://www.zaber.com/products/xy-xyz-gantry-systems/T-G-LSM/details/T-G-LSM200A200A},
  2020.

\bibitem{doi:10.1063/1.1715998}
H.~O. Anger.
\newblock {Scintillation Camera}.
\newblock {\em Review of Scientific Instruments}, 29(1):27--33, 1958.

\bibitem{4324123}
H.~O. {Anger}.
\newblock {Sensitivity, Resolution, and Linearity of the Scintillation Camera}.
\newblock {\em IEEE Transactions on Nuclear Science}, 13(3):380--392, 1966.

\bibitem{SHI2019117}
R.~Shi et~al.
\newblock {Experimental evaluation of reconstruction algorithms for
  scintillation crystal array based on charge projection readout}.
\newblock {\em Nuclear Instruments and Methods in Physics Research Section A:
  Accelerators, Spectrometers, Detectors and Associated Equipment}, 937:117 --
  124, 2019.

\bibitem{GOTOH1971485}
H.~Gotoh and H.~Yagi.
\newblock {Solid angle subtended by a rectangular slit}.
\newblock {\em Nuclear Instruments and Methods}, 96(3):485 -- 486, 1971.

\bibitem{Mikiko2010}
M.~Ito et~al.
\newblock {Design and simulation of a novel method for determining
  depth-of-interaction in a PET scintillation crystal array using a
  single-ended readout by a multi-anode PMT}.
\newblock {\em Physics in medicine and biology}, 55:3827--41, 07 2010.

\bibitem{10.1145/1365490.1365500}
J.~Nickolls et~al.
\newblock {Scalable Parallel Programming with CUDA}.
\newblock {\em Queue}, 6(2):40–53, March 2008.

\bibitem{PRZYBYLSKY2017}
K.-F. Przybylski~A., Thiel~B. and J.~et~al.
\newblock {Gpufit: An open-source toolkit for GPU-accelerated curve fitting}.
\newblock {\em Science Report}, 7:15722, 2017.

\bibitem{LERCHE2009359}
C.~Lerche et~al.
\newblock {Maximum likelihood positioning for gamma-ray imaging detectors with
  depth of interaction measurement}.
\newblock {\em Nuclear Instruments and Methods in Physics Research Section A:
  Accelerators, Spectrometers, Detectors and Associated Equipment}, 604(1):359
  -- 362, 2009.
\newblock PSD8.

\bibitem{BRUYNDONCKX2007304}
P.~Bruyndonckx et~al.
\newblock {Investigation of an in situ position calibration method for
  continuous crystal-based PET detectors}.
\newblock {\em Nuclear Instruments and Methods in Physics Research Section A:
  Accelerators, Spectrometers, Detectors and Associated Equipment}, 571(1):304
  -- 307, 2007.
\newblock Proceedings of the 1st International Conference on Molecular Imaging
  Technology.

\bibitem{MATEO2009366}
F.~Mateo et~al.
\newblock {High-precision position estimation in PET using artificial neural
  networks}.
\newblock {\em Nuclear Instruments and Methods in Physics Research Section A:
  Accelerators, Spectrometers, Detectors and Associated Equipment}, 604(1):366
  -- 369, 2009.
\newblock PSD8.

\bibitem{tensorflow2015-whitepaper}
M.~Abadi et~al.
\newblock { {TensorFlow}: Large-Scale Machine Learning on Heterogeneous
  Systems}, 2015.
\newblock Software available from tensorflow.org.

\bibitem{chollet2015keras}
F.~Chollet et~al.
\newblock {Keras}.
\newblock \url{https://keras.io}, 2015.

\bibitem{kingma2017adam}
D.~P. Kingma and J.~Ba.
\newblock {Adam: A Method for Stochastic Optimization}, 2017.

\bibitem{RobustCovarianceRousseeuw}
P.~J. Rousseeuw.
\newblock {Least Median of Squares Regression}.
\newblock {\em Journal of the American Statistical Association},
  79(388):871--880, 1984.

\bibitem{Chi2Wilson}
H.~M. Wilson~EB.
\newblock {The Distribution of Chi-Square.}
\newblock {\em Proc Natl Acad Sci U S A.}, 17:684--688, 1931.

\bibitem{scikit-learn}
F.~Pedregosa et~al.
\newblock {Scikit-learn: Machine Learning in {P}ython}.
\newblock {\em Journal of Machine Learning Research}, 12:2825--2830, 2011.

\bibitem{sklearn_api}
L.~Buitinck et~al.
\newblock "{API} design for machine learning software: experiences from the
  scikit-learn project".
\newblock In {\em ECML PKDD Workshop: Languages for Data Mining and Machine
  Learning}, pp. 108--122, 2013.

\bibitem{Morrocchi_2016}
M.~Morrocchi et~al.
\newblock {Evaluation of event position reconstruction in monolithic crystals
  that are optically coupled}.
\newblock {\em Physics in Medicine and Biology}, 61(23):8298--8320, nov 2016.

\bibitem{CABELLO2013148}
J.~Cabello et~al.
\newblock {High resolution detectors based on continuous crystals and SiPMs for
  small animal PET}.
\newblock {\em Nuclear Instruments and Methods in Physics Research Section A:
  Accelerators, Spectrometers, Detectors and Associated Equipment}, 718:148 --
  150, 2013.
\newblock Proceedings of the 12th Pisa Meeting on Advanced Detectors.

\bibitem{Agiaz2019IEEE}
A.~{Giaz} et~al.
\newblock {3''$\times$ 3'' LaBr$_{3}$:Ce position sensitivity with multi-anode
  PMT readout}.
\newblock In {\em 2014 IEEE Nuclear Science Symposium and Medical Imaging
  Conference (NSS/MIC)}, pp. 1--5, 2014.

\bibitem{Smola04atutorial}
A.~J. Smola and B.~Schölkopf.
\newblock {A tutorial on support vector regression }, 2004.

\bibitem{10.5555/1162264}
C.~M. Bishop.
\newblock {\em {Pattern Recognition and Machine Learning (Information Science
  and Statistics)}}.
\newblock Springer-Verlag, Berlin, Heidelberg, 2006.

\bibitem{Seifert_2013}
S.~Seifert et~al.
\newblock {First characterization of a digital {SiPM} based time-of-flight
  {PET} detector with 1 mm spatial resolution}.
\newblock {\em Physics in Medicine and Biology}, 58(9):3061--3074, apr 2013.

\bibitem{Borghi_2016}
G.~Borghi et~al.
\newblock {\em Physics in Medicine and Biology}, 61(13):4929--4949, jun 2016.

\end{thebibliography}

\end{document}